%% file: article_DAM_v_finale_joli.tex
\documentclass[12pt,a4paper,fleqn]{article}

\usepackage[left=1.5cm,right=1.5cm,top=2cm,bottom=2cm]{geometry}
\usepackage [utf8]{inputenc}
\usepackage[english]{babel}
\usepackage[T1]{fontenc}
\usepackage{amsmath,amssymb}
\usepackage{amsfonts}
\usepackage{stmaryrd}
\usepackage{xspace}
\usepackage{thmbox_aef}
\usepackage{array,multirow,tabularx,ltablex,hhline}
\usepackage{cancel}
\usepackage{bbold} 
\usepackage{tikz}
\usetikzlibrary{decorations.pathreplacing,patterns}
\usepackage{lineno}

\DeclareMathAlphabet{\mathsfit}{T1}{\sfdefault}{\mddefault}{\sldefault}

\definecolor{turquoise}{RGB}{0, 204, 203} 
\definecolor{violet}{RGB}{128, 0, 128} 
\definecolor{rose}{RGB}{255, 0, 127} 
\definecolor{vert}{RGB}{194, 247, 50}
\definecolor{orange}{RGB}{250, 164, 1}
\definecolor{jaune}{RGB}{255, 195, 0 }
\colorlet{expli}{gray}

\input{notations.tex}

\input{dominances.tex}
\newcommand{\imp}[1]{\textit{#1}}
\newcommand{\ie}{\textit{i.e.}\xspace}
\newcommand{\cf}{\textit{cf.}\xspace}

\begin{document}
\newcommand{\blue}[1]{#1}
\newcommand{\modif}[1]{#1}
\newcommand{\red}[1]{\textcolor{red}{#1}}

\newtheorem{df}{Definition}
\newtheorem[cut=false]{pte}[df]{Property}
\newtheorem[cut=false]{cor}[df]{Corollary}
\newtheorem[cut=false]{lm}[df]{Lemma}
\newtheorem[cut=false]{thm}[df]{Theorem}
\renewcommand{\claimname}{Claim}

\input{blocs_inegalites.tex}

\input{figures_ordos.tex}
\input{figures_autres.tex}

\title{ 
Mixed integer formulations using natural variables \\
for single machine scheduling around a common due date
}

\author{Anne-Elisabeth Falq $^a$,
Pierre Fouilhoux $^a$, 
Safia Kedad-Sidhoum$^b$\\[0.2cm]
{\normalsize
$^a$ Sorbonne Universit\'e, CNRS, LIP6, 4 Place Jussieu, 75005 Paris, France}\\
{\normalsize
$^b$ CNAM, CEDRIC, 292 rue Saint Martin, 75141 Paris Cedex 03, France}
\\
}

%
%
%
%

\date{} 

\maketitle

\begin{abstract}
While almost all existing works
which optimally solve just-in-time scheduling problems
propose dedicated algorithmic approaches,
we propose in this work mixed integer formulations.
We consider a single machine scheduling problem that aims at minimizing
the weighted sum of earliness tardiness penalties around a common due date.
Using natural variables,
we provide one compact formulation for the unrestrictive case and,
for the general case, a non-compact formulation based on non-overlapping inequalities.
We show that the separation problem related to the latter formulation is solved polynomially.
In this formulation, solutions are only encoded by extreme points.
We establish a theoretical framework to show the validity of such a formulation using
non-overlapping inequalities, which could be used for other scheduling problems.
A Branch-and-Cut algorithm together with
an experimental analysis are proposed to assess the practical
relevance of this mixed integer programming based methods.
\end{abstract}

\textbf{Keywords}:
Just-in-time scheduling,
Mixed integer programming formulation,
polyhedral approaches

\section{Introduction}
In the most general statement, single-machine scheduling
is to process a set $J$ of tasks non-preemptively on a single machine.
Each task $j\!\in\! J$  is ready for processing at time zero 
and has a processing time $p_j$, 
that is neither time-dependent nor sequence-dependent
(w.l.o.g. we assume that $p_j \!\geqslant\!1$).

A schedule can be then encoded by
the vector of its completion times $(C_j)_{j\in J}$.
Such an encoding allows us to express
a wide range of criteria,
particularly the so-called regular criteria,
which are decreasing functions of $C_j$ for each task $j$.
Using these continuous variables,
Queyranne~\cite{Queyranne} provided useful polyhedral tools 
for minimizing one of the most studied regular criteria: 
$\sum \omega_j C_j$.
To the best of our knowledge,
the scheduling literature lacks  
similar results for non-regular criteria.
The contribution of this work falls within this scope.
Our focus is on minimizing a non-regular criterion
occurring in just-in-time scheduling.

We consider a single machine scheduling problem where
all tasks share a common due date $d$. 
A task $j\!\in\! J$ is early (resp. tardy) 
if  $C_j\!\leq\! d$ (resp.  $C_j\!>\! d$).
Using $[x]^+$  to denote the positive part of $x \!\in\! \mathbb{R}$,
the earliness (resp. tardiness) of any task $j\!\in\! J$ 
is given by $[d\!-\!C_j]^+$ (resp. $[C_j\!-\!d]^+$).
Given unit earliness penalties $(\alpha_j)_{j\in J}$ 
(resp. tardiness penalties $(\beta_j)_{j\in J}$),
the problem aims at finding a schedule 
that minimizes the total penalty defined as follows.\\[-0.3cm]
$$
\fab(C) = \sum\limits_{j\in J}
\left( \alpha_j\, [d\!-\!C_j]^+ +\beta_j\, [C_j\!- \!d]^+ \right)
$$

When $d \!\geqslant\! \sum p_j$, 
the common due date is called \imp{unrestrictive} 
since the due date does not restrict 
the total duration of early tasks~\cite{Hall_et_Posner}.
In this case, 
the so-called V-shaped dominance property~\cite{Hall_et_Posner}
ensures that there exists 
an optimal solution such that early tasks are scheduled 
by increasing ratio $\alpha_j/p_j$ 
while tardy tasks are scheduled 
by decreasing ratio $\beta_j/p_j$. 
In addition,
according to some strong dominance properties~\cite{Hall_et_Posner}, 
there exists an optimal schedule without idle time 
and with an \imp{on-time} task, \ie completing exactly at $d$.
\modif{For the common due date setting}, idle time only refers to an idle time between tasks,
 regardless of the interval between 0 and the starting time of the schedule.
The problem with an unrestrictive common due date is NP-hard 
even if $\alpha_j\!=\!\beta_j$ for any task $j\!\in\! J$~\cite{Hall_et_Posner}.
However, if $\alpha_j\!=\!\beta_j \!= \!1$ for any task $j\!\in\!J$, 
the problem is solvable in polynomial time~\cite{Kanet}. 

In the general case, there might be a \imp{straddling} task,
\ie a task starting before $d$ and completing after $d$,
in all optimal schedules:
the problem is shown to be NP-hard,
even if $\alpha_j\!=\!\beta_j=1$ 
for all $j\!\in\! J$~\cite{HVDV,Hall_Kubiak_et_Sethi}.

In addition to these fundamental results of the common due date problem,
the just-in-time field scheduling benefits from a rich literature.
These problems have been solved by several approaches: 
with heuristics 
(\textit{e.g.}~\cite{benchmark},~\cite{survey_17}),
with branch-and-bound algorithms 
(\textit{e.g.}~\cite{Sourd_09}),
and with dynamic programming methods 
(\textit{e.g.}~\cite{HVDV},~\cite{Tanaka_Araki_13}).
The reader can refer to the seminal surveys 
of~\cite{Baker_Scudder_90},~\cite{survey_17} and~\cite{survey_2000}
for the early results in this field.\\

Furthermore, there exist several ways to encode a single machine schedule 
leading to distinct formulations.
Such encodings can be based on completion times,
time-indexed variables,
linear ordering,
positional date and assignement variables~\cite{Queyranne_Schulz_94}.
Some of these encodings allow to formulate 
just-in-time scheduling problems as Mixed Integer Program (MIP).
However, few solving approaches based on these formulations have been proposed 
for just-in-time scheduling problems~\cite{benchmark}.

We focus in this article on natural variables,
similar to completion times variables.
To the best of our knowledge,
no linear formulation with such variables
has been considered for just-in-time scheduling,
in contrast with scheduling problems
dealing with regular criteria.
Since tasks have to be processed on a single machine,
a schedule is feasible if it satisfies the task non-overlapping,
\ie if they are executed on disjoint time slots.
Providing a linear formulation of non-overlapping is
an important issue to solve a single-machine scheduling problem
using linear programming.
Studying the polyhedron defined as the convex hull
of the feasible completion times vectors
provides LP or MIP formulations.
\cite{Balas} and~\cite{Queyranne}
propose seminal works in this research line.
The authors consider the problem of minimizing $\sum \omega_j C_j$.
Other works consider the same problem with additional constraints:
release dates
(\textit{e.g.}~\cite{Dyer_Wolsey_90})
or precedence constraints
(\textit{e.g.}~\cite{Correa_Schul_2005},~\cite{Queyranne_Wang_91}).

A particularity of an encoding based on such natural variables 
is the non connectivity of the feasible vectors set.
Therefore, a vector in the convex hull of feasible vectors
can correspond to an infeasible schedule.
In this context, providing a linear formulation 
describing this polyhedron is not sufficient.
\cite{Queyranne} describes
the convex hull of feasible completion times vectors
by linear inequalities, 
and shows that the extreme points of this polyhedron encode
feasible schedules.
He deduces a formulation which can be solved by LP algorithms.
This formulation is an LP with an additional constraint:
the solution must be an extreme point.
This constraint will be called an \imp{extremality constraint}.
\\[-0.1cm]

In this article,
we provide MIP based methods
to solve a core problem in just-in-time scheduling.
Such approaches can be easily extended 
to tackle other variants embedding this core structure,
in contrast with the dedicated methods commonly used 
in scheduling field.
We use natural variables
to handle the common due date problem,
dealing with a non-regular criterion.
Using few additional binary variables,
we describe a polyhedron 
containing the convex hull of dominant vectors
for the unrestrictive case,
and another one for the general case.
We show that, in both cases,
extreme points of this polyhedron 
correspond to feasible schedules.
\modif{Thanks to these theoretical results,}
we derive two non-compact MIP formulations 
with an additional extremality constraint.
We explain how both formulations can be solved
using a branch-and-cut algorithm.
We also propose a compact MIP formulation 
for the unrestrictive case,
which is more efficient 
but cannot be adapted to the general case.
Finally we provide an experimental analysis
to assess the practical relevance of the proposed approaches.
The analysis is based on the reference benchmark proposed by~\cite{benchmark}
and also used by~\cite{Sourd_09}, \modif{as well as a new benchmark covering larger processing times. For sake of comparison, MIP formulations of the literature are also considered.}
\\[-0.1cm]

This article is organized as follows.
Section~\ref{sec_non-ch} presents basic tools
to express the task non-overlapping.
We recall Queyranne's linear inequalities
for the non-overlapping \cite{Queyranne}.
We also provide two lemmas,
which permit to extend the framework
in which those inequalities can be used.
In Sections~\ref{sec_formul1},~\ref{sec_formul2} and \ref{sec_formul3},
we provide new formulations 
for the unrestrictive case and the general one.
In each section we first enunciate dominance properties,
then we give the formulation,
and finally we prove its validity.
All separation algorithms for these formulations
are gathered in Section~\ref{sec_sepa}.
In Section~\ref{sec_exp} we present some experimental results
and compare the different formulations.

\section{Linear inequalities for non-overlapping}
\label{sec_non-ch}
\setcounter{equation}{-1}

For a single-machine problem,
a schedule must only satisfy two conditions to be \imp{feasible}:
each task must begin after time 0 and two tasks must not be executed at the same time.
In the sequel, the first condition will be called \imp{positivity}
and the second one will be called \imp{non-overlapping}.
Given the processing times $p\!\in\! \RpeJ$, a vector $y \!\in\! \mathbb{R}^J$
encodes a feasible schedule by its completion times
if and only if it satisfies the two following constraints.\\[-0.6cm]
\begin{align}
\text{ positivity }\hspace{1.3cm}
\forall j\!\in\! J, \enskip &
y_j \!\geqslant\! p_j \label{pos}\\
\label{non-ch} \text{ non-overlapping }\hspace{0.7cm}
\forall (i,j) \!\in\! J^2,\enskip &
y_j\!\geqslant\! y_i + p_j \text{ or } y_i\!\geqslant\! y_j + p_i
\end{align}

The set $Q$ will denote the set of all vectors 
encoding a feasible schedule by its completion times,
\ie all vectors satisfying constraints~(\ref{pos}) and~(\ref{non-ch}).
Completion times allow an easy way to express feasibility 
at the expense of the non-linearity of constraints~(\ref{non-ch}).
However,
 \cite{Queyranne} introduces linear inequalities
using completion times to handle the non-overlapping.
We first recall notations and results proposed by \cite{Queyranne}
as we will generalize them to a larger framework.
To this end, we use vector $y$ to represent more than completion times.
In the next sections, 
$y$ will be either the earliness or tardiness of tasks.
For $S\subseteq J$ and $y \!\in\! \mathbb{R}^J$,\\[-0.4cm]
$$
S^< \!=\! \big\{ (i,j) \!\in\! S^2 \,|\,  i\!<\!j \big\}
,\enskip\enskip
{y(}S{)} \!=\!
\sum\limits_{i\in S}y_i 
,\enskip\enskip
p{*}y{(}S{)} 
\!=\!
\sum\limits_{i\in S}p_i y_i
,\enskip \text{ and } \enskip
{g_p(} S {)} \!=\!
\frac{1}{2}\left(\sum\limits_{i\in S}p_i\right)^2 \!+
\frac{1}{2}\sum\limits_{i\in S} p_i^2  
.$$
We give some properties about the function $g_p$, 
useful for the next proofs.\\[-0.6cm]
\begin{align}
\forall S\!\subseteq\!J,&\enskip g_p(S)\!=
\!\! \sum\limits_{(i,j)\in S^<}\!\! p_i p_j +\! \sum\limits_{j\in S}\!p_j^2 
\\
\forall S\!\subseteq\!J,&\enskip \forall i\!\in\!J{\setminus S},\enskip 
g_p\big(S\!\sqcup\! \{i\}\big)=g_p(S)+p_i\big( p(S)\!+\! p_i \big)
\label{lm_fonda_g}
\end{align}
The \Q are defined as follows.\\[-0.6cm]
\begin{align}
\forall S\subseteq J,\, \pc{S} \geqslant g_p(S)
\tag{Q0}\label{Q}
\end{align}
We denote by $P^Q$ the polyhedron defined by inequalities~(\ref{Q}).
The following property establishes that
these inequalities are valid for all vectors of $Q$,
inducing $\text{conv}(Q)\!\subseteq\!P^Q$.

\begin{pte}
\textbf{If} $y$ satisfies constraints (\ref{pos}) and (\ref{non-ch}),
\textbf{then} $y$ satisfies inequalities~(\ref{Q}).
\label{lm_valide}
\end{pte}

\begin{proof}[]
Let $S \!\subseteq\! J$.
If $S\!=\!\emptyset$, inequality~(\ref{Q}) is satisfied.
If $S\!=\!\{j\}$, then inequality~(\ref{Q}) is $p_j y_j\!\geqslant\! p_j^2$,
that is $y_j\!\geqslant p_j$ since $p_j\!>\!0$.
So constraints (\ref{pos}) ensure that the inequalities~(\ref{Q}) 
associated to the singletons are all satisfied. 
If $|S| \geqslant 2$, we need to exhibit an order on $J$.
Since processing times are strictly positive, 
the constraints~(\ref{non-ch}) ensure that $(y_j)_{j\in J}$ are distinct 
and so that there exists a (single) total order $\prec$ on $J$ such that 
$i\!\prec\! j \Leftrightarrow y_i\!<\!y_j$.
Then constraints~(\ref{non-ch}) translate into
$\forall (i,j)\!\in\!J^2,\, i\!\prec\!j \Rightarrow y_j\!\geqslant y_i+p_j$.
Using inequalities~(\ref{pos}) we deduce
that $y_j\!\geqslant\! p(I) +p_j$ for $I \!\subseteq\! J$ and $j\!\in\! J$ 
such that $i\!\prec\!j$ for all $i \!\in\! I$.\\
This allows to prove by induction on
the cardinality of $S$ that all inequalities (\ref{Q}) are satisfied.
Indeed let us assume that they are satisfied for all sets of cardinality $k$ 
where $k\!\geqslant\! 1$ and let $S\!\subseteq\!J$ with $|S| \!=\! k\!+\!1$. 
By setting $j\!=\!\max_\prec S$ and $U\!=\!S{\setminus\{j\}}$,
then, on one hand, by induction $p*y(U)\!\geqslant\!g_p(U)$,
and, on the other, by previous arguments $y_j\!\geqslant\!p(U)+p_j$.
Consequently $p*y(S)\!=\!
p*y(U)+p_j y_j \geqslant g_p(U) + p_j\big(\,p(U)\!+\!p_j\,\big) 
\!=\! g_p(S)$ 
using (\ref{lm_fonda_g}),
hence $y$ satisfies the inequality~(\ref{Q}) associated to $S$.
\end{proof} 

%
%
%

\begin{figure}[h]
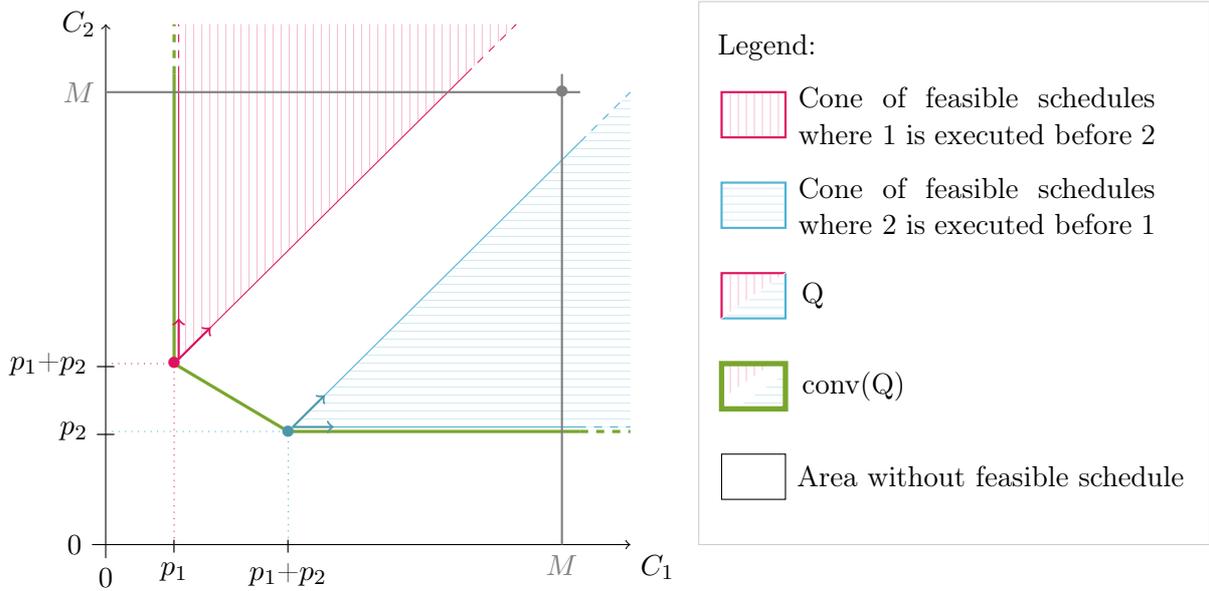

\small
\centering
\figConesBorne\\[-0.4cm]
\caption{$Q$ and conv$(Q)$ in the case of two tasks}
\label{fig_cones_bornes}
\end{figure}

Some points in conv$(Q)$ correspond to infeasible schedules
due to the disjunction inherent to the problem.
Figure \ref{fig_cones_bornes} illustrates $Q$ and $P^Q$ 
for an instance with only two tasks.
The two cones represent the set of feasible schedules:
each corresponding to an order in the task execution.
Vectors in between correspond to schedules where the tasks overlap.
By definition of conv$(Q)$, 
these vectors are in conv$(Q)$,
so they cannot be cut by the \Q.
Note that there are only two extreme points 
and that they correspond to feasible schedules.
This observation is true in general.
Indeed, \cite{Queyranne} shows that
the extreme points of $P^Q$ correspond to feasible schedules.
This inclusion (extr$(P^Q) \!\subseteq\! Q  \!\subseteq\! \text{conv}(Q)$)
and the previous one (conv$(Q) \!\subseteq\! P^Q$) are sufficient to say 
that $\min_{ x\in Q} f(x) = \min_{ x\in P^Q} f(x) $
for any given linear function $f$,
but not sufficient to conclude that $P^Q$ is exactly conv$(Q)$.
\cite{Queyranne} shows this equality using a geometrical argument,
that is the equality of the two recession cones.
The following theorem sums up these results.

\begin{thm} [\cite{Queyranne}]
  (i) extr$(P^Q)\subseteq Q$\\
  (ii) $P^Q=\text{conv}(Q)$
  \label{thm_Q}
\end{thm}

Moreover, \cite{Queyranne} shows that 
each extreme point of $P^Q$ encodes
a \imp{left-tight} schedule,
\ie a feasible schedule without idle time
starting at time zero.
Conversely each left-tight schedule is encoded by
an extreme point of $P^Q$
since, according to the Smith rule~\cite{Smith},
it is the only point in $Q$ 
(and then in conv$(Q) \!=\! P^Q$) minimizing
$\omega \!*\! C (J)$ 
for $\omega \!\in\! \RpJ$ such that
the tasks are scheduled 
by strictly decreasing ratio $\omega_j / p_j$.\\

\begin{figure}[h]
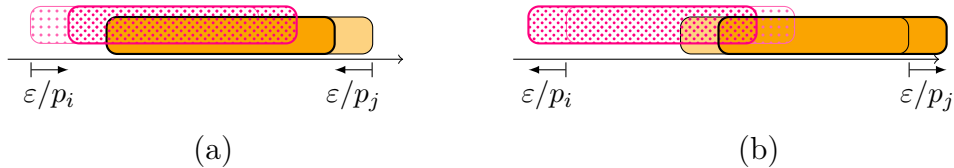

\centering
\figResserre \hspace{0.8cm} \figEcarte\\[-0.4cm]
\caption{Illustration of the schedules disruption 
between $y$ and $y^{+-}$ (a) (resp. $y^{-+}$ (b))}
\label{fig_perturbations}
\end{figure}

We now provide two lemmas which will be the key 
for showing the validity of our formulations.
The first one gives a new proof of Theorem~\ref{thm_Q}(i).
In this lemma, we explain how a vector of $P^Q$
can be slightly disrupted in two directions without leaving $P^Q$
if an overlap is observed in the schedule it encodes.
Figure \ref{fig_perturbations} illustrates the two ways of disrupting the overlapping tasks
so that the corresponding vectors stay in $P^Q$.

\begin{lm}
Let us assume that $y$ satisfies inequalities~(\ref{Q}).\\[.1cm]
\textbf{If} there exists $(i,j)\!\in\! J^2$ with $i\!\neq\! j$
such that $y_i\!\leqslant\! y_j \!<\! y_i + p_j$,\\
\textbf{then} there exists $\varepsilon \!\in\!\Rpe$ 
such that ${y^{+-}}\!=\!y+\!
\frac{\varepsilon}{p_i}\mathbb{1}_i -\! 
\frac{\varepsilon}{p_j}\mathbb{1}_j$
and ${y^{-+}}\!=\!y-\!
\frac{\varepsilon}{p_i}\mathbb{1}_i +\!
\frac{\varepsilon}{p_j}\mathbb{1}_j$
also satisfy~(\ref{Q}). \\[-0.4cm]
\label{lm_cle_1}
\end{lm}

\begin{proof}
Let $\varepsilon\!= \!\min(m_1,m_2)$ 
\begin{tabular}[t]{@{}r@{ }l}
where &
$m_1\!=\!\min\left\lbrace\pc{S}-g_p(S)\, | \, S\!\subseteq\!J, i\!\not\in\!S, j\!\in\!S\right\rbrace$ \\[0.2cm]
and &
$m_2\!=\!\min\left\lbrace\pc{S}-g_p(S)\, | \, S\!\subseteq\!J, i\!\in\!S, j\!\not\in\!S\right\rbrace$.
\end{tabular}\\[0.2cm]
Since $y$ satisfies inequalities~(\ref{Q}), 
$m_1 \!\geqslant \!0$ and $m_2 \!\geqslant \!0$,
thus $\varepsilon \!\geqslant \!0$.\\
Let $S\!\subseteq\!J$.
We first check that vector $y^{+-}$ defined by $\varepsilon$
satisfies inequality~(\ref{Q}) associated to $S$.
\\
If $i \!\not\in\! S$ and $j \!\not\in\! S$ 
then $p*y^{+-}(S) = p*y(S) \geqslant g_p(S)$.\\
If $i \!\in\! S$ and $j \!\in\! S$
then $p*y^{+-}(S) 
= p*y(S) + p_i \frac{\varepsilon}{p_i} - p_j \frac{\varepsilon}{p_j}
= p*y(S) \geqslant g_p(S)$.\\
If $i \!\not\in\! S$ and $j \!\in\! S$
then $p*y^{+-}(S) = p*y(S) - p_j \frac{\varepsilon}{p_j} \geqslant g_p(S)$ 
since $\varepsilon \!\leqslant\! m_1$.\\
If $i \!\in\! S$ and $j \!\not\in\! S$
then $p*y^{+-}(S) = p*y(S) + p_i \frac{\varepsilon}{p_i} 
\geqslant p*y(S)
\geqslant\!g_p(S)$.

In each case $p*y^{+-}(S)\!\geqslant\!g_p(S)$,
then $y^{+-}$ satisfies~(\ref{Q}).
Similarly we can check that
$y^{-+}$ satisfies~(\ref{Q}) using that 
$\varepsilon \!\leqslant\! m_2$.
Finally, we have to check that $\varepsilon \!>\! 0$.
For this purpose we use the next two claims.

\begin{claim}
Let $(i,j)\!\in\! J^2$.
\renewcommand{\arraystretch}{0.3}
\textbf{If} $y_i\!\leqslant\! y_j$,
\textbf{then} $\forall S\!\subseteq\!J,\, i \!\not\in\! S, j\!\in\! S$
$\!\Rightarrow\!\pck{S} \!>\! g_p(S)$.
\end{claim}

\begin{proofclaim}

Let us assume on the contrary that there exists $S\!\subseteq\!J$ 
such that $i\!\not\in\! S$, $j \!\in\! S$  and $\pck{S} \!=\! g_p(S)$.\\
Setting $U\!=\!S{\setminus\{j\}}$, we have 
$\pck{S}\!=\!\pck{U} +p_j y_j$  and 
$g_p(S)\!=\!g_p(U)+p_j\,p(S)$  by~(\ref{lm_fonda_g}).
Since we assume that these two terms are equal,
and since $\pc{U}\! \geqslant\!g_p(U)$ from inequalities~(\ref{Q}),
we deduce that $p_j y_j \!\leqslant\! p_j p(S)$ 
and even $y_j \!\leqslant\! p(S)$ since $p_j\!>\! 0$.\\
Moreover $p\!*\!y \big(S \!\sqcup\! \{i\}\big) 
= \pck{S}+p_i y_i 
= g_p(S)+ p_i y_i 
\leqslant g_p(S)+ p_i y_j$
by assumption.\\
Using these two inequalities, we get 
$p\!*\!y \big(S \!\sqcup\! \{i\}\big) 
\leqslant g_p(S)+ p_i p(S) 
< g_p(S)+ p_i \big [p(S) \!+\! p_i \big]$ 
since $p_i\!>\! 0$.
Furthermore,
$g_p(S)+ p_i \big[p(S)  \!+\! p_i\big] = g_p\big(S \!\sqcup\! \{i\}\big) $
from~(\ref{lm_fonda_g}) and 
$ g_p\big(S \!\sqcup\! \{i\}\big)  \leqslant p\!*\!y \big(S \!\sqcup\! \{i\}\big)  $ from inequality~(\ref{Q}).
We finally get $ p\!*\!y \big(S\!\sqcup\!\{i\}\big) \!<\!  p\!*\!y \big(S\!\sqcup\!\{i\}\big)$,
a contradiction.
\end{proofclaim}
\vspace*{-0.4cm}
This first claim ensures that $m_1 \!>\!0$.

\begin{claim}
Let $(i,j)\!\in\! J^2$.
\renewcommand{\arraystretch}{0.4}
\textbf{If} $y_j \! <\!  y_i+p_j$,
\textbf{then} $\forall S\!\subseteq\!J,\, i \!\in\! S, j\!\not\in\! S$
$\!\Rightarrow\!\pck{S} \!>\! g_p(S)  $ 
\end{claim}

\begin{proofclaim}
Let us assume on the contrary that there exists 
$S\!\subseteq\!J$ such that 
$i \!\in\! S$, $j\!\not\in\! S$ 
and $\pck{S} \!=\! g_p(S)$.\\
Like in the previous proof we can show that
$y_i \!\leqslant\! p(S)$.\\
Moreover $p\!*\!y \big(S \!\sqcup\! \{j\}\big)  
= \pck{S}+p_j y_j = g_p(S)+ p_j y_j 
< g_p(S)+ p_j \big[ y_i \!+\! p_j\big]$
by assumption.\\
Using these two inequalities, we can write  
$\pck{S\!\sqcup\!\{j\}} < g_p(S)+ p_j \big[p(S) \!+\!p_j\big]$ 
since $p_j\!>\! 0$.
Furthermore,
$g_p(S)+ p_j \big[p(S)+p_j\big] = g_p \big(S \!\sqcup\! \{j\}\big)$
from~(\ref{lm_fonda_g}) and 
$ g_p \big(S \!\sqcup \! \{j\}\big) \leqslant p\!*\!y \big(S \!\sqcup\! \{j\}\big) $ from inequality~(\ref{Q}). 
We finally get 
$p\!*\!y \big(S \!\sqcup\! \{j\}\big) \!<\! \pc{S\!\sqcup\!\{j\}}$,
a contradiction.
\end{proofclaim}
\vspace*{-0.4cm}
This second claim ensures that $m_2 \!>\!0$, we can deduce that $\varepsilon \!>\! 0$.
\end{proof}

To obtain an alternative proof of  Theorem~\ref{thm_Q}(i),
Lemma~\ref{lm_cle_1} can be reformulated as follows.
If $C$ is a vector of $P^Q$ 
that gives the completion times of a schedule with an overlap,
then $C$ is the middle of two other vectors of $P^Q$,
$C^{+-}$ and $C^{-+}$ .
That implies that $C$ is not an extreme point of $P^Q$.
By contraposition, we deduce that 
an extreme point of $P^Q$ encodes a schedule without overlap,
and since inequalities~(\ref{Q}) associated to singletons ensure the positivity,
an extreme point of $P^Q$  encodes a feasible schedule,
\ie extr$(P^Q) \!\subseteq\! Q$.

This way of proving that
the extreme points correspond to feasible schedules
can be adapted to a more complex polyhedron,
that is a polyhedron defined by inequalities~(\ref{Q}) 
and additional inequalities.
Indeed, it is then sufficient to check that 
the two vectors $C^{+-}$ and $C^{-+}$ 
also satisfy these additional inequalities.
However, for some extreme points,
the two vectors introduced by Lemma~\ref{lm_cle_1}
may not satisfy the additional inequalities.
For example, if the completion times of the tasks
are limited by a constant $M$ 
(with $M \!\geqslant\! p(J)$),
the additional inequalities are the following.\\[-0.25cm]
\begin{equation} 
\forall j \!\in\! J,\, C_j \!\leqslant\! M
\label{borne_M}
\end{equation}

Note that inequalities~\eqref{borne_M} induce extreme points encoding infeasible schedules as depicted in Figure \ref{fig_cones_bornes} for a 2-task instance.
Adding the inequalities 
$C_1 \!\leqslant\! M$ and $C_2 \!\leqslant\! M$ 
leads to the extreme point $(M,M)$ 
which encodes a schedule with an overlap.
We can see that this point will never be proposed 
as an optimum during the minimization of 
$\omega_1 C_1 \!+\! \omega_2 C_2$ 
if $\omega \! \in\! {(\Rpe)}^2$.  
In general, the aim is to minimize a non-negatively weighted sum of variables.
For any given polyhedron $P$ of $\mathbb{R}^n$,
we consider the following set of extreme points
which are unique minimizer of such function.
The unicity is required 
to deal with some zero weights.
\\[-0.2cm]
$$\extre{P}=
\left\lbrace
\begin{tabular}{@{ }c|l@{ }}
$x^* \!\in\! P $&
$\exists\, \omega \!\in\! \Rp^n,\,  
\{x^*\} \!=\! \argmin\limits_{x\in P} 
\textstyle\sum\limits_{i=1}^{n}\! \omega_i x_i$
\end{tabular}
\right\rbrace 
$$
Since the extreme points are exactly the points that can  be written
as the unique minimizer of a linear function, 
\hbox{$\extre{P}\!\subseteq\!\extr{P}$}.
Let $P^{Q,M}$ denote the polytope defined 
by inequalities~(\ref{Q}) and~(\ref{borne_M}).
Let us assume that $C \!\in\! P^{Q,M}$
is the completion time vector of
a schedule with an overlap.
If one of the two overlapping tasks
has a completion time equal to  $M$,
applying Lemma~\ref{lm_cle_1} to $C$
provides a vector $C^{-+}$ which 
does not satisfy  inequalities~(\ref{borne_M})
and therefore is not in $P^{Q,M}$.
Point $y$ cannot be proved to not be extreme in $P^{Q,M}$.
In order to prove that such point
is not a unique minimizer,
we provide the following lemma.


\begin{lm}
Let us assume that $y$ satisfies inequalities~(\ref{Q}).\\[0.1cm]
\textbf{If} there exist $(i,j) \!\in\!J^2$ 
with $i \!\neq\! j$ such that
$ y_j \!< y_i \!+\! p_j$,
and $y_j \!\geqslant\! p(J) $,\\
\textbf{then} there exists $\varepsilon \!\in\! \Rpe$ 
such that  $y\!-\! \frac{\varepsilon}{p_j} \mathbb{1}_j$ also satisfies inequalities~(\ref{Q}).\\[-0.3cm]
\label{lm_cle_2}
\end{lm}

\begin{proof}
Since $y$ satisfies inequalities~(\ref{Q}),
setting $\varepsilon\!=\! \min\{\, \pc{S}-g_p(S) \,|\, S \!\subseteq\! J ,\, j \!\in\! S\,\}$ 
suffices to ensure that $y\!-\! \frac{\varepsilon}{p_j} \mathbb{1}_j$ 
also satisfies  inequalities~(\ref{Q}) 
and that  $\varepsilon \!\geqslant\! 0$. 
It remains to show that $\varepsilon \!>\! 0$, 
that is for any subset $S \!\subseteq\! J$ containing $j$,
the associated inequality~(\ref{Q}) is not tight.\\
Let $S \!\subseteq\! J$ such that $j \!\in\! S$ 
and let $U \!=\! S\setminus\{j\}$.
First remark the following equivalent inequalities.\\[-0.4cm]
$$
\pc{S} > g_p(S) 
\Leftrightarrow \pc{U} + p_j y_j  > g_p(U) + p_j \big[ p(U) \!+\! p_j\big]\\
\Leftrightarrow \pc{U} -  g_p(U) >  p_j \big[ p(S) - y_j\big]
$$
If $S \!\varsubsetneq\! J$,
then $p(S) \!<\! p(J) \!\leqslant\! y_j $,
thus $p_j \big[ p(S) - y_j\big] \!<\! 0$. 
Moreover $\pc{U} -  g_p(U)  \!\geqslant\! 0$ 
since $y$ satisfies the inequality~(\ref{Q}) associated to $T$.
We deduce that $\pc{S} > g_p(S)$ in this case.
\\[0.1cm]
If $S \!=\! J$, 
then $p_j \big[ p(S) - y_j\big] \!\leqslant\! 0$
since $y_j \!\geqslant\! p(J) $.
In this case,
$p_j \big[ p(S) - y_j\big]$ can be equal to zero 
if $y_j \!=\! p(J) $,
but we prove that $\pc{U} -  g_p(U) \!>\! 0$ as follows.\\[-0.6cm]
\begin{align*}
\pc{U} -  g_p(U) > 0 
&\Leftrightarrow
p \!*\! y \big(J \!\setminus\! \{j\} \big) >  g_p\big(J \!\setminus\! \{j\} \big)\\
&\Leftrightarrow
p \!*\! y \big(J \!\setminus\! \{i,j\} \big) + p_i y_i
> g_p\big(J \!\setminus\! \{i,j\} \big) 
+ p_i \big[ p\big(J \!\setminus\! \{i,j\}\big) \!+\! p_i\big]\\
&\Leftrightarrow
p \!*\! y \big(J \!\setminus\! \{i,j\} \big) - g_p\big(J \!\setminus\! \{i,j\} \big) 
> p_i \big[ p\big(J \!\setminus\! \{j\}\big) - y_i\big]
\end{align*}
By assumption 
$y_i > y_j \!-\! p_j  \geqslant p(J)  \!-\! p_j  =  p\big(J \!\setminus\! \{j\}\big)$,
thus $p_i \big[ p\big(J \!\setminus\! \{j\}\big) - y_i\big] < 0$
and since $y$ also satisfies  
the inequality~(\ref{Q}) associated to $J \!\setminus\! \{i,j\}$,
we have
$p\!*\!y \big(J\!\setminus\!\{i,j\} \big) - g_p\big(J\!\setminus\!\{i,j\} \big) \geqslant 0 $.
We deduce that $\pc{U} -  g_p(U) > 0 $ in this case, 
and finally that  $\pc{S} > g_p(S)$. 
\end{proof}

Combining Lemmas~\ref{lm_cle_1} and~\ref{lm_cle_2},
we prove that a vector $C$ in $\extre{P^{Q,M}}$ is in $Q$,
that is it encodes a feasible schedule by its completion times.
Indeed, since such a vector $C$ satisfies inequalities~(\ref{Q}),
an overlap between tasks $i$ and $j$ 
such that $C_i \!\leqslant\! C_j \!<\! C_i \!+\! p_j$ 
contradicts either the extremality of $C$ or its minimality.
If $C_j\!<\! p(J)$, 
we can construct $C^{+-}$ and $C^{-+}$ 
as proposed in Lemma \ref{lm_cle_1}
for $\varepsilon$ set in $]\,0,p(J) \!-\!C_j\,[$,
so that $C^{+-}$ and $C^{-+}$  
satisfy inequalities~(\ref{Q}) and~(\ref{borne_M}).
Thus, $C$ can be written 
as the middle of two other vectors of $P^{Q,M}$,
then it is not an extreme point.
If conversely $C_j \!\geqslant\! p(J)$,
we can construct a vector $C^{-}$  
as proposed in Lemma~\ref{lm_cle_2},
so that $C^{-}$ is component-wise smaller
than $C$ and satisfies inequalities~(\ref{Q}).
Thus, $C^-$ is another point of $P^{Q,M}$,
which has a smaller value than $C$ 
for any linear function with positive (or zero) coefficients, 
then $C$ cannot be the single minimizer 
of such a function on $P^{Q,M}$.
Moreover, using the same argument as for $P^Q$,
we can say that every left-tight schedule is encoded 
by an extreme point of $P^{Q,M}$, 
and even by a vector of $\extre{P^{Q,M}}$.

For the common due date problem, an encoding by completion times
does not lead to a linear objective function
(except in the very particular case where $d \!=\! 0$,
since the tardiness are then equal to the completion times). 
Therefore, we propose in the next sections
a schedule encoding
together with a set of inequalities ensuring that
every minimum extreme point corresponds to a feasible schedule.

\newpage

\section{A first formulation for the unrestrictive common due date problem\\[-0.6cm]}
\label{sec_formul1}
In this section, we consider the common due date problem 
when the due date is unrestrictive, \ie $d\!\geqslant\! p(J)$.
Before providing  the formulation,
we recall some well known dominance properties
which allow not only to reduce the search space
but also to restrict the instances set.\\[-0.6cm]

\subsection{Dominance properties}
We say that a set of solutions is \imp{dominant} 
if it contains (at least) one optimal solution,
and that it is \imp{strictly dominant} 
if it contains all optimal solutions.
In both cases, the search of an optimal solution 
can be limited to the dominant set.

For the common due date scheduling problem,
we define a \imp{\blocSing} as a feasible schedule 
without idle time, 
a \imp{\dSing} as a feasible schedule with an on-time task,
and a \imp{\dblocSing} as a \blocSing which is also a \dSing.
The following lemma gives dominance properties 
for the common due date problem,
already known for symmetric penalties~\cite{Hall_et_Posner}.
These results can be extended to asymmetric penalties, 
using the same task shifting arguments.

\begin{lm}
Let $ \alpha \!\in\!\RpJ,\,
\beta \!\in\!\mathbb{R}_+^J$.\\[0.1cm]
(i) In the general case,
the \blocPlur are dominant 
when minimizing $\fab$.\\
Moreover, if $\alpha \!\in\! (\Rpe)^J$ 
and $\beta \!\in\! (\Rpe)^J,\,$ 
the \blocPlur are strictly dominant.\\[0.1cm]
(ii) In the unrestrictive case, the \dPlur are dominant
when minimizing $\fab$.\\[-0.4cm]
\label{dom_unres}
\end{lm}

Thanks to these dominance properties, 
only \blocPlur will be considered in the sequel,
and  only \dblocPlur in the unrestrictive case.

From Lemma~\ref{dom_unres}, 
in the unrestrictive case we only have to consider  
instances with strictly positive earliness and tardiness penalties, 
\ie with  $\alpha \!\in\! (\Rpe)^J$ 
and $\beta \!\in\! (\Rpe)^J$.
Indeed,
if the tardiness penalty of a task $j \!\in\! J$ is zero,
solving the instance obtained by removing task $j$ 
provides a \dblocSing, 
which is optimal for $J\setminus\{j\}$.
Placing task $j$ at the end of the \dblocSing
does not increase the cost,
since $j$ is then tardy.
Thus, 
the obtained schedule is an optimal \dblocSing.
Conversely, 
if the earliness penalty of a task $j$ is zero,
placing task $j$ at the beginning of 
an optimal \dblocSing for $J\setminus\{j\}$,
which is always possible when $d$ is unrestrictive,
provides an optimal \dblocSing.
Hence,
for the unrestrictive case, we will set 
$\alpha \!\in\! \RpeJ$ and
$\beta \!\in\! \RpeJ$.

\subsection{A natural formulation for the unrestrictive case}
\label{sub_formul_etdx}
~\indent$\bullet$\textit{ A linear objective function using $e$ and $t$ variables}\\
Since earliness and tardiness are not linear 
with respect to completion times,
the objective function $f_{\alpha,\beta}$ is not linear.
Therefore, 
we propose an encoding by earliness and tardiness of each task,
by introducing the corresponding variables:
$(e_j)_{j\in J}$ for the earliness of the tasks,
and  $(t_j)_{j\in J}$ for their tardiness.
In this way,
the total penalty of 
a schedule encoded by vector $(e,t)$ is 
$\gab(e,t) = \sum_{j\in J} \left(\alpha_j\, e_j +\beta_j\, t_j \right)$
which is linear.
\noindent
If~$C$ encodes a schedule by its completion times,
the encoding by earliness and tardiness of this schedule is given by
$\theta (C) = \left(\, \big ([d \!-\! C_j]^+ \big)_{j \in J}, \,
\big ([C_j \!-\! d]^+ \big)_{j \in J} \,\right)$.
Using function $\theta$, 
we have
$\fab \!=\! \gab \circ \theta$.\\[-0.2cm]


$\bullet$ \textit{Consistency between $e$ and $t$ using $\delta$ variables }\\
A vector $(e,t)$ in $\RpJ \!\times\! \RpJ$
is \imp{consistent}
if $ \forall j \!\in\! J$, 
either ($e_j \!\geqslant\! 0$ and $t_j \!=\! 0$)  
or ($e_j \!=\!0$ and $t_j \!\geqslant\! 0$).
There exists $C$ in $\mathbb{R}^J$ 
such that $\theta(C) \!=\! (e,t)$ 
if and only $(e,t)$ is consistent.
In order to ensure consistency,
we introduce 
the following inequalities
using new boolean variables $(\delta_j)_{j\in J}$. 
For each task $j$, $\delta_j$ indicates if $j$ is early.\\[-0.6cm]
\begin{minipage}[t]{0.4\linewidth}
\begin{eqnarray}
&&\forall j \!\in\! J,\enskip e_j \geqslant 0 \label{e0}\\
&&\forall j \!\in\! J,\enskip e_j \leqslant \delta_j \,(p(J)\!-\!p_j) \label{e1}
\end{eqnarray}
\end{minipage}
\hspace*{1.4cm}
\begin{minipage}[t]{0.4\linewidth}
\begin{eqnarray}
&&\forall j \!\in\! J,\enskip t_j \geqslant 0 \label{t0}\\
&&\forall j \!\in\! J,\enskip t_j \leqslant (1 \!-\! \delta_j) \,p(J) \label{t1}
\end{eqnarray}\\[-0.8cm]
\end{minipage}

Inequalities~(\ref{e0}) and (\ref{e1}) 
force $e_j$ to be zero when $\delta_j \!=\! 0$.
Since we only consider \dblocPlur,
$p(J) \!-\! p_j$ is an upper bound 
on the earliness of task $j$.
Thus, inequality~(\ref{e1}) 
does not restrict $e_j$ when $\delta_j \!=\! 1$.
Note that in the unrestrictive case,
$p(J) \!-\! p_j$ is tighter than $d \!-\! p_j$. of 
Similarly, 
inequalities~(\ref{t0}) and (\ref{t1}) force 
$t_j$ to be zero when $\delta_j \!=\! 1$,
without restricting $t_j$ when $\delta_j \!=\! 0$,
since $p(J)$ is an upper bound on the tardiness in a \dblocSing.
Consequently,
we have the following lemma.

\begin{lm}
Let $(e,t,\delta) \!\in\! \mathbb{R}^J \!\times\! \mathbb{R}^J \!\times\! \{0,1\}^J$.\\
\textbf{If} $e,t,\delta$ satisfy inequalities~\eqref{e0}-\,\eqref{t1},
\textbf{then} $(e,t)$ is consistent and  
$C \!=\! \big(d \!-\! e_j \!+\! t_j\big)_{j\in J}$ 
satisfies $\theta(C) \!=\! (e,t)$.\\[-0.4cm]
\label{lm_theta-1}
\end{lm}
For a consistent $(e,t)$ vector, we define
$\theta^{-1}(e,t)=(d \!-\! e_j \!+\! t_j)_{j\in J}$.
Besides,
inequalities~\eqref{e0}-\,\eqref{t1}, ensure 
the positivity of the encoded schedule.
Indeed,
for any $j$ in $J$,
inequalities~(\ref{e1}) and~(\ref{t0}) ensure that 
$d \!-\! e_j \!+\! t_j \geqslant d \!-\! e_j \geqslant d \!-\! p(J) \!+\! p_j$.
Since $d$ is unrestrictive, we deduce that 
$ d \!-\! e_j \!+\! t_j \geqslant p_j$.
Hence,
we obtain the following lemma.
\begin{lm}
Let $(e,t,\delta) \!\in\! \mathbb{R}^J \!\times\! \mathbb{R}^J \!\times\! \{0,1\}^J$.
\textbf{If} $e,t,\delta$ satisfy~\eqref{e0}-\,\eqref{t1},
\textbf{then} $\theta^{-1}(e,t)$ satisfies~(\ref{pos}).\\[-0.4cm]
\label{lm_pos}
\end{lm}

$\bullet$ \textit{Handling the non-overlapping}\\
To ensure the non-overlapping,
it suffices that 
early tasks are fully processed before $d$ 
and do not overlap each other,
and that tardy tasks are fully processed after $d$ 
and do not overlap each other either.
Note that for a \dSing, 
the non-overlapping reduces to these two constraints
related to early and tardy tasks respectively.\\
In order to use the partition between early and tardy tasks
induced by the completion times $C$,
we introduce the following notations.\\[-0.6cm]
$$\Ec (C) \!=\! \{\, j\!\in\! J\,|\, C_j \!\leqslant\! d \,\}
\enskip   \text{ and }\enskip
\Tc(C) \!=\! \{\, j\!\in\! J\,|\, C_j \!>\! d \,\}$$

\begin{figure}[h]
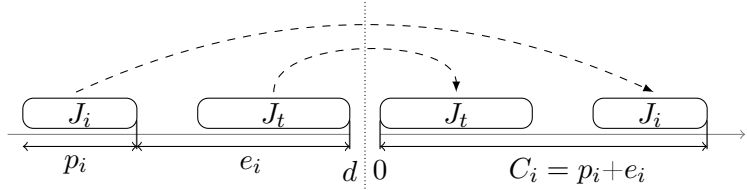

\centering
\figMiroir
\vspace*{-0.4cm}
\caption{Illustration of the role of $p_i \!+\!e_i$ for an early task $i$ }
\label{fig_miroir} 
\end{figure}

For a tardy task,
the tardiness can be seen
as a completion time with respect to $d$.
Therefore,
ensuring that the tardy tasks are fully processed before $d$ 
(resp. they do not overlap each other) 
is equivalent to imposing
positivity constraints for tardy tasks
(resp. the non-overlapping constraint for tardy tasks).
As shown on Figure~\ref{fig_miroir},
for an early task $j$,
the value $e_j \!+\! p_j$ can be seen as a completion time.
Using $x_{\slash S}$ to denote $\big(x_j\big)_{j\in S}$ 
for any subset $S$ of $J$ 
and for any vector $x$ in $\mathbb{R}^J$, 
the following lemma sums up these observations.

\begin{lm}
Let $C \!\in\! \mathbb{R}^J$
and set $(e,t)=\theta(C)$.
\textbf{If} there is no  $j \!\in\! J$ 
such that $C_j \!-\! p_j \!<\! d$ and $C_j \!>\! d$,\\
\textbf{then}
$C$ satisfies (\ref{non-ch})
$\Leftrightarrow (e \!+\! p)_{\slash \Ec(C)}$
and $t_{\slash \Tc(C)}$ satisfy  (\ref{pos}) and (\ref{non-ch}).\\[-0.4cm]
\label{lm_ct_non-ch_local}
\end{lm}
In the formulation,
$\delta$ describes
the partition between early and tardy tasks,
denoted as follows.
\\[-0.3cm]
$$
\Ed(\delta) \!=\! \{\, j\!\in\! J\,|\, \delta_j \!=\! 1 \,\}
\enskip   \text{ and }\enskip
\Td(\delta) \!=\! \{\, j\!\in\! J\,|\, \delta_j \!=\! 0 \,\}
$$

According to Section~\ref{sec_non-ch},
we want to apply Queyranne's inequalities~(\ref{Q}) to the vectors 
$(e \!+\! p)_{\slash E(\delta)}$ 
and $t_{\slash T(\delta)}$ respectively,
so that they satisfy (\ref{pos}) and (\ref{non-ch}).
Therefore, we consider the following inequalities.
\begin{eqnarray}
\forall S \!\subseteq\! J ,&& p*(e \!+\! p)\big(S \!\cap\! E(\delta) \big) \geqslant g_p(S) 
\label{S1.0}\\
\forall S \!\subseteq\! J,&& p*t\big(S \!\cap\! T(\delta) \big) \geqslant g_p(S) 
\label{S2.0} 
\end{eqnarray}

These inequalities are not linear 
as $E(\delta)$ and $T(\delta)$
depend on $\delta$ variables.
Replacing \hbox{$S \!\cap\! E(\delta)$ }
(resp. $S \!\cap\! T(\delta)$ ) by $S$
raises non valid inequalities.
Indeed, inequality~(\ref{S2.0}) 
for $S \!=\! \{i,j\}$ where $i\!\in\!E$,
would become $p_j t_j \geqslant p_i^2 \!+\! p_j^2 \!+\! p_i p_j$
since $t_i\!=\! 0$ by~(\ref{t0}) and~(\ref{t1}).
This implies that $t_j \!>\! p_j$,
which is not valid for all the feasible schedules.

To ensure that only the terms corresponding to early (resp. tardy) tasks 
are involved in~(\ref{S1.0}) (resp. in (\ref{S2.0})),
we multiply each term of index $j$ in $S$ 
by $\delta_j$ (resp. by $(1\!-\!\delta_j)$).
If $\delta_j\!\in\!\{0,1\}$,
then $(1\!-\!\delta_j)^2 = (1\!-\!\delta_j)$,
$e_j\,\delta_j \!=\! e_j$ from inequality~\eqref{e1}
and $t_j\,(1\!- \!\delta_j) \!=\!t_j$ from inequality~\eqref{t1}.
We obtain the following quadratic inequalities.
\begin{eqnarray}
\forall S\subseteq J,&&\enskip \sum\limits_{j\in S}p_j e_j
\geqslant \hspace{-0.2cm}
\sum\limits_{(i,j)\in S^<}\hspace{-0.2cm}p_ip_j\delta_i\delta_j
\label{S1.1} \\
\forall S \!\subseteq\! J,&&\enskip \sum\limits_{j\in S} p_j t_j (1 \!-\! \delta_j)
\geqslant \hspace{-0.2cm}
\sum\limits_{(i,j)\in S^<}  p_i p_j \, (1 \!-\! \delta_i)(1 \!-\! \delta_j) +
\sum\limits_{j\in S} p_j^2 (1 \!-\! \delta_j) 
\label{S2.1} 
\end{eqnarray}

$\bullet$ \textit{Linearization of the quadratic terms using $x$ variables }\\
In order to remove the quadratic terms,
we introduce a new variable $x_{i,j}$ 
representing whether $\delta_i$ is different than $\delta_j$
for each $(i,j)$ in $J^<$.
Since the quadratic terms are the products of boolean variables,
the following inequalities ensure their consistency with respect to $\delta$.\\[-0.6cm]
\begin{eqnarray}
\forall (i,j) \!\in\! J^<,\enskip && x_{i,j} \geqslant \delta_i \!-\! \delta_j  
\label{x.1}\\
\forall (i,j) \!\in\! J^<,\enskip && x_{i,j} \geqslant \delta_j \!-\! \delta_i 
\label{x.2}\\
\forall (i,j) \!\in\! J^<,\enskip && x_{i,j} \leqslant \delta_i \!+\! \delta_j 
\label{x.3}\\
\forall (i,j) \!\in\! J^<,\enskip && x_{i,j} \leqslant 2 \!-\! (\delta_i \!+\! \delta_j)
\label{x.4}
\end{eqnarray}

The following lemma provides the correspondence between quadratic and linear terms.

\begin{lm}[\cite{Fortet}]
\textbf{If} $\delta\!\in\!\{0,1\}^J$ 
\textbf{then} for all $(i,j)\!\in\!\K$:\\  
(i) $\delta$ and $x$ satisfy~\eqref{x.1}-\eqref{x.4} associated with $(i,j)
\enskip \Leftrightarrow \enskip x_{i,j}\!=\!0$ 
if $\delta_i \!=\! \delta_j$ 
and $ x_{i,j}\!=\!1$ otherwise.\\[0.1cm]
(ii) In case (i) holds, then
$\enskip \delta_i \delta_j \!=\! 
\dfrac{\delta_i\!+\!\delta_j\!-\!x_{i,j}}{2} \enskip$
and
$\enskip(1\!-\!\delta_i) (1\!-\!\delta_j) \!=\! 
\dfrac{2\!-\!(\delta_i\!+\!\delta_j)\!-\!x_{i,j}}{2}$.\\[-0.4cm]
\label{lm_x}
\end{lm}
The proof can be easily done by considering the two cases 
$\delta_i \!=\! \delta_j$ and $\delta_i \!\neq\! \delta_j$.\\

$\bullet$ \textit{Non-overlapping inequalities}\\
Using Lemma~\ref{lm_x}(ii), 
we obtain the following inequalities.\\[-0.6cm]
\begin{align}
\forall S \!\subseteq\! J,\enskip &
\sum\limits_{j\in S} p_j e_j 
\geqslant 
\sum\limits_{(i,j)\in S^<} \hspace{-0.2cm} p_i p_j \,
\frac{\delta_i \!+\!\delta_j \!-\! x_{i,j}}{2}
\tag{Q1}
\label{S1}\\
\forall S \!\subseteq\! J,\enskip &
\sum\limits_{j\in S} p_j t_j 
\geqslant 
\sum\limits_{(i,j)\in S^<} \hspace{-0.2cm} p_i p_j \,
\frac{2 \!-\! (\delta_i \!+\!\delta_j) \!-\! x_{i,j}}{2}+
\sum\limits_{j\in S} p_j^2 (1 \!-\! \delta_j)
\tag{Q2}
\label{S2} 
\end{align}

\noindent
The following lemma summarizes the relationship 
between the inequalities~(\ref{S1}), (\ref{S2}) and (\ref{Q}).

\begin{lm}
Let $(\delta,x)\!\in\! \{0,1\}^J \!\times\!  \mathbb{R}^{J^<}$ 
satisfying inequalities~\eqref{x.1}-\eqref{x.4}.\\[0.1cm]
(i) \textbf{If} $e \!\in\! \mathbb{R}^J$ satisfies inequalities
(\ref{e0}) and (\ref{e1}) for all $j\!\in\! \Ed(\delta)$,\\
\hspace*{0.4cm}\textbf{then} 
$e,\delta,x$ satisfy inequalities (\ref{S1}) 
for all $S\!\subseteq\! J \Leftrightarrow \left(e\!+\!p\right)_{\slash \Ed(\delta)}$ 
satisfies inequalities~(\ref{Q}).\\[0.1cm]
(ii) \textbf{If} $t \!\in\! \mathbb{R}^J$ satisfies inequalities
(\ref{t0}) and (\ref{t1}) for all $j\!\in\! \Td(\delta)$,\\
\hspace*{0.4cm}\textbf{then} 
$t,\delta,x$ satisfy inequalities (\ref{S2}) 
for all $S\!\subseteq\! J \Leftrightarrow t_{\slash \Td(\delta)}$
satisfies inequalities~(\ref{Q}).\\[-0.4cm]
\label{lm_S_Q}
\end{lm}

The following lemma allows to make the bridge between  
$(e \!+\! p)_{\slash E(C)}$ from Lemma~\ref{lm_ct_non-ch_local} 
and $(e \!+\! p)_{\slash E(\delta)}$ from Lemma~\ref{lm_S_Q}
(resp. between $t_{\slash T(C)}$ and $t_{\slash T(\delta)}$).

\newpage
\begin{lm}
Let $(e,t,\delta) \!\in\! \mathbb{R}^J \!\times\! \mathbb{R}^J \!\times\! \{0,1\}^J$.\\
\textbf{If} $e,t,\delta$ satisfy \eqref{e0}-\eqref{t1} and (\ref{S2})
\textbf{then} 
$\Ed(\delta) \!=\! \Ec\big(\theta^{-1}(e,t)\big)$ and
$\Td(\delta) \!=\! \Tc\big(\theta^{-1}(e,t)\big)$.\\[-0.4cm]
\label{lm_E_T}
\end{lm}

\begin{proof}
Let $C \!=\! \theta^{-1}(e,t)$. 
If $j \!\in\! \Tc(C)$,
then $C_j \!>\! d$ by definition.
That is $ t_j \!>\! e_j$ since $C_j \!=\! d \!-\!e_j \!+\! t_j$.
From inequality~(\ref{e0}),
we deduce that $t_j \!>\! 0$
and from inequality~(\ref{t1}),
that $\delta_j \!\neq\!1$.
Since $\delta_j$ is an integer,
$\delta_j \!=\!0$.
That proves $\Tc(C) \!\subseteq\! \Td(\delta)$.
Conversely,
if $j \!\in\! \Td(\delta)$,
inequalities~(\ref{e0}) and (\ref{e1}) 
ensure that $e_j \!=\! 0$,
since $\delta_j \!=\!0$ by definition. 
Thus,
$C_j \!=\! d \!+\! t_j$.
Since $ t_j\!\geqslant\!p_j \!>\! 0$ 
from inequality~(\ref{S2}) for $S \!=\! \{j\}$,
we deduce that  $C_j \!>\! d$,
that proves $\Td(\delta) \!\subseteq\! \Tc(C)$.\\
Similarly,
we can prove the equality for the early tasks
(without using~(\ref{S1})). 
\end{proof}

$\bullet$ \textit{Formulation \Fetdx}\\
Let us define
$\boldsymbol{\Petdx} \!=\! 
\left\lbrace\begin{tabular}{@{ }l|l@{ }}
$(e,t,\delta,x) \!\in\! \mathbb{R}^J \!\times\! \mathbb{R}^J \!\times\!  
[0,1]^J \!\times\!  \mathbb{R}^{J^<}$
& \eqref{e0}-\eqref{t1}, \eqref{x.1}-\eqref{x.4}, \eqref{S1} and \eqref{S2} are satisfied
\end{tabular}\right\rbrace$.
Note that this polyhedron does not depend on
either $\alpha$, $\beta$, or even $d$,
but is  only defined from $p$.
Moreover, this polyhedron is defined 
by an exponential number of inequalities,
inducing the use of a separation algorithm, 
this subject will be the purpose of Section \ref{sec_sepa}.\\
Since $\delta$ are boolean variables,
we are only interested in vectors 
for which $\delta$ is an integer,
that are integer points.
Therefore, we introduce the operator \imp{$\intd$},
which only keeps the integer points of a set.
For $V$ included in 
$\mathbb{R}^J \!\times\! \mathbb{R}^J \!\times\! \mathbb{R}^J \!\times\!\mathbb{R}^{J^<}\!,
\enskip \intd (V) \!=\!
\big\{ (e,t,\delta,x) \!\in\! V \,|\, \delta \!\in\! \{0,1\}^J \big\}.$
However,
the formulation is not a classical MIP formulation,
since some integer points do not encode feasible schedules.
The same observation holds for $P^{Q,M}$,
as discussed in Section~\ref{sec_non-ch}
(apart from the integrity constraints on $\delta$).
Therefore, we need to add an extremality condition 
(and consider the minimality condition)
to ensure the feasibility.
Finally,
our formulation for the unrestrictive common due date problem
defined by the unit penalties $(\alpha,\beta)$ is the following.\\[-0.2cm]
$$\text{\Fetdx}
\begin{tabular}[]{l}
$\min \gab (e,t)$\\
\hspace*{0.2cm}
s.t. $(e,t,\delta,x) \!\in\!\intd\big(\extr{\Petdx}\big) $
\end{tabular}
$$

\subsection{Validity of Formulation~\Fetdx}

The following theorem establishes that
a feasible schedule, under some assumptions,
is encoded by an integer point of $\Petdx$.
In particular a \dblocSing is encoded by an integer point of $\Petdx$.

\begin{thm}
\textbf{If} vector $C$ gives the completion times of  
a feasible schedule without  any straddling task 
such that 
tasks are processed between $d \!-\!p(J)$ and $d \!+\!p(J)$, \ie 
$\forall j\!\in\! J,\,d \!-\!p(J) \!\leqslant\! C_j \!-\! p_j $
and $C_j \!\leqslant\!d \!+\!p(J) $\\ 
\textbf{then} there exists 
$X \!=\!(e,t,\delta,x) \!\in\! \PetdxE$,
such that $\theta(C) \!=\!(e,t)$.\\[-0.4cm]
\label{thm_etdx_complet}
\end{thm}

\begin{proof}
From $C$, let us set: 
$(e,t) \!=\! \theta(C),\enskip 
\delta \!=\! \mathbb{1}_{E(C)},\enskip 
x \!=\! \left(\mathbb{1}_{\delta_i \neq \delta_j}\right)_{(i,j)\in J^<}
\, \text{ and } \enskip
X \!=\! (e,t,\delta ,x)$.\\
Note that the definition of $\delta$ ensures that
$\delta \!\in\! \{0,1\}^J \!\subseteq[0,1]^J$,
and that $\Ed(\delta) \!=\! \Ec(C)$ (resp. $\Td(\delta)\!=\!\Tc(C)$),
which allows the notation $E$ (resp $T$)
for sake of brevity.
Inequalities~(\ref{e0}) and~(\ref{t0}),
as well as~(\ref{e1}) for $j$ in $T$ and~(\ref{t1}) for $j$ in $E$,
are automatically satisfied
by construction of $e$, $t$ and $\delta$.
The assumption that 
$\forall j\!\in\! J,\,d \!-\!p(J) \!\leqslant\! C_j \!-\! p_j $
(resp. $C_j \!\leqslant\!d \!+\!p(J)$)
ensures that inequalities~(\ref{e1}) for $j$ in $E$
(resp. inequalities~(\ref{t1}) for $j$ in $T$)
are satisfied.\\
Using Lemma~\ref{lm_x}(i), 
$x$ and $\delta$ satisfy inequalities~\eqref{x.1}-\eqref{x.4}.\\
Since $C$ encodes a feasible schedule, 
$C$ satisfies~(\ref{pos}) and~(\ref{non-ch}).
Using Lemma~\ref{lm_ct_non-ch_local},
$(e \!+\! p)_{\slash E}$  (resp. $t_{\slash T}$)
satisfies~(\ref{pos}) and~(\ref{non-ch}).
Applying Property~\ref{lm_valide} to these two vectors,
we deduce that they satisfy~(\ref{Q}), 
and using Lemma~\ref{lm_S_Q}, 
that $e,\delta,x$ satisfy~(\ref{S1}) and $t,\delta,x$ satisfy~(\ref{S2}).
Thus, $X$ belongs to $\Petdx$, 
and even to $\PetdxE$ 
since $\delta \!\in\! \{0,1\}^J$.
\end{proof}

The following theorem establishes that
an optimal solution of formulation \Fetdx is a solution 
for the unrestrictive common due date problem.

\begin{thm}
\textbf{Let} $X^* \!=\!(e,t,\delta,x) \!\in\! \PetdxE $.\\
\textbf{If} $X^*\!\in\!\text{extr}(\Petdx)$
and $(e,t)$ minimizes $\gab$ 
\textbf{then} $X^*$ encodes a \dblocSing.\\[-0.4cm]
\label{thm_etdx_correct}
\end{thm}

\begin{proof}
The first step is to show that $X^*$ encodes a feasible schedule.\\
From Lemma~\ref{lm_theta-1},
$(e,t)$ is consistent 
and we can set $C^*\!\!=\!\theta^{-1}(e,t)$.
Then $X^*$ encodes a schedule defined by the completion times $C^*$.
This schedule will be denoted by \Sopt.
Proving that \Sopt is feasible consists then 
in showing that $C^*$ satisfies~(\ref{pos}) and (\ref{non-ch}).
From Lemma~\ref{lm_pos}, $C^*$ satisfies~(\ref{pos}). 
From Lemma~\ref{lm_E_T}, 
$\Ed(\delta) \!=\! \Ec(C^*)$ (resp. $\Td(\delta)\!=\!\Tc(C^*)$),
which allows the notation $E$ (resp. $T$)
for sake of brevity.
Using Lemma~\ref{lm_ct_non-ch_local},
to show that $C^*$ satisfies~(\ref{non-ch}),
it remains to show that 
$(e \!+\! p)_{\slash E}$ 
(resp. $t_{\slash T}$) 
satisfies (\ref{pos}) and (\ref{non-ch}).\\
From Lemma~\ref{lm_S_Q}, we know that 
$(e \!+\! p)_{\slash E}$ (resp. $t_{\slash T}$)
satisfies inequalities~(\ref{Q}).\\
On one hand, 
using these inequalities for the singletons,
ensures that 
$(e \!+\! p)_{\slash E}$ (resp. $t_{\slash T}$)
satisfies~(\ref{pos}).
We deduce that no straddling task occurs in \Sopt.\\
On the other hand,
inequalities~(\ref{Q})
will allow us to show that 
$(e \!+\! p)_{\slash E}$ (resp. $t_{\slash T}$) satisfies~(\ref{non-ch})
in the same way that we have shown that a vector in  $\extre{P^{Q,M}}$
encodes a schedule without overlapping in Section~\ref{sec_non-ch}.

Let us assume that $(e \!+\! p)_{\slash E}$ does not satisfy~(\ref{non-ch}).
Then there exists $(i,j) \!\in\! E^2$ such that
$e_i \!+\! p_i \!\leqslant\! e_j \!+\! p_j \!<\!(e_i \!+\! p_i) \!+\! p_j$.
Two cases have to be considered:

$\enskip \rightarrow$ If $e_j \!+\! p_j \!<\! p(J)$,
then from Lemma~\ref{lm_cle_1}
on  $(e \!+\! p)_{\slash E}$ 
there exists $\varepsilon \!\in\!\Rpe$
such that setting
$e^{+-} \!=\! e 
+ \frac{\varepsilon}{p_i}\mathbb{1}_i 
- \frac{\varepsilon}{p_j}\mathbb{1}_j 
\enskip \text{ and } \enskip
e^{-+} \!=\! e 
- \frac{\varepsilon}{p_i}\mathbb{1}_i 
+ \frac{\varepsilon}{p_j}\mathbb{1}_j,$
both $(e^{+-} \!+\! p)_{\slash E}$
and $(e^{-+} \!+\! p)_{\slash E}$
satisfy~(\ref{Q}).
Using Lemma~\ref{lm_S_Q},
both $e^{+-},\delta,x$ and $e^{-+},\delta,x$ satisfy~\eqref{S1}.
Since changing the value of $\varepsilon$ 
for $\min\big( \varepsilon,\, p(J)\!-\!p_j\!-\!e_j \big)$
does not affect the satisfaction of~\eqref{S1},
we can assume $\varepsilon \!\leqslant\! p(J)\!-\!p_j\!-\!e_j$,
while ensuring $\varepsilon \!>\! 0$.
Since 
$e^{+-}_i \!\!=\! e_i \!+\! \frac{\varepsilon}{p_i} \!\leqslant\! e_i \!+\! \varepsilon$,
using this latter assumption
and $e_j \!+\! p_j \!\geqslant\! e_i \!+\! p_i$,
we obtain $e^{+-}_i \!\leqslant\! p(J) \!-\! p_i$.
For $k$ in $J \!\setminus\! \{i\}$, $e^{+-}_k \!\!\leqslant\! e_k$,
and since $e$ satisfies~\eqref{e1}, 
we deduce that 
$e^{+-}_k \!\leqslant\! p(J) \!-\! p_k$.
Thus $e^{+-}$ satisfies inequalities~\eqref{e1}.\\
Besides, since $(e^{+-} \!+\! p)_{\slash E}$
satisfies inequalities~\eqref{Q} for the singletons,
$e^{+-}_k \!+\! p_k  \!\geqslant\! p_k$ for all $k$ in $E$.
Since $e^{+-}_k \!=\! e_k$ for all $k$ in $T$ 
and $e$ satisfies ~\eqref{e0},
we deduce that $e^{+-}$ satisfies inequalities~\eqref{e0}.
Similarly,
$e^{-+}$ satisfies inequalities~\eqref{e0} and~\eqref{e1}.
Finally, $X^{+-} \!=\! (e^{+-},t,\delta,x)$ 
and $X^{-+} \!=\! (e^{-+},t,\delta,x)$,
are two points of $\Petdx$ 
whose middle point is $X^*$.
A contradiction, since $X^*$ is extreme.

$\enskip \rightarrow$ If $e_j \!+\! p_j \!\geqslant\! p(J)$,
then $e_j \!+\! p_j \!\geqslant\! p(E)$,
and from Lemma~\ref{lm_cle_2} 
on $(e \!+\! p)_{\slash E}$
there exists $\varepsilon \!\in\!\Rpe$
such that setting
$e^{-} \!=\! e 
- \frac{\varepsilon}{p_j}\mathbb{1}_j$,
$(e^{-} \!+\! p)_{\slash E}$
satisfies~(\ref{Q}).
Using Lemma~\ref{lm_S_Q},
$e^{-},\delta,x$ satisfy~(\ref{S1}).
Since $e^-$ is  component-wise smaller than $e$,
$e^-$ also satisfies inequalities~(\ref{e1}).
Besides, the inequality~(\ref{Q}) for the singleton $\{j\}$
ensures that $e^-_j \!\geqslant\! 0$,
thus $e^-$ satisfies inequalities~(\ref{e0}).
Finally, setting $X^{-} \!=\! (e^{-},t,\delta,x)$,
we exhibit a point of $\Petdx$,
which has a smaller value than $X^*$ according to $ \gab$.
A contradiction, since $(e,t)$ minimizes $ \gab$.

Finally, $(e \!+\! p)_{\slash E}$ satisfies~(\ref{non-ch}).
In the same way, we can prove that $t_{\slash T}$ satisfies~(\ref{non-ch}).
We deduce that \Sopt is a feasible schedule.
The second step consists in showing that
\Sopt is a \dblocSing.

Since we already know that \Sopt
does not hold a straddling task,
it suffices to show that it is a \blocSing 
with at least one early task
to conclude that is a \dblocSing.
Let us assume that \Sopt 
holds an idle time or has no early task.
Let \Smieux denotes the schedule obtained by
tightening tasks around $d$ to fill idle times between tasks
and, if there is no early task, shifting backward all the tasks
such that the first one becomes on-time.
Since the due date is unrestrictive,
no task is scheduled before $0$ 
despite the backward shifting,
then \Smieux is a \dblocSing by construction.
If $\widehat{C}$ denotes 
the completion times defining \Smieux, then
$\forall j\!\in\! J,\,d \!-\!p(J) \!\leqslant\! \widehat{C}_j \!-\! p_j $
and $\widehat{C}_j \!\leqslant\!d \!+\!p(J)$.
Then using Theorem~\ref{thm_etdx_complet}, 
there exists 
$\widehat{X} \!=\!(\hat e,\hat t,\hat \delta, \hat x) \!\in\! \PetdxE$,
such that $\theta(\widehat C) \!=\!(\hat e, \hat t)$.
Moreover, 
$\fab(\widehat{C}) \!<\! \fab(C^*)$,
since the early tasks stay early but with a smaller earliness,
and the tardy tasks,
except the first tardy task which becomes eventually on-time,
stay tardy with a smaller tardiness.

Then $ \gab(\hat e,\hat t) \!=\! \fab(\widehat{C}) \!<\! \fab(C^*)\!=\! \gab(e,t)$,
which contradicts the minimality of $(e,t)$.\\
Finally, $X^*$ encodes a \dblocSing.
\end{proof}

The following theorem establishes that
the unrestrictive common due date problem 
reduces to solving formulation \Fetdx.
\begin{thm} 
(i) Any optimal \dblocSing
is encoded by a vector
minimizing $ \gab$ 
on $\intd\big(\extr{\Petdx}\big)$.\\
(ii) Conversely, any vector minimizing $ \gab$ 
on  $\intd\big(\extr{\Petdx}\big)$
encodes an optimal \dblocSing.\\[-0.4cm]
\label{thm_etdx_final}
\end{thm}

\begin{proof}
Let us consider an optimal \dblocSing \Sopt.
From Theorem~\ref{thm_etdx_complet},
there exists a vector $X^*\!=\!(e^*,t^*,\delta^*,x^*)$ 
in $\text{int}_{\delta}(\Petdx)$ encoding \Sopt.
We introduce
$\Pdelta \!\!= \big\{(e,t) \,|\, (e,t,\delta^*,x^*) \!\in\! \Petdx \big\},$
which is the slice of $\Petdx$  according to $\delta^*$,
\ie the projection of set of points of $\Petdx$
satisfying $\delta \!=\!\delta^*$
and therefore $x \!=\! x^*$.\\
To show that $X^*$ is an extreme point of $\Petdx$,
it suffices to prove that
$(e^*,t^*)$ is an extreme point of $\Pdelta$.
Indeed, if there were $X^1 \!=\! (e^1,t^1,\delta^1,x^1)$ 
and $X^2 \!=\! (e^2,t^2,\delta^2,x^2)$ in $\Petdx$
such that $X^* \!=\! \frac{1}{2} (X^1 \!+\! X^2)$,
$\delta^1$ and $\delta^2$ would necessarily be equal to $\delta^*$
since $\delta^* \!\in\! \{0,1\}^J$,
$\delta^1 \!\in\! [0,1]^J$
and $\delta^2 \!\in\! [0,1]^J$.
By Lemma~\ref{lm_x}, we deduce that $x^1 \!=\! x^*$ (resp. $x^2 \!=\! x^*$),
and thus $(e^1,t^1)$ (resp. $(e^2,t^2)$) is in $\Pdelta$.
Yet $(e^*,t^*) \!=\! \frac{1}{2} \big( (e^1,t^1) \!+\! (e^2,t^2)\big)$,
and $(e^*,t^*)$ would not be an extreme point of $\Pdelta$.

Let $(E,T)$
denote the partition of tasks given by $\delta^*$,
\ie $E \!=\! \Ed(\delta^*)$ and $ T \!=\! \Td(\delta^*)$.
Using Lemma~\ref{lm_S_Q},
we decompose $\Pdelta$ 
as a Cartesian products of polyhedra as follows.
\\[-0.2cm]
$$\Pdelta \!\!= \PdeltaE \!\times \{0\}^T \!\times \PdeltaT \!\times  \{0\}^E 
\text{ where }
\left\lbrace\begin{tabular}{@{}l}
$\PdeltaE \!=\! 
\left\lbrace \tilde{e} \!\in\! \mathbb{R}^E \,|\, 
\tilde{e} \!+\! p_{\slash E} \text{ satisfies } (\ref{Q}) 
\text{ and }\forall j\!\in\! E,\, \tilde{e}_j \!+\! p_j \!\leqslant\! p(J)\, \right\rbrace$\\[0.2cm]
$\PdeltaT \!=\! 
\big\{ \tilde{t} \!\in\! \mathbb{R}^T \,|\, 
\tilde{t}\text{ satisfies } (\ref{Q}) 
\text{ and }\forall j\!\in\! T,\, \tilde{t}_j \leqslant p(J) \big\}$
\end{tabular}\right.
$$

Knowing that the extreme points set of a Cartesian product is exactly
the Cartesian product of the extreme points sets, 
it remains to show that $e^*_{\slash E} \!\in\! \text{extr}(\PdeltaE)$ 
and that $t^*_{\slash T} \!\in\! \text{extr}(\PdeltaT)$.
Note that $\PdeltaT$ is the polyhedron
called $P^{Q,M}$ in Section~\ref{sec_non-ch}, 
\blue{where the index set $J$ 
is replaced by $T$
while keeping $M \!=\!p(J) \!\geqslant \! p(T)$.}
Similarly,
$\PdeltaE$ is a translation according to $-p_{\slash E}$
of $P^{Q,M}$, 
\blue{where $J$ 
is replaced by $E$
while keeping $M \!=\!p(J) \!\geqslant \! p(E)$.}
Then it suffices that $t^*_{\slash T}$ (resp. $e^*_{\slash E} \!+\! p_{\slash E}$)
encodes a left-tight schedule of tasks in $T$ (resp. $E$)
to ensure its extremality in $ \PdeltaT$ (resp. $\PdeltaE$).
Both conditions are satisfied since $X^*$ encodes a \dblocSing.
We deduce that $(e^*,t^*)$ belongs to extr$(\Pdelta)$.
Thus $X^*$ belongs to $\intd\big(\extr{\Petdx}\big)$.

\blue{To prove item \textsl{(i)},
it remains to show that 
$X^*$, or more precisely $(e^*,t^*)$, 
is a minimizer of $\gab$. 
By contradiction,
let us assume that there exists 
$\widehat{X} \!=\!(\hat{e},\hat{t},\hat{\delta},\hat{x})
\!\in\! \intd\big(\extr{\Petdx}\big)$
such that $(\hat{e},\hat{t})$ minimizes $\gab$
and $\gab(\hat{e},\hat{t}) \!<\! \gab(e^*,t^*)$.
According to Theorem~\ref{thm_etdx_correct},
$\widehat{X}$ encodes a schedule 
inducing a total penalty $\gab(\hat{e},\hat{t})$, 
which is lower than the total penalty of \Sopt 
a contradiction.
}


The \blue{second item \textsl{(ii)} }
is then a direct corollary of Theorem~\ref{thm_etdx_correct}.
The schedule encoded by a vector $X^*$ minimizing $ \gab$ 
on $\intd\big(\extr{\Petdx}\big)$
is a \dblocSing, and if it \blue {is} not optimal,
there would exist a strictly better \dblocSing,
and a vector in $\intd\big(\extr{\Petdx}\big)$
with a smaller value according to $ \gab$, a contradiction.
\end{proof}

\subsection{Dealing with formulation~\Fetdx}
The aim of this section is to show that formulation~\Fetdx
can be solved by a classical branch-and-cut algorithm.
Let us consider three relaxations of \Fetdx.
$$
\text{\FetdxLP}
\begin{tabular}[]{r@{ }l}
$\min$ & $ \gab(e,t)$\\
s.t.  & $(e,t,\delta,x) \!\in\! \Petdx $
\end{tabular}
\enskip\enskip
\text{\Fetdxextr}
\begin{tabular}[]{r@{ }l}
$\min$ & $ \gab(e,t)$\\
s.t. & $(e,t,\delta,x) \!\in\! \text{extr}(\Petdx)$
\end{tabular}
\enskip \enskip
\text{\Fetdxint}
\begin{tabular}[]{r@{ }l}
$\min$ & $ \gab(e,t)$\\
s.t. & $(e,t,\delta,x) \!\in\! \intd(\Petdx)$
\end{tabular}
$$

The formulation \FetdxLP is obtained
by relaxing the integrity and the extremality conditions.
It is a linear program defined 
by an exponential number of inequalities.
We will explain in Section~\ref{sec_sepa} that
the separation problem associated \blue{with} 
the non-overlapping inequalities defining $\Petdx$
is solvable in polynomial time.
Then \FetdxLP can be solved in polynomial time
using a cutting plane algorithm~\cite{GLS}.\\[-0.2cm]

Using the simplex algorithm 
for each LP-relaxation of a cutting plane algorithm,
the extremality of the solution is ensured.
Then in this case,
solving \FetdxLP is equivalent to solving \Fetdxextr.
A classical way to manage the integrity constraint 
is to use a branch-and-bound algorithm,
and even in this case a branch-and-cut algorithm.
Using an algorithm which provides an extreme point to solve each LP-relaxation,
a branch-and-bound algorithm directly computes a solution of \Fetdx.

\begin{pte}
Let us consider a branch-and-bound algorithm $\mathcal{A}$,
where the LP-relaxation at each node
provides an extreme point.
Using  $\mathcal{A}$ to solve \Fetdxint
by branching on $\delta$ variables solves \Fetdx.\\[-0.4cm]
\label{pte_bb}
\end{pte}

\begin{proof}
By assumption, the solution provided 
at each node of the branch-and-bound tree
is an extreme point of the polyhedron defined 
by the decisions previously taken,
and we will prove that this solution is also an extreme point of $\Petdx$.\\
Formally,
if variables $\delta_j$ for $j \!\in\! J_0$ (resp. for $j \!\in\! J_1$)
have been fixed to $0$ (resp. to $1$),
the polyhedron considered is $\Petdx \!\cap F^{J_0,J_1}$ where:
\\[-0.4cm]
$$F^{J_0,J_1} \!=\! \{\, 
(e,t,\delta,x) \!\in\! \mathbb{R}^J \!\!\times \mathbb{R}^J \!\!\times [0,1]^J \!\!\times [0,1]^{J^<}
\enskip|\enskip
\forall j\!\in\!J_0,\, \delta_j \!=\!0 \text{ and }
\forall j\!\in\!J_1,\, \delta_j \!=\!1 \, \}$$
We consider an arbitrary node defined by $J_0$ and $J_1$, 
and a vector $X \!=\!(e,t,\delta,x) \!\in\! \text{extr}\big(\Petdx \cap F^{J_0,J_1}\big)$.\\
By definition of $\Ed(\delta)$ and $\Td(\delta)$,
\hbox{$X \!\in\! \Petdx\! \cap F^{\Td(\delta),\Ed(\delta)}$}.
Moreover,
$J_1\!\subseteq\! \Ed(\delta)$ and $J_0 \!\subseteq\! \Td(\delta)$,
thus we have
\hbox{$\Petdx \cap F^{\Td(\delta),\Ed(\delta)} \subseteq \Petdx \cap F^{J_0,J_1}$}.
Recall that if $A\subseteq B$,
then $\text{extr}(B)\cap A \subseteq \text{extr}(A)$,
we deduce that 
$X \!\in\! \text{extr}\big(\Petdx\! \cap F^{T(\delta),E(\delta)}\big)$.
Since $\Petdx \cap F^{\Td(\delta),\Ed(\delta)}$ is exactly 
the set \blue{denoted by} $P^\delta$ in the previous proof,
we get extr$(\Petdx \cap  F^{\Td(\delta),\Ed(\delta)}) \subseteq \text{extr}(\Petdx)$.
We deduce that $X \!\in\! \text{extr}(\Petdx)$.
\end{proof}

Note that in general,
such an algorithm $\mathcal{A}$
is  not sufficient
to minimize a linear function
under both integrity and extremality constraints in a polyhedron.
To illustrate this observation, 
let us consider the following formulation.
where int$_y$ denotes 
the operator keeping only \blue{the points $(y,z)$
such that $y$ is an integer}.\\[-0.2cm]
$$
\text{(F)}
\begin{tabular}[]{r@{ }l}
$\max$ & $z$\\
s.t.  & $(y,z) \!\in\! \text{int}_y\big(\text{extr}(P)\big) $
\end{tabular}
\text{ with }
P \!=\! \Big\{ (y,z) \!\in\! 
\Rp \!\times \Rp \,|\enskip
z\!\leqslant\! \frac{2}{3} y \!+\! 2,\enskip
z\!\leqslant\! -2 y \!+\! 6
\Big\},$$
$\mathcal{A}$ provides a solution 
which does not belong to extr$(P)$.
Indeed, since $(\frac{3}{2},3)$ 
is the solution at the root node, 
the search space is divided into 
$P \,\cap\,\, ]-\!\!\infty, 1] \!\times\! \mathbb{R}$ 
and $P \cap [2,+\!\infty[ \times \mathbb{R}$,
and the extreme points maximizing $z$ 
in these \blue{polyhedra} are respectively 
$(1,2+\frac{2}{3})$, and $(2,2)$.
The provided point is then $(1,2+\frac{2}{3})$,
with a value of $2+\frac{2}{3}$
whereas the best value for an integer extreme point is $2$,
reached by $(0,2)$.\\
\blue{
The particularity of formulation~\Fetdx 
is that the integrity constraint on $\delta_j$
can be rewritten as 
\hbox{$\delta_j \!\in\! \text{extr}\big([0,1]\big)$},
for any $j\!\in\!J$.
Therefore, the integrity of $\delta$ 
and the extremality in $\Pdelta$
induce the extremality in $\Petdx$.
}

%

For any formulation (F),
let us denote by \val{(F)}
the value of any optimal solution 
for the optimization problem F.
Using any algorithm to solve each LP-relaxation,
a branch-and-bound algorithm can solve \Fetdxint,
that gives \val{\Fetdx}, 
but not directly a solution of \Fetdx.
Indeed,
if $X \!=\!(e,t,\delta,x)$ denotes the provided vector,
$\delta$ is $0$-$1$ 
and $(e,t)$ minimizes $g_{\alpha,\beta}$ on $P^\delta$ by construction.
Then,
there exists $(e^*,t^*)$ in $\text{extr}(P^\delta)$
such that $g_{\alpha,\beta}(e^*,t^*) \!=\! g_{\alpha,\beta}(e,t)$.
Since $X^* \!=\! (e^*,t^*,\delta,x)
\!\in\!\intd \big( \text{extr}(\Petdx) \big)$,
we get $ \gab(e^*,t^*) 
\!\geqslant\! \text{value\Fetdx}
\!\geqslant\! \text{value\Fetdxint}
\!=\! \gab(e,t)$.

In addition to this theoretical way to
come down to an extreme point,
and then to a feasible solution,
there is a computational way to do that
from the partition between early and tardy tasks defined by $\delta$.
It will be the purpose of the next section.

\newpage
\section{A second formulation for the unrestrictive case}
\label{sec_formul2}

The unrestrictive common due date problem is NP-hard,
so the problem associated \blue{with} \Fetdx is  NP-hard.
In contrast, 
\Fetdxextr is solvable in polynomial time.
We deduce that the hardness
of the formulation \Fetdx is only due to 
the integrity constraints on $\delta$ variables
This suggests that the main difficulty
of the unrestrictive common due date problem
lies in choosing which tasks are early
and which ones are tardy.
This observation is corroborated by
the following dominance property
known in the just-in-time scheduling field,
which ensures in the unrestrictive case 
once the partition between early and tardy tasks is fixed,
it suffices to sort tasks to obtain an optimal schedule.
A question is then:
how to exploit the strength of this property in a linear way?
This issue leads to a compact formulation 
for the unrestrictive case, presented in this section.

\subsection{Dominance properties}
We recall some dominance properties 
known for the symmetric penalties case~\cite{HVDV}, 
but given here in their most general statement.\\[-0.4cm]

\begin{lm}
Let $ \alpha \!\in\!\RpJ,\,
\beta \!\in\!\RpJ$.\\[0.1cm]
In the general case, the schedules where
the tasks ending before or at $d$  (resp. starting at or after $d$) 
are in order of nondecreasing  $\frac{\alpha_j}{p_j}$ 
(resp. nonincreasing  $\frac{\beta_j}{p_j}$) 
are strictly dominant when minimizing $\fab$.\\[-0.4cm]
\label{dom_V}
\end{lm}

For given unit penalties $\alpha$ and $\beta$,
a feasible schedule is said \imp{V-shaped} 
if the early tasks \blue{ are scheduled }
in \blue{nondecreasing}  order of $\alpha_j/p_j$
and the tardy ones in \blue{nonincreasing} order of $\beta_j/p_j$.
Since the tasks ending before or at $d$
are exactly the early ones in any schedule,
and the tasks starting after or at $d$
are exactly the tardy ones in a \dSing,
we deduce from Lemmas~\ref{dom_unres} and~\ref{dom_V},
that V-shaped \dblocPlur are dominant 
in the unrestrictive case.

In case of equality between two ratios 
$\alpha_i/p_i$ and $\alpha_j/p_j$ 
(resp. $\beta_i/p_i$ and $\beta_j/p_j$),
swapping tasks $i$ and $j$ does not change the total penalty 
of a schedule if both are early (resp. tardy).
Thus in this case,
there exist different optimal V-shaped \dblocPlur 
with the same partition between early and tardy tasks.
To ensure there is only one way 
to decode a partition between early and tardy tasks 
into a dominant schedule, 
we fix \textit{a priori} two orders on tasks :
one by decreasing $\alpha_j/p_j$,
and one by decreasing $\beta_j/p_j$.
Let $\rho$ and $\sigma$ denote two functions 
\blue{from $\unn$ to $J$} such that:\\[-0.3cm]
$$\left( 
\dfrac {\alpha_{\rho(k)}}{p_{\rho(k)}} \right)_{k\in\unn}
\text{ and }\hspace{0.2cm}
\left( \dfrac {\beta_{\sigma(k)}}{p_{\sigma(k)}} \right)_{k\in\unn}
\text{ are \blue{nonincreasing.} }
$$

We say that a feasible schedule is \imp{$\rho$-$\sigma$-shaped} 
when early (resp. tardy) tasks are processed in decreasing order of
\blue{$\rho^{-1}$} (resp. increasing order of \blue{$\sigma^{-1}$}).
These schedules are dominant 
in the unrestrictive case,
and will only be considered 
\blue{in} the remainder of this section.
Note that there is a one-to-one correspondence between 
the $\rho$-$\sigma$-shaped \dblocPlur and the vectors $\delta \!\in\! \{0,1\}^J$.

\subsection{A compact formulation for the unrestrictive case}

If the partition between early and tardy tasks 
of a $\rho$-$\sigma$-shaped \dblocSing is given by $\delta$,
then the earliness and tardiness are given by:\\[-0.3cm]
$$
e^\rho (\delta) \!=\!
\left(  \delta_j 
\sum\limits_{k=1}^{\rho^{-1}(j) -1 } \!\!
 p_{\rho(k)} \, \delta_{\rho(k)}
\right)_{j\in J}
\text{and } \enskip \enskip
t^\sigma (\delta) \!=\!
\left(   (1\!-\!\delta_j)
\sum\limits_{k=1}^{ \sigma^{-1}(j) } \!\!
p_{\sigma(k)} \, (1 \!-\!\delta_{\sigma(k)})
\right)_{j\in J}.
$$

\noindent
Using the same $x$ variables
as those in Section~\ref{sec_formul1}
to linearize these terms,
we consider\\[-0.3cm]
$$
e^\rho (\delta,x) \!=\!\!
\left( 
\sum\limits_{k=1}^{\rho^{-1}(j) -1 } \!\!\!
 p_{\rho(k)} \, 
\dfrac{\delta_{j}\!+\!\delta_{\rho(k)}- x_{j,\rho(k)} }{2}
\right)_{\hspace*{-0.15cm}j\in J}
\hspace*{-0.1cm}\text{and } \enskip 
t^\sigma (\delta,x) \!=\!\!
\left( 
\sum\limits_{k=1}^{\sigma^{-1}(j) -1 } 
\hspace{-0.2cm}  p_{\sigma(k)} \,  
\dfrac{ 2 \!-\! (\delta_{j}\!+\!\delta_{\sigma(k)}) - x_{j,\sigma(k)} }{2}
+ p_j (1\!-\! \delta_j)
\right)_{\hspace*{-0.15cm}j\in J}
$$
where we use $x_{i,j}$ without carrying if $i\!<\!j$, 
that is to denote the variable $x_{\min(i,j),\max(i,j)}$.\\

Therefore,
the total penalty is simply expressed by
$h_{\alpha,\beta}^{\rho,\sigma} (\delta,x) \!=\! 
\fab \big(\, e^\rho (\delta,x) ,\, t^\sigma (\delta,x) \,\big),$
which is linear.
We then consider the polyhedron
$
\boldsymbol{\Pdx} \!=\! 
\left\lbrace\begin{tabular}{@{ }l|l@{ }}
$(\delta,x) \!\in\! [0,1]^J \!\times\!  \mathbb{R}^{J^<}$
& \eqref{x.1}-\eqref{x.4} are satisfied
\end{tabular}\right\rbrace.
$\\
By definition of $e^\rho (\delta,x) $ and $t^\sigma (\delta,x)$,
a vector $(\delta,x)$ in $\Pdx$
cannot encode an \blue{infeasible} schedule.
So there is no need to add non-overlapping inequalities,
and hence we do not have to provide a separation algorithm 
or to only consider the extreme points of $\Pdx$.

Finally,
a compact formulation for the unrestrictive common due date problem
defined by the penalties $(\alpha,\beta)$ is \\[-0.5cm]
$$\text{\Fdx}
\begin{tabular}[]{l}
$\min h_{\alpha,\beta}^{\rho,\sigma}(\delta,x)$\\
\hspace*{0.2cm}
s.t. $(\delta,x) \!\in\!\intd \big(\Pdx\big) $
\end{tabular},
$$
where $\rho$ and $\sigma$ are pre-computed.

Note that polyhedron $\Pdx$ does not depend on $\rho$ or $\sigma$.
Indeed,
it is an extended polytope 
of the classical cut polytope
for the complete undirected graph on $J$~\cite{Barahona_Mahjoub_86}. 
A linear transformation of this polytope
has been studied in \cite{Padberg_89}.
From this work we can directly derive that 
$\Pdx$ is a full-dimensional polytope
and that inequalities~\eqref{x.1}-\eqref{x.4}
define facets of $\Pdx$.\\[-0.4cm]

\newpage
\section{General case}
\label{sec_formul3}
In this section, we provide a formulation 
for the general case
based on the ideas of the formulation \Fetdx.
In the general case,
we have to consider arbitrary earliness unit penalties, 
\blue{that is positive or zero unit earliness penalties}.
We can no longer derive an optimal solution
from the one obtained for the instance 
which does not include \blue{zero unit earliness penalty} tasks.
Indeed,
the due date could not allow to add 
these tasks at the beginning of the schedule.
For some instances, such tasks
are tardy in all optimal schedules.
For example if $J\!=\!\llbracket 1 , 3\rrbracket$, $d \!=\! 6$,
$ p_1\!=\!5$, $p_2\!=\! 3$, $p_3\!=\! 2$, 
$\alpha_1\!=\!0$, $\beta_1\!=\!1$
and $\alpha_2\!=\! \beta_2\!=\!\alpha_3\!=\! \beta_3\!=\!2$,
then the optimal schedule is given by 
$C_2 \!=\! 4 \!\leqslant d$,
$C_3 \!=\! 6 \!=\! d$,
$C_1 \!=\! 11 \!\geqslant d$.
Note that, conversely,
tasks with a zero unit tardiness penalty
can still be added at the end of  
an optimal schedule obtained
for the instance reduced to 
the non-zero earliness penalty tasks
in order to obtain
an optimal schedule for the original instance.
Hence, for the general case, we will set 
$\alpha \!\in\! \RpJ$ and
$\beta \!\in\! \RpeJ$.


\subsection{Dominance properties}
In the general case,
the dominance of the \dblocPlur 
is no longer valid.
Let us define a \imp{\dZblocSing} as 
a block which is a \dSing or which starts at time $0$, or both,
to enunciate the following dominance property
\cite{Hall_Kubiak_et_Sethi}.\\[-0.4cm]
\begin{lm}
In the general case,
\dZblocPlur are dominant when minimizing $\fab$.\\[-0.4cm]
\label{dom_gen}
\end{lm}
In the sequel, only \dZblocPlur will be considered.\\

Due to the potential occurrence of a straddling task 
in all optimal schedules for some instances,
the partition between early and tardy tasks is
no longer sufficient 
to deduce an optimal schedule.
As explained in Section~\ref{sec_formul2},
we can compute the best \dblocSing
with respect to this partition.
Conversely,
computing the best left-block
(\ie the best \blocSing starting at time 0)
with respect to this partition
is not straightforward, 
since we cannot say \textit{a priori} 
which is the straddling task
among the tardy ones.

\begin{figure}[h]
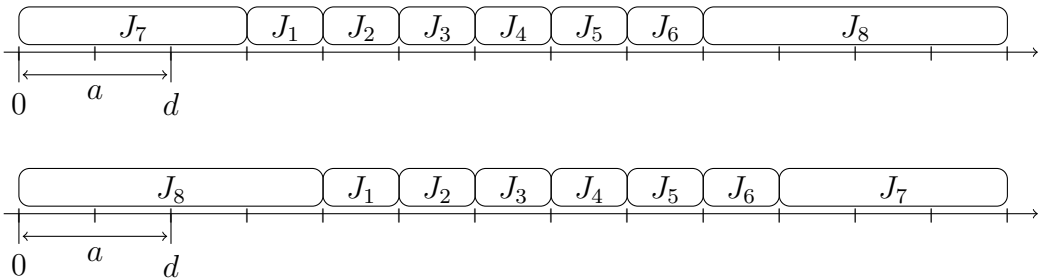

\centering
\figCEdompur\\[0.1cm]
\figCEdombispur\\[-0.5cm]
\caption{The two types of dominant schedules for $E \!=\! \emptyset$ and $T \!=\! J$.}
\label{fig_c-ex_ratio_max}
\end{figure}

Let us consider the best left-block 
with respect to a given partition.
Then the time $a$ between the beginning of the straddling task and $d$ 
is equal to $d \!-\! p(E)$
and the straddling task belongs to
$\{\,  j \!\in\! T \,|\,\, p_j \!>\! a \,\}$,
where $E$ (resp. $T$) 
denotes the set of early (resp. tardy) tasks
given by the partition.
One can conjecture that
the straddling task maximizes $\beta_j/p_j$ over this set.
However,
it is not the case,
as we shown by the following instance:
$J \!=\! [1..8],\enskip
\forall i\!\in\![1..6],\, 
p_i\!=\!1,\,
\alpha_i\!=\!40,\,
\beta_i\!=\!4 ,\,
p_7\!=\! 3,\,
\alpha_7\!=\! 20,\,
\beta_7\!=\! 8,\enskip
p_8\!=\! 4,\enskip
\alpha_8\!=\! 20,\,
\beta_8\!=\! 11,\,$
and $ d \!=\! 2$.
%
%
We can easily verify that 
the optimal partition is
$E \!=\! \emptyset$,
$T \!=\! J$ and $a\!=\!2$.
According to Lemma~\ref{dom_V},
an optimal schedule can be found among the left-blocks  
starting by task $7$ and ending by task $8$, or
starting by task $8$ and ending by task $7$.
The order of the other tasks is arbitrary,
since they all have the same ratio.
Figure~\ref{fig_c-ex_ratio_max} 
represents one optimal schedule of each type.\\[-0.4cm]

\noindent
The best ones are those starting by task $7$ and ending by task $8$.
Nevertheless the ratio $\beta_7/p_7 \!=\! 8/3$
is smaller than the ratio $\beta_8 /p_8 \!=\! 11/14$.
This example can be extended 
to an example where $E \!\neq\! \emptyset$
by adding tasks with zero unit earliness penalty 
and large unit tardiness penalty.\\[-0.2cm]

\begin{figure}[h]
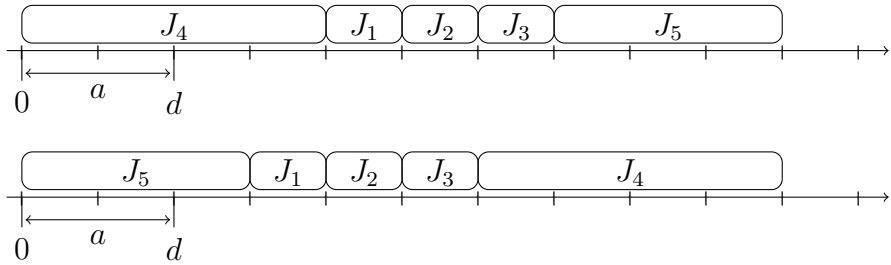

\centering
\figCEdomterpur\\[0.1cm]
\figCEdomquaterpur\\[-0.4cm]
\caption{The two types of dominant schedules
for $E \!=\! \emptyset$ and $T \!=\! J$.}
\label{fig_c-ex_ratio_max_bis}
\end{figure}

In this example,
the non optimality seems to be
induced by an incorrect ratio choice:
if we consider the ratio $\beta_j/(p_j \!-\! a)$ 
instead of $\beta_j/p_j $,
task $7$ has a greater ratio than task $8$. 
Then one can conjecture
that the straddling task $j$ maximizes
$\beta_j/(p_j \!-\!a)$ over tardy tasks 
with a processing time larger than $a$.
Unfortunately,
this is also false,
as shown by the following instance:
$ J \!=\! \llbracket 1 , 5\rrbracket,\enskip
\forall i\!\in\![1..3],\, p_i\!=\!1,\,
\alpha_i\!=\!10,\,
\beta_i\!=\!2,\enskip$
$p_4\!=\! 4,\,
\alpha_4\!=\! 10,\,
\beta_4\!=\! 5,\enskip
p_5\!=\! 3,\,
\alpha_5\!=\! 10,\,
\beta_5\!=\! 3$
and $d \!=\! 2$
We can easily verify that 
the optimal partition is
$E \!=\! \emptyset$,
$T \!=\! J$ and $a\!=\!2$.
According to Lemma~\ref{dom_V},
an optimal schedule can be found among the left-blocks  
starting by task $4$ and ending by task $5$, or
starting by task $5$ and ending by task $4$.
The order of the other tasks is arbitrary,
since they all have the same ratio.
Figure~\ref{fig_c-ex_ratio_max_bis} 
represents one optimal schedule of each type.

The best ones are 
those starting by task $4$ and ending by task $5$.
Nevertheless the ratio $\beta_4/(p_4 \!-\! a) \!=\! 2.5$
is smaller than the ratio  $\beta_5/(p_5 \!-\! a) \!=\! 3$.
This example can also be extended 
to an example where $E \!\neq\! \emptyset$.\\

The idea of the compact formulation \Fdx for the unrestrictive case
was to obtain the value $b(E,T)$ of a best schedule
for a fixed partition between early and tardy tasks $(E,T)$.
In the general case,
to derive $b(E,T)$ from a partition $(E,T)$ which is feasible
(\ie such that $p(E) \!\leqslant\! d$ ),
we have to consider several cases 
before using the dominance property.\\[-0.4cm]

Firstly, if we assume that $b(E,T)$ is achieved by
a schedule having an on-time task.
then we simply obtain $b(E,T)$ 
as for the unrestrictive case.
Secondly, if we assume that $b(E,T)$ is achieved by
a schedule having a straddling task,
then we can also assume, without lost of generality, 
that the schedule starts at time $0$
(using Lemma~\ref{dom_gen}). 
We have then to consider the case where the straddling task is $j$
for each $j \!\in\! T$ such that $p_j \!\geqslant\! d\!-\! p(E)$.
In each case,
Lemma~\ref{dom_V} allows to derive the optimal schedule
and we obtain the value $b(E,T)$ 
in a similar way as for the unrestrictive case.
It seems difficult to derive 
a linear function from this observation.
Therefore, we adapt the first formulation and not the second
for the general case.

\subsection{A natural formulation for the general case}
\label{sub_formul_aetdx}
~\indent
$\bullet$ \textit{An encoding based on a new reference point}\\
In case of a schedule with a straddling task $\js$,
\ie $C_\js \!-\! p_\js \!<\! d\!<\! C_\js$,
the tardiness of tardy tasks do not satisfy 
the non-overlapping constraints,
\ie $t_{\slash  T}$ does not satisfy inequalities~(\ref{Q}),
particularly $t_\js \!>\! p_\js$.
Indeed,
these tardiness no longer play 
the same role as completion times.
Therefore, we will use  variables
describing the schedule 
with respect to a new reference point,
which is the starting time of $j_s$ 
instead of the due date $d$.

\begin{figure}[h]
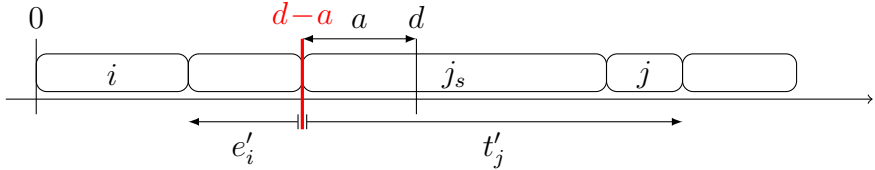

\centering
\figastr\\[-0.5cm]
\caption{The $(a,e',t')$ encoding 
for a schedule holding a straddling task $\js$}
\label{fig_aet_str}
\end{figure}

We introduce a new variable $a$,
so that $\dma$ is the starting time of $\js$.
The schedule is then a $(\dma$)-schedule.
For each task $j$ in $J$,
we consider a  variable $e'_j$ (resp. $t'_j$) 
instead of $e_j$ (resp. $t_j$), 
representing the earliness (resp. the tardiness)
according to the new reference point $\dma$.
Figure~\ref{fig_aet_str} illustrates this encoding
for a schedule holding a straddling task.\\[-0.6cm]

Since we do not know \textit{a priori} 
if there is a straddling task in the optimal schedule,
our formulation must also handle \dblocPlur.
Hence,
we also need to encode
\dblocPlur by variables $a,e',t'$.\\

\begin{figure}[h!]
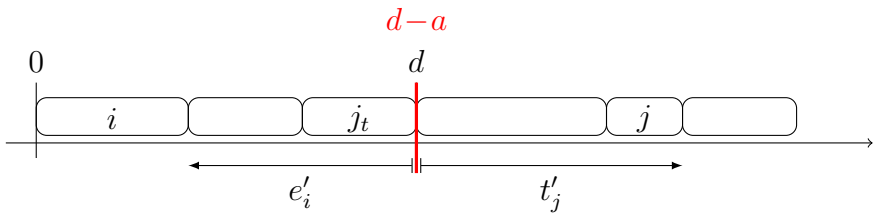

\vspace*{-0.2cm}
\centering
\figaotun\\[-0.5cm]
\caption{The first $(a,e',t')$ encoding 
for a schedule holding an on-time task $\jt$}
\label{fig_aet_ont}
\end{figure}

\begin{figure}[h!]
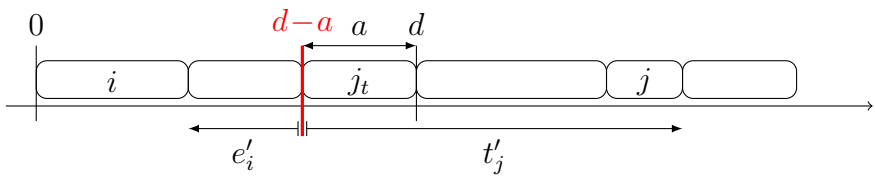

\centering
\figaotdeux\\[-0.4cm]
\caption{The second $(a,e',t')$ encoding 
for a schedule holding an on-time task $\jt$}
\label{fig_aet_ont_bis}
\end{figure}

In case of a schedule holding an on-time task $\jt$,
we can keep $d$ as the reference point,
since we can use earliness and tardiness
as proposed in formulation \Fetdx.
Hence, the first encoding consists in setting $a\!=\!0$,
and using $e'$ (resp. $t'$)
to represent earliness (resp. tardiness).
Figure~\ref{fig_aet_ont} illustrates this encoding
for a schedule holding an on-time task.
Unfortunately,
to ensure that $a$ takes the expected value 
in case of a schedule holding a straddling task,
we will introduce a boolean variable 
to identify the task $\jj$ beginning at $\dma$.
It force to have in every schedule
a task beginning at $\dma$.
Therefore,
this first encoding is not valid
in case of a \dblocSing without tardy task.
We then propose a second encoding for the \dblocPlur.
It consists in choosing the starting time of $\jt$
as the new reference point,
which is setting $a \!=\! p_\jt$,
This second encoding can be also used
for a schedule holding an on-time task and having tardy tasks,
as illustrated by Figure~\ref{fig_aet_ont_bis}.\\


%

To sum up, 
the first encoding, with  $a \!=\! 0$,
is suitable for  \dblocPlur,
except those without tardy tasks, 
\blue{and} 
the second encoding, with $a \!=\! p_\jt$,
is suitable for any \dblocSing.
Fortunately,
the three encodings proposed in this section
can be decoded in the same way :
$C \!=\!\big( \dma \!-\! e'_j \!+\! t'_j \big)_{j\in J}$
gives the completion times of the encoded schedule.\\

$\bullet$ \textit{Consistency between $e'$ and $t'$ using $\delta$ variables}\\
To ensure consistency between $e'$ and $t'$,
we use again variables $\delta$.
\blue{In} the previous formulation,
$\delta_j$ indicated if task $j$ completes before or at $d$.
\blue{In} this formulation 
$\delta_j$ indicates if the task completes before or at $\dma$.
We also use inequalities~\eqref{e0}-\eqref{t1} 
where $e$ (resp. $t$) 
are replaced by $e'$ (resp. $t'$).
These inequalities will be \blue{denoted by}~(\ref{e0}')-(\ref{t1}') in the sequel.

Note that $\delta_j$ no longer necessarily indicates
if task $j$ is early or not.
Keeping the previous notations 
$\Ed(\delta) \!=\!\{ j\!\in\! J \,|\, \delta_j \!=\!1\}$
and 
$\Td(\delta) \!=\!\{ j\!\in\! J \,|\, \delta_j \!=\!0\}$,
$\big(\Ed(\delta),\Td(\delta)\big)$ is not the partition between early and tardy tasks 
as soon as we use the second encoding for a \dblocSing.
Therefore,
we introduce a new partition of tasks:
if $C$ encodes a schedule by its completion times, 
we define
$\Etc(C) \!=\! \{\, j\!\in\! J\,|\, C_j \!<\! d \}$
and 
$\Ttc(C) \!=\! \{\, j\!\in\! J\,|\,  \blue{C_j \!\geqslant\! d}\, \}$.

Note that if there is a straddling task in the schedule, 
then $\Ec(C) \!=\! \Etc(C)$ and $\Tc(C) \!=\! \Ttc(C)$.
\blue{Moreover},
the only encoding in this case is such that
$\Ed(\delta) \!=\! \Ec(C)\!=\! \Etc(C)$ 
and $\Td(\delta) \!=\! \Tc(C)\!=\! \Ttc(C)$.\\
In the case of a \dblocSing, 
using the first encoding we also have 
$\Ed(\delta) \!=\! \Ec(C)$ 	and $\Td(\delta) \!=\! \Tc(C)$,
but using the second one we have
$\Ed(\delta) \!=\! \Etc(C)$ and $\Td(\delta) \!=\! \Ttc(C)$.\\[-0.2cm]

$\bullet$ \textit{Handling the positivity\\}
Since the due date can be smaller than $p(J)$,
avoiding \blue{overlaps} and idle times 
does not ensure the positivity constraint.
Therefore,
we add the following inequalities 
ensuring that $e'_j \!+\! p_j \!\leqslant\! \dma$ 
for each task $j$ completing before $\dma$.
They are valid since $d$ is an upper bound of $a$.\\[-1cm]

\begin{eqnarray}
\forall j \!\in\! J,\enskip && e'_j \!+\! p_j \delta_j  \leqslant d \!-\! a \label{a1}
\end{eqnarray}\\[-0.7cm]

$\bullet$ \textit{Handling the non-overlapping}\\
To ensure the non-overlapping, 
we use again variables $x$,
satisfying~(\ref{x.1}-\ref{x.4}) 
and the inequalities~(\ref{S1}) and~(\ref{S2}),
where $e$ (resp. $t$) 
are replaced by $e'$ (resp. $t'$).
These inequalities will be \blue{denoted by}~(\ref{S1}') and (\ref{S2}') in the sequel.\\[-0.4cm]

In order to ensure that tasks completing before or at $\dma$
do not overlap using inequalities~(\ref{S1}'),
inequalities~(\ref{a1}) must not restrict too much $e'_j$ from above.
Indeed, an inequality of the form $C_j \!\leqslant\! M$
is compatible with the non-overlapping inequalities~(\ref{Q})
only if $M \!\geqslant p(J)$.
If $M \!<\! p(J)$,
adding such an inequality makes appear 
extreme points which can be reached by minimization,
whereas they do not correspond to feasible schedules.

\blue{
For example,
let us consider the instance defined by 
$J \!=\! \llbracket 1 , 2\rrbracket$, 
$d \!=\! 5$, 
$p_1 \!=\! p_2 \!=\! 3,\, 
\alpha_1 \!=\! \alpha_2 \!=\! 1,\,
\beta_1 \!=\! \beta_2 \!=\! 10$,
and the polyhedron 
$P \!=\! 
\left\lbrace\begin{tabular}{@{ }l|l@{ }}
$(e,t,\delta,x,a) \!\in\! \mathbb{R}^J \!\times\! \mathbb{R}^J \!\times\!  
[0,1]^J \!\times\!  \mathbb{R}^{J^<} \!\times\! \mathbb{R}$
& (\ref{e0}')-(\ref{t1}'), \eqref{x.1}-\eqref{x.4}, (\ref{S1}'),(\ref{S2}'), \eqref{a1}
\end{tabular}\right\rbrace$
defined by the inequalities introduced for the general case so far.
The vector $X\!=\!(2,2,0,0,1,1,0,0)$, 
is an integer extreme point of $P$.
It corresponds to the schedule $\mathcal{S}$
where both tasks complete at time $3$,
since $e'_1 \!=\! e'_2 \!=\!2$ and $a\!=\!0$.
The induced penalty is 4,
which is the minimal penalty over $P$.
However,
$\mathcal{S}$ 
is infeasible since the two tasks overlap.
This overlapp occurs in spite of inequalities~(\ref{S1}') 
because  $\dma \!=\!5 \!<\! 6 \!=\! p(E)$,
implying that $\dma$ is a too restrictive upper bound 
in inequalities~\eqref{a1}. 
To prevent this restrictiveness,
we introduce the following inequality.
}
\begin{eqnarray}
 && \sum\limits_{j\in J} p_j \delta_j \leqslant \dma\label{a3}
\end{eqnarray}\\[-0.8cm]

\blue{
To ensure that inequalities~(\ref{S1}') (resp.~(\ref{S2}'))
prevent overlaps of tasks completing before (resp. after) $\dma$ do not overlap,
the total penalty must be a \blue{nonincreasing} function
of variable $e'_j$ (resp. $t'_j$)  
for each task $j$ such that $\delta_j \!=\!1$ 
(resp. $\delta_j \!=\!0$).
We have to  provide linear inequalities
ensuring that the variable $a$ takes a value
such that the objective function fulfils these two conditions.
If $a$ is such that $\dma$ is the starting time of
the straddling task, the on-time task or the first tardy task 
as proposed by the previous encodings,
then these two conditions are ensured.}\\[-0.2cm]

$\bullet$ \textit{Ensuring that $a$ takes the expected value}\\
In spite of their apparent symmetry,
the two \blue{conditions} 
are completely different.\\[-0.4cm]

To ensure the first one,
it suffices to ensure that 
any task completing before or at $\dma$
completes before or at $d$.
Indeed, reducing $e'_j$ for such a task $j$ 
while satisfying the inequality~(\ref{S1}') 
associated \blue{with} $\{j\}$,
\ie $e'_j \!\geqslant\! 0$,
task $j$ remains early and its tardiness decreases,
which reduces the induced penalty.
Therefore, the first constraint 
is guaranteed by the following inequality.
\begin{eqnarray}
 && a \geqslant 0 \label{a0} 
\end{eqnarray}\\[-0.9cm]

\begin{figure}[h]
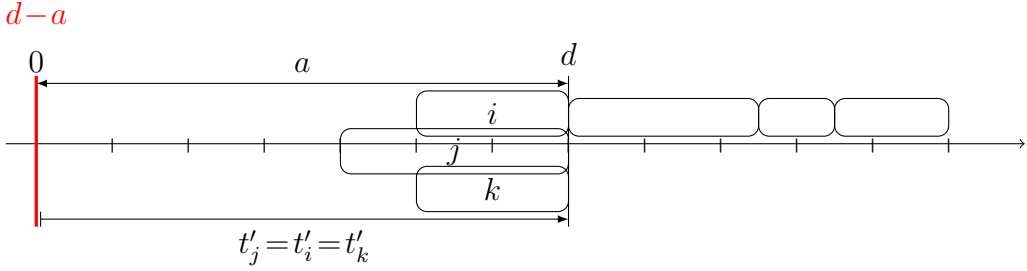

\centering
\figpba\\[-0.4cm]
\caption{An \blue{infeasible} schedule when $a \!=\!d$}
\label{fig_pb_a}
\end{figure}

To ensure the second one,
ensuring that any task completing after $\dma$ completes after or at $d$
is not sufficient.
Indeed, reducing $t'_j$ for such a task $j$ 
while satisfying the inequality~(\ref{S2}') 
associated \blue{with} $\{j\}$,
\ie $t'_j \!\geqslant\! p_j$,
task $j$ can become early,
so the induced penalty does not necessarily decrease.
Figure~\ref{fig_pb_a} illustrates 
the extreme case of this phenomenon,
that is when $a \!=\! d$, 
$\Ed(\delta) \!=\! \emptyset$,
and all early tasks overlap each others to be on-time.\\[-0.6cm]

\vspace*{-0.3cm}
Note that this case appears even if we add inequalities 
$\forall j \!\in\! J,\, t'_j \!\geqslant\! a (1 \!-\! \delta_j)$.
Adding inequalities 
$\forall j \!\in\! J,\, p_j \!\geqslant\! a (1 \!-\! \delta_j)$,
could avoid this issue,
but unfortunately they are not valid,
since a task completing after $\dma$ 
can be shorter than $a$,
as longer it is not the first one.\\
In order to ensure that 
the first task $\jj$ completing after $\dma$
completes after or at $d$,
we introduce a boolean variable $\gamma_j$ for each task $j$,
representing if $j$ is $\jj$,
and the following inequalities.
\begin{eqnarray}
&& \sum\limits_{j\in J} \gamma_j = 1 \label{g1}\\
\forall j \!\in\! J,\enskip && \delta_j \leqslant 1 \!-\! \gamma_j \label{g2}\\
\forall j \!\in\! J,\enskip && t'_j \leqslant p_j + (1 \!-\! \gamma_j)\, (p(J) \!-\! p_j) \label{g3}
\end{eqnarray}

Whereas inequalities~\eqref{g1}-\eqref{g2} ensure that 
$\gamma$ designates one and only one task $\ii$
among those completing after $\dma$,
inequalities~(\ref{g3}) ensure that $\ii$
is the first one, \ie $\ii \!=\! \jj$.
Indeed,
they ensure that $t'_\ii \!\leqslant\! p_\ii$,
and since $t'_\ii \!\geqslant\! p_\ii$
by inequality~(\ref{S2}') 
associated to the singleton $\{\ii\}$,
we deduce that $t'_\ii \!=\! p_\ii$.
Then, the inequality~(\ref{S2}') 
associated to a pair $\{\ii,j\}$ with $\delta_j \!=\!0$,
suffices to prove that task $j$ 
completes after $\ii$.

\begin{lm}
Let  $(t',\delta,x,\gamma) \!\in\! \mathbb{R}^J \!\times\!\{0,1\}^J\!\times\! [0,1]^{J^<}\!\times\! [0,1]^{J}$.\\
(i) $\gamma \!\in\!\{0,1\}^J$ and  ($\gamma$,$\delta$) satisfies~\eqref{g1}-\eqref{g2}
$ \Leftrightarrow \exists \ii \!\in\! T(\delta),\, \gamma \!=\! \mathbb{1}_\ii $ \\
(ii) \blue{\textbf{If}} (i) holds and $t',\delta,x$ satisfy 
\eqref{x.1}-\eqref{x.4}, \eqref{g3} and ($\ref{S2}'$),
\textbf{then} $t'_\ii \!=\! p_\ii$
and $\forall j \!\in\! T(\delta),\, j \!\neq\! \ii,\, t'_j \!\geqslant\! t'_\ii \!+\!p_j $.\\[-0.4cm]
\label{lm_gamma}
\end{lm}

Using $\gamma$, which identifies $\jj$,
we add the following valid inequalities
to ensure that 
$a \!\leqslant\! p_\jj \!=\! t'_\jj$.\\[-0.5cm]
\begin{eqnarray}
\forall j \!\in\! J,\enskip && a \leqslant p_j \!+\! ( 1 \!-\! \gamma_j)\, d \label{a2}
\end{eqnarray}\\[-0.4cm]

$\bullet$ \textit{A linear objective function using $e'$,$t'$, $a$ and $b$ variables}\\
Using $e'$ and $t'$ variables instead of $e$ and $t$ 
offers an easy way to ensure positivity, consistency and non-overlap
at the expense of a linearization of the product $a \delta_j$.
Indeed, in the objective function,
we need a linear expression 
for the earliness (resp. the tardiness)
of any task $j$ in $J$,
which is equal to $e'_j \!+\! a \delta_j$ 
(resp. to $t'_j \!-\! a (1 \!-\! \delta_j)$).

Then we introduce a variable $b_j$ for each task $j$ in $J$
to replace the product $a \delta_j$.
We add the following inequalities 
to ensure that $b$ variables take the expected values.
\begin{eqnarray}
\forall j \!\in\! J,\enskip && b_j \geqslant 0 \label{ap0}\\
\forall j \!\in\! J,\enskip && b_j \leqslant a \label{ap1}\\
\forall j \!\in\! J,\enskip && b_j \leqslant \delta_j  d \label{ap2}\\
\forall j \!\in\! J,\enskip && b_j \geqslant a -  (1 \!-\! \delta_j) \,d \label{ap3}
\end{eqnarray}

\noindent
Since $d$ is an upper bound of $a$ by construction,
we get the following lemma.
\begin{lm}
Let  $(a,b,\delta) \!\in\! \mathbb{R} \!\times\! \mathbb{R}^J \!\times\! \{0,1\}^J$.\\
$a,b$ and $\delta$ satisfy inequalities~\eqref{ap0}-\eqref{ap3}
$ \Leftrightarrow  b\!=a\,\delta $.\\[-0.4cm]
\label{lm_ap}
\end{lm}

Then the total penalty of a schedule encoded by $(e',t',a,b)$
is \\[-0.3cm]
$$\hab (e',t',a,b) = 
\sum\limits_{j\in J} \alpha_j e'_j + \beta_j t'_j 
+(\alpha_j\!+\! \beta_j)\, b_j  - \beta_j a$$
which is linear.
If $C$ encodes a schedule by its completion times,
the two possible vectors $(e',t',a,b)$ encoding this schedule are
\blue{the following}.
$$ \theta' (C) = \left( 
\big( [\dma \!-\! C_j]^+ \big)_{j \in J},\,
\big ([C_j \!-\! (\dma)]^+ \big)_{j \in J},\,
a,\,
a \,\mathbb{1}_{E(C)} 
\right)
\text{ where }
a = d -\!\! \min\limits_{i \in T(C)} C_i \!-\! p_i$$ 
$$ \widetilde{\theta'} (C) = \left( 
\big( [\dmat \!-\! C_j]^+ \big)_{j \in J},\,
\big ([C_j \!-\! (\dmat)]^+ \big)_{j \in J},\,
\ati,\,
\ati \,\mathbb{1}_{\Et(C)} 
\right)
\text{ where }
\ati = d -\!\! \min\limits_{i \in \Tt(C)} C_i \!-\! p_i$$

Note that if the schedule holds a straddling task, 
then $ \theta (C) \!=\! \widetilde{\theta'} (C)$,
since $\Ec(C) \!=\! \Etc(C)$
and $\Tc(C) \!=\! \Ttc(C)$.
Even for a schedule admitting two different encodings,
(\ie for a \dSing with at least one tardy task)
the function $\hab$ holds the total penalty,
as long as the schedule satisfies 
the non-overlapping constraint.


\begin{lm}
Let $C \!\in\! \mathbb{R}^J$.
\imp{If} $C$ satisfies (\ref{non-ch}),
\imp{then}
$\hab\big(\theta'(C)\big) \!=\! \hab\big(\widetilde{\theta'}(C)\big) \!=\! \fab\big(\theta'(C)\big)$.\\[-0.4cm]
\label{lm_h}
\end{lm}
\vspace{0.2cm}

$\bullet$ \textit{Formulation \Faetdx}\\
Let us define the polyhedron\\[-0.3cm]
$$
\boldsymbol{\Paetdx} \!=\! 
\left\lbrace\begin{tabular}{@{ }l|l@{ }}
$(e',t',\delta,x,a,b,\gamma) \!\in\! 
\mathbb{R}^J \!\times\! \mathbb{R}^J \!\times\!  
[0,1]^J \!\times\!  \mathbb{R}^{J^<} \!\times\! 
\mathbb{R} \!\times\! \mathbb{R}^J \!\times\! 
[0,1]^J$
& (\ref{e0}')-(\ref{t1}'), \eqref{x.1}-\eqref{x.4},
\eqref{a1}-\eqref{a0}, \eqref{a2}-\eqref{ap0},\\
& \eqref{g1}-\eqref{a2}, (\ref{S1}')  and (\ref{S2}') are satisfied
\end{tabular}\right\rbrace.
$$

\noindent
Note that this polyhedron depends on $d$, in addition to $p$.
Inequalities (\ref{S1}') and (\ref{S2}')
require the same separation algorithm 
as for (\ref{S1}) and (\ref{S2}),
which will be developed in Section \ref{sec_sepa}.
\blue{
We introduce the operator \imp{$\intdg$},
which only keeps points with integer $\delta$ and $\gamma$.
For $V \!\subseteq\! \mathbb{R}^J \!\times\! \mathbb{R}^J \!\times\!  
\mathbb{R}^J \!\times\!  \mathbb{R}^{J^<} \hspace{-0.2cm }\times\! 
\mathbb{R} \!\times\! \mathbb{R}^J \!\times\! 
\mathbb{R}^J$,
the set 
$\intdg (V) \!=\!
\big\{ (e',t',\delta,x,a,b,\gamma) \!\in\! V \,|\, 
\delta \!\in\! \{0,1\}^J,\,
\gamma \!\in\! \{0,1\}^J \big\} $.
}
Finally,
our formulation for the general common due date problem is
$$\text{\Faetdx}
\begin{tabular}[]{l}
$\min \hab(e',t',a,b)$\\[0.1cm]
\hspace*{0.2cm}
s.t. $(e',t',\delta,x,a,b,\gamma) \!\in\!\intdg\big(\extr{\Paetdx}\big) $
\end{tabular}.
$$

\subsection{Validity of Formulation~\Faetdx}

Thanks to the natural variables $e'$ and $t'$,
ensuring the non-overlapping constraint
reduces to ensuring the positivity and  non-overlapping constraints
for two subsets of tasks.
In contrast with Formulation \Fetdx
where these two subsets are the early and the tardy tasks
(\cf Lemma~\ref{lm_ct_non-ch_local}),
in Formulation \Faetdx, 
the subsets to consider depend on
the occurrence of a straddling or an on-time task,
as detailed in the following lemma.\\[-0.4cm]

\begin{lm}
Let $C \!\in\! \mathbb{R}^J$.\\[0.1cm]
(i) \textbf{If} there exists $\js \!\in\! J$ such that 
$C_\js \!\!-\! p_\js \!<\! d \!<\! C_\js $ and
$(e',t',a,b) \!=\! \theta'(C) \!=\! \widetilde{\theta'}(C)$,\\
\hspace*{0.4cm}\textbf{then} 
$C$ satisfies (\ref{non-ch})
$\Leftrightarrow 
(e' \!+\! p)_{\slash E(C)}\text{ and  }
\enskip t'_{\slash T(C)}$satisfies  (\ref{pos}) and (\ref{non-ch}).
\\[0.1cm]
(ii) \textbf{If} there exists $j_t \!\in\! J$ such that
 $C_\jt \!=\! d$ and 
 $(e',t',a,b) \!=\! \widetilde{\theta'}(C)$,\\
\hspace*{0.4cm}\textbf{then} 
$C$ satisfies (\ref{non-ch})
$\Leftrightarrow 
(e' \!+\! p)_{\slash E(C)}\text{ and  }
\enskip t'_{\slash T(C)}$satisfies  (\ref{pos}) and (\ref{non-ch}).
\\[-0.4cm]
\label{lm_8bis}
\end{lm}


The following theorem establishes that
a feasible schedule, under some assumptions,
is encoded by an integer point of $\Paetdx$.
In particular a \dZblocSing is encoded by an integer point of $\Paetdx$.\\[-0.4cm]


\begin{thm}
Let $C \!\in\! \mathbb{R}^J$ 
satisfying~(\ref{pos}) and~(\ref{non-ch}).\\[0.1cm]
(i) \textbf{If} there exists $\js \!\in\! J$ such that 
$C_\js \!\!-\! p_\js \!<\! d \!<\! C_\js $, 
$\forall j\!\in\! J,\,d \!-\!p(J) \!\leqslant\! C_j \!-\! p_j $
and $C_j \!\leqslant\!C_\js\!\!-p_\js\!+p(J) $,\\ 
\hspace*{0.4cm}\textbf{then} there exists 
$X \!=\!(e',t',\delta,x,a,b,\gamma) \!\in\! \intdg(\Paetdx)$
such that $\theta'(C) \!=\!(e',t',a,b)$.\\[0.1cm]
(ii) \textbf{If} there exists $j_t \!\in\! J$ such that
 $C_\jt \!=\! d$, 
$\forall j\!\in\! J,\,d \!-\!p(J) \!\leqslant\! C_j \!-\! p_j $
and $C_j \!\leqslant\!C_\jt\!\!-p_\jt\!+ p(J) $,\\ 
\hspace*{0.4cm}\textbf{then} there exists 
$X \!=\!(e',t',\delta,x,a,b,\gamma) \!\in\! \intdg(\Paetdx)$
such that $\widetilde{\theta'}(C) \!=\!(e',t',a,b)$.
\\[-0.4cm]
\label{thm_aetdx_complet}
\end{thm}

\begin{proof}
Let us start by proving (i).\\
From $C$, let us set:
$(e',t',a,b) \!=\! \theta'(C),\enskip 
\delta \!=\! \mathbb{1}_{E(C)},\enskip 
x \!=\! \left(\mathbb{1}_{\delta_i \neq \delta_j}\right)_{(i,j)\in J^<},\enskip 
\gamma \!=\! \mathbb{1}_\js$
and  $X \!=\! (e',t',\delta,x,a,b,\gamma)$.
We will prove that $X \!\in\! \intdg(\Paetdx)$.\\
Note that the definition of $\delta$ ensures that
$\delta \!\in\! \{0,1\}^J$,
and that $\Ed(\delta) \!=\! \Ec(C)$ and $\Td(\delta)\!=\!\Tc(C)$,
which allows the notation $E$ and $T$
for sake of brevity.
By Lemma~\ref{lm_x}(i), 
the definition of $x$ ensures that 
inequalities~\eqref{x.1}-\eqref{x.4} are satisfied.
By Lemma~\ref{lm_gamma}(i), 
the definition of $\gamma$ ensures that 
inequalities~\eqref{g1}-\eqref{g2} are satisfied,
since $\js \!\in\! T$.
By Lemma~\ref{lm_ap}, 
inequalities~\eqref{ap0}-\eqref{ap3} are satisfied,
since $b \!=\! a\, \mathbb{1}_{\Ec(C)} \!=\! a\, \delta$.

For the straddling task $\js$,
we have 
$C_\js \!\!- p_\js =\! \min\limits_{j \in \Tc(C)} \left( C_j \!-\! p_j \right)$,
so $a \!=\! d - (C_\js \!\!- p_\js)$, by definition of $\theta'$.

Since task $\js$ starts at or after $0$ and before $d$,
\ie $0 \!\leqslant\! C_\js \!\!- p_\js \!<\! d$,
we have $0 \!<\! a \!\leqslant\! d$.
Thus inequality~(\ref{a0}) is satisfied,
and for any task $j \!\neq\! \js$,
$a \leqslant d \!+\! p_j = (1 \!-\!\gamma_j)\,d + p_j $.
More precisely, task $\js$ starts after all early tasks,
and since they do not overlap,
$p(E) \!\leqslant\! C_\js \!\!- p_\js \!= \dma$,
thus inequality~(\ref{a3}) holds.
Since task $j_s$ completes after $d$,
\ie $C_\js \!>\! d\,$,
we get $a = p_\js + (d \!-\! C_\js) < p_\js = p_\js + (1 \!-\! \gamma _\js)\, d$.
We deduce that inequalities~(\ref{a2}) are satisfied.

Inequalities~(\ref{e0}') and~(\ref{t0}')
are satisfied by construction of $e'$ and $t'$.

For a task $j$ in $E$,
$C_j \!\leqslant\! C_\js \!\!-p_\js \!= \dma$
since $j$ and $j_0$ do not overlap,
then $e'_j \!=\! \dma - C_j$ and $t'_j \!=\! 0$.
The corresponding inequality~(\ref{t1}') is thus satisfied,
as well as~(\ref{g3}) since 
$p_j +(1 \!-\! \gamma_j) \, \big(p(J) \!-\! p_j\big) =p(J) \geqslant 0$.
By assumption $C_j \!\geqslant\! d\!-\! p(J) \!+\! p_j$,
thus $e'_j \!=\! \dma - C_j  \!\leqslant\! p(J) - p_j$,
and inequality~(\ref{e1}') is also satisfied  for $j$.
Moreover, $d - e'_j \!-p_j \delta_j = a + C_j\! -p_j$,
and by positivity constraint $C_j \!-p_j \geqslant 0$,
thus $d - e'_j \!-p_j \delta_j \geqslant a$ 
and inequality~(\ref{a1}) is satisfied  for $j$.

For a task $j$ in $T$,
$C_j \!\geqslant\! d  \geqslant  \dma$,
then  $e'_j \!=\!0$ and $t'_j \!=\!  C_j \!- (\dma)$.
The corresponding inequality~(\ref{e1}') is thus satisfied.
Moreover, $d - e'_j \!-p_j \delta_j = d \geqslant a$,
then inequality~(\ref{a1}) is satisfied for $j$.
By assumption $C_j \!\leqslant\! C_\jj\!\!-p_\jj\!+\!p(J)  \!\leqslant\! d\!+\! p(J)$,
thus $t'_j \!\leqslant\!  \big(d\!+\! p(J)\big) - (\dma) $
and then $t'_j\!\leqslant\! p(J)$.
We deduce that 
the corresponding inequality~(\ref{t1}') is also satisfied,
as well as inequality~(\ref{g3}), since
$p_j +(1 \!-\! \gamma_j) \, \big(p(J) \!-\! p_j\big)$
is equal to $p(J)$ (resp.  to $p_\js \!= C_\jj \!- (\dma) = t'_\js$)
if $j \!\neq\! \js$ (resp. if $j \!=\! \js$).

Since $C$ encodes a feasible schedule, 
$C$ satisfies~(\ref{non-ch}).
Using Lemma~\ref{lm_8bis},
$(e' \!+\! p)_{\slash E}$ ,
as well as $t'_{\slash T}$,
satisfies~(\ref{pos}) and~(\ref{non-ch}).
Applying Property~\ref{lm_valide} to these two vectors,
we deduce that they satisfy~(\ref{Q}), 
and using Lemma~\ref{lm_S_Q}, 
that $e',\delta,x$ satisfy~(\ref{S1}') and $t',\delta,x$ satisfy~(\ref{S2}').
Thus, $X$ belongs to $\intdg(\Paetdx)$.

Rewriting the proof by replacing 
$\theta'$  by $\widetilde{\theta'}$,
$\Ec(C)$ by $\Etc(C)$,
$\Tc(C)$ by $\Ttc(C)$,
and the straddling task $\js$ by the on-time task $\jt$
provides almost the proof of (ii).
The only difference lies in 
the justification of inequality~(\ref{a2}) for $\jt$:
in this case $C_\jt \!\!=\!d$,
then $a = p_\js + (d \!-\! C_\js) = p_\js = p_\js + (1 \!-\! \gamma _\js)\, d$.
\end{proof}

The following theorem establishes that
an optimal solution of formulation \Faetdx is a solution 
for the general common due date problem. 
\blue{The proof is given in Appendix.}

\begin{thm}
\textbf{Let} $X^* \!=\!(e',t',\delta,x,a,b,\gamma) \!\in\! \PaetdxE $.\\
\textbf{If} $\alpha \!\in\! \RpeJ\!\!$, 
$X^*\!\in\!\text{extr}(\Paetdx)$
and $(e',t',a,b)$ minimizes $\hab$
\textbf{then} $X^*$ encodes a \dZblocSing,
by $\theta'$ or $\widetilde{\theta'}$.
\\[-0.4cm]
\label{thm_aetdx_correct}
\end{thm}

If some tasks have a zero unit earliness penalty,
formulation \Faetdx provides a vector
$X^* \!=\!(e',t',\delta,x,a,b,\gamma)$
which partially encodes an optimal schedule.
Indeed, except for early tasks
having a zero unit earliness penalty,
the completion time of a task $j$
is given as previously by
$C^*_j = (\dma) - e'_j\!+t'_j$.
Conversely, for an early task $j$
such that $\alpha_j \!=\!0$,
$e'_j $ could be $d  \!-\!p_j$ for instance
and the previous encoding would give $C^*_j \!=\! p_j$.
If there are several early tasks having zero unit earliness penalty,
an \blue{overlap} would appear at time $0$.\\
Since their unit earliness penalty is zero,
the minimality of $X^*$ does not ensure
that these tasks are well spread out
(in this context Lemma~\ref{lm_cle_2} cannot be applied).
However,
the minimality of $X^*$ ensures that
the other early tasks
(\ie having a non-zero unit earliness penalty)
are right-tight with respect to $d$.
Hence, using inequality~\eqref{a3},
there is enough time between $0$ and their processing duration
to process the overlapping tasks.
Thus, it suffices to schedule these tasks
in an arbitrary order from time $0$
to obtain a feasible schedule \S.
Since these tasks do not induce any penalty,
the total penalty of \S is $\hab(X^*)$,
regardless of their order.
We deduce that \S is an optimal schedule.\\[-0.2cm]

The following theorem establishes that
the general common due date problem 
reduces to solving formulation \Faetdx.
We omit the proof since 
it follows the same lines as 
the one of Theorem~\ref{thm_etdx_final}.
\begin{thm} 
Any optimal \dZblocSing,
is encoded by a vector
minimizing $ \hab$ 
on $\intdg\big(\extr{\Paetdx}\big)$.\\
Conversely, any vector minimizing $\hab$ 
on  $\intdg\big(\extr{\Paetdx}\big)$,
encodes an optimal \dZblocSing.\\[-0.4cm]
\label{thm_aetdx_final}
\end{thm}

\newpage
\section{Separation algorithms\\[-0.6cm]}
\label{sec_sepa}

In this section, 
we explain how to separate
inequalities~(\ref{S1}),(\ref{S2}),
(\ref{S1}') or (\ref{S2}'),
by solving a min-cut problem in a suitable graph.
We write the following development
for inequalities~(\ref{S1}) and~(\ref{S2}),
but a rewriting exercise suffices 
to obtain the equivalent results for
inequalities~(\ref{S1}') and~(\ref{S2}').

Let $X \!=\! (e,t,\delta,x)  
\!\in\! \mathbb{R}^J \!\times\! \mathbb{R}^J \!\times\!  
[0,1]^J \!\times\!  \mathbb{R}^{J^<}$ a vector satisfying 
inequalities~(\ref{e0}-\ref{t1}) and (\ref{x.1}-\ref{x.4}).
The separation problem for inequalities~(\ref{S1}) 
is to find a subset $S$ of $J$ such that 
$X$ does not satisfy the associated inequality~(\ref{S1}) 
or to guarantee that $X$ satisfies all inequalities~(\ref{S1}).\\
We will first show that this separation problem reduces
to the maximization of a set function $\gamcq$ defined 
from parameters $(c,q) \!\in\! \mathbb{R}^J \!\times\! \mathbb{R}^{J^<}$ as 
$\forall S \!\subseteq\! J,\enskip \gamcq(S)  =\!\!
\sum\limits_{(i,j)\in S^<} \!\! q_{i,j}+
\sum\limits_{i\in S}c_i$.\\[0.4cm]
Indeed we have:\\[-1.4cm]
\begin{align*}\hspace*{1.8cm}
X \text{ satisfies }(\ref{S1}) 
& \Leftrightarrow \forall S \subseteq J,
\sum\limits_{(i,j)\in S^<} \!\! p_i p_j \,
\frac{\delta_i \!+\!\delta_j \!-\! x_{i,j}}{2} \,
\leqslant  
\sum\limits_{i\in S} p_i e_i 
 \\
& \Leftrightarrow \forall S \subseteq J,
\sum\limits_{(i,j)\in S^<} \!\! p_i p_j \,
(\delta_i \!+\!\delta_j \!-\! x_{i,j}) 
- 2 \sum\limits_{i\in S} p_i e_i 
\leqslant 0\\
& \Leftrightarrow \forall S \subseteq J,\enskip 
\gamun(S)\leqslant 0.
\end{align*}
where $c^1 \!=\! -2\, \Big( p_j e_j \Big)_{j\in J}$
and $q^1 \!=\! \Big(p_i p_j \, (\delta_i \!+\!\delta_j \!-\! x_{i,j}) \Big)_{(i,j)\in J^<}$.
Then it suffices to maximize $\gamun$ over the subsets of $J$.
Indeed, if the obtained value is negative or zero,
then $X$ satisfies all inequalities~(\ref{S1}),
conversely if the obtained value is positive,
then the maximizing set is not empty and
corresponds to an inequality~(\ref{S1})
that $X$ does not satisfy.
Similarly, the separation problem of inequalities~(\ref{S2}),
is equivalent to the maximization of $\gamde$ where
$c^2 \!=\! 2\,\Big((1\!-\!\delta_j)p_j^2 - p_jt_j \Big)_{j\in J}$
and $q^2 \!=\!\Big( p_i p_j \, 
(2 \!-\! \big( \delta_i \!+\!\delta_j) \!-\! x_{i,j}\big)
\Big)_{(i,j)\in J^<}$.

\begin{figure}[h]
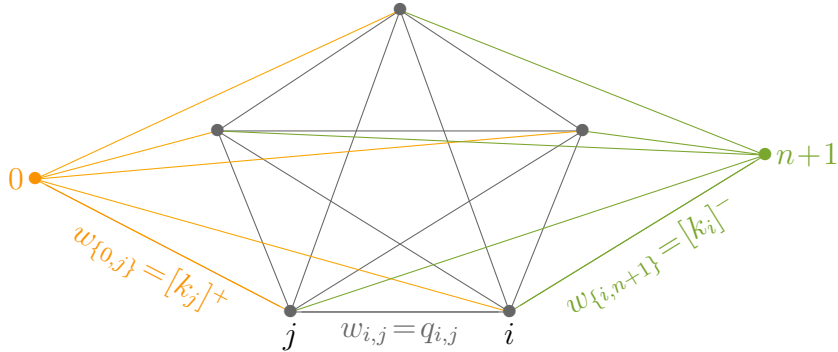

\centering
\figGraphSepa\\[-0.4cm]
\caption{Illustration of the weighted undirected graph $G$ for $n \!=\! 5$}
\label{fig_Graph_Sepa}
\end{figure}

Note that in both definitions of $\gamun$ and $\gamde$,
the parameter $q$ is non-negative since
$\delta$ and $x$ satisfy inequalities~(\ref{x.3}-\ref{x.4}).
Therefore, let us now explain 
how to reduce the maximization of $\gamcq$ 
for  $(c,q) \!\in\! \mathbb{R}^J \!\!\times\! (\Rpe)^{J^<}$
to a min-cut problem in an undirected graph
as proposed by \cite{min_cut}.
Let us assume that $J \!=\! \unn$ for sake of brevity.
We consider the weighted undirected graph $G \!=\! (V,\blue{A},w)$,
where $V \!=\! \llbracket 0, n\!+\!1 \rrbracket$,
$\blue{A} \!=\! \big\{ \{i,j\} \,|\, (i,j) \!\in\! V^2, \{i,j\} \!\neq\! \{0, n\!+\!1 \} \big\}$,
$\forall j \!\in\! J,\ 
w_{\{0,j\}} \!=\! \left[ k_j \right]^+ \!\!,\,
w_{\{j, n+1\}} \!=\! \left[ k_j \right]^-$
where
$k_j \!=\!  2 c_i + \!\! \sum\limits_{i=1}^{j-1} q_{i,j} + \hspace{-0.2cm}\sum\limits_{k=j+1}^{n}\!\! q_{j,k}$,
and
$\forall (i,j) \!\in\! J^<,\, w_{\{i,j\}} \!=\! q_{i,j}$.
Figure~\ref{fig_Graph_Sepa} gives an illustration of such a graph 
for $n \!=\! 5$.\\[-0.6cm]

Note that $V$ and $\blue{A}$ only depend on $J$,
and $w$ only depends on parameters $c$ and $q$.
For a \blue{\imp{cut}} $(W,\widebar{W})$,
\ie$W \!\cap \widebar{W} \!=\! \emptyset$ and $W \!\cup \widebar{W} \!=\! V$,
let $w(W,\widebar{W})$ denote its weight according to $w$,
\ie $w(W,\widebar{W}) \!=\! 
\sum\limits_{i\in W,j\in \widebar{W}} w_{\{i,j\}}$.\\
Let us introduce three constants:
$Q \!=\!\!\! \sum\limits_{(i,j) \in J^<} \!\! q_{i,j}$,
$C \!=\!\sum\limits_{j \in J}  c_j$
and $K \!=\!\sum\limits_{j \in J}  |k_j|$.
\\[0.5cm]
Hence, for any $S \!\subseteq\! J$:
\\[-0.9cm]
$$\hspace{0.3cm}\gamcq(S) = -\frac{1}{2}\, 
w \big(S \!\cup\! \{0\}, \llbracket1, n \!+\!1\rrbracket\backslash S\big)
+\frac{Q \!+\! C}{2} + \frac{K}{4}.$$

Since $Q,C,K$ do not depend on $S$,
finding a subset $S$ maximizing $\gamcq$
is equivalent to finding a minimum cut 
separating the additional vertices $0$ and $n\!+\!1$.
Since $w$ is positive,
this problem is solvable in polynomial time, 
using the \cite{Gomory_et_Hu_61} algorithm,
as it will be explained in the next section.\\[-0.6cm]

\input{tables_article.tex}
\section{Experimental results\\[-0.6cm]}
\label{sec_exp}

The experiments are conducted on a single thread on
a machine with 
Intel(R) Xeon(R) CPU E5-2630 v2 @2.60GHz,
and 16Gb RAM.
We use the solver CPLEX version 12.6.3.0,
and the open source C++ optimization library LEMON \cite{lemon}.
The branching scheme and the management of the current bounds
is done by CPLEX.
\modif{The time limit is set to 3600 seconds. For sake of comparison,
all the formulations use CPLEX Default.}
\\[-0.2cm]

$\bullet$ \textit{Implementation of the separation algorithm}\\
The separation of inequalities~(\ref{S1}) and~(\ref{S2})
is implemented using the so-called Callback functions proposed by CPLEX.
The separation algorithm consists in \blue{the following steps}.\\
\begin{tabular}{r@{ }p{12.5cm}}
1.&Computing the weights $w_{\{i,j\}}$ introduced in Section~\ref{sec_sepa}
according to the value of variables $e,t,\delta,x$ (resp. $e',t',\delta,x$)
in the solution provided by CPLEX. \\
2.&Running the \cite{Gomory_et_Hu_61} algorithm
provided by LEMON to obtain the Gomory-Hu tree rooted in $0$.\\
3.& Finding all minimum cost edges 
along the path between $0$ and $n\!+\!1$
in the Gomory-Hu tree.\\
4.& Testing for any of such edges 
if the related cut $W/\widebar{W}$ such that $0 \!\in\! W$ 
corresponds to a negative value.
\\
5.& Adding in the model
the inequality (\ref{S1}) (resp. (\ref{S2}))
associated to $S$, where $S \!=\! W \setminus\!\{0\}$,
if there exists.
\end{tabular}\\
\modif{Due to these Callback functions, some CPLEX features are disabled in \Fetdx and \Faetdx.}\\

\modif{$\bullet$ \textit{ Biskup and Feldmann's benchmark}}\\
We test our three formulations
on the benchmark proposed by~\cite{benchmark},
available online on \blue{OR-Library~\cite{benchmark_url}}.
For each number of tasks 
$n \!\in\! \{10,20,50\}$,
ten triples $(p,\alpha,\beta)$ of $\big(\mathbb{N}_*^n\big)^3$ are given.
For each one,
setting \modif{$d \!=\! \left\lfloor h\, p(J) \right\rfloor$}
for $h\!\in\!\{0.2,\, 0.4,\, 0.6,\, 0.8,\, 1\}$,
gives five instances, 
including one with an unrestrictive due date
corresponding to $h\!=\!1$.  
We obtain $30$-task and $40$-task instances,
by considering only the first tasks of
$50$-task instances.
In the following,
the average values considered are computed 
over the ten instances proposed by this benchmark 
for fixed values of $n$ and $h$,
unless otherwise specified.

\cite{Sourd_09} succeeded 
in solving instances of this benchmark having up to $1000$ tasks.
The running time does not exceed $1400$ seconds,
and the average running time for $1000$-tasks instances is between $611$ and $918$ seconds 
depending on the value of $h$.
He obtained these results thanks to a dedicated branch-and-bound algorithm
using Lagrangian relaxation and dynamic programming.
\modif{However, Sourd's approach is based on a time-indexed formulation which involves $O(np(J))$ variables and hence nodes in the graph used for computing the Lagrangian lower bound. The  Biskup and Feldmann's benchmark considers small values for the job processing times which ensures a fast computation time of the Lagrangian lower bound. \\
}

\newcommand{\pmax}{p_{max}}
$\bullet$ \modif{\textit{New benchmark with long processing times}}\\
In the Biskup and Feldmann's benchmark, 
processing times range is $[1,20]$.
We propose a benchmark where processing times 
are randomly drawn from the uniform distribution
$\mathcal{U}\left[\frac{\pmax}{10},\pmax \right]$
for $\pmax \!\in\! \{100,200,300\}$.
For each  $\pmax \!\in\! \{100,200,300\}$ 
and each $n \!\in\! \{10,20,30,40,50\}$,
we randomly generate ten triples $(p,\alpha,\beta)$ of $\big(\mathbb{N}_*^n\big)^3$.
For each task $j$, $\alpha_j$ and $\beta_j$ 
are randomly drawn from the uniform distribution
$\mathcal{U}[1..20]$.
By setting \modif{$d \!=\! \left\lfloor h\, p(J) \right\rfloor$}
for $h\!\in\!\{0.2,\, 0.4,\, 0.6,\, 0.8,\, 1\}$,
each triple gives five instances, 
including one unrestrictive,
which results in 750 instances.\\

$\bullet$ \textit{MIP formulations from the literature}\\
In order to assess our formulation efficiency,
we implement two other MIP formulations proposed in the literature:
the formulation \Fbf based on linear-ordering variables
proposed by~\cite{benchmark}
and the time-indexed formulation \Fti 
used in~\cite{Sourd_09}.\\

\newpage
\newcommand{\nd}{\#nd\! 1}
\newcommand{\ndd}{\#nd\! 2}
\renewcommand{\arraystretch}{1}
$\bullet$ \textit{Entries of the following tables}\\
\begin{tabular}{r@{ : }p{12cm}}
$n$& the number of tasks\\
$h$& the parameter setting the due date $d$ to 
\modif{$\left\lfloor h\, p(J) \right\rfloor$} 
(in the general case only)\\
\#opt & number of instances optimally solved 
among the ten proposed by the benchmark
\modif{under the 3600 seconds time limit}\\
avg-T& the average running time in seconds 
over the optimally solved  instances \\
gap & the average gap over the instances not solved to optimality, 
that is the relative gap between the best lower and upper bounds\\
\end{tabular}\\

\newcommand{\default}{Default\xspace}
\newcommand{\featurePlus}{Feature+\xspace}

\newcommand{\aef}[1]{\textcolor{magenta!50!blue}{#1}}

\subsection{Formulations for the unrestrictive case }
In this section the problem is solved using formulations 
\Fetdx and \Fdx, as well as formulations \Fbf 
and \Fti.
Table~\ref{tab_comp_bf_unres} presents the results obtained
on Biskup and Feldmann's benchmark,
while Table~\ref{tab_comp_pi_long_unres} 
presents those obtained on long processing times instances,
for $\pmax \!\in\! \{100,200,300\}$.
\tblCompBFunres

As shown in Table~\ref{tab_comp_bf_unres},
\Fbf is unable to solve any $20$-task instance
within the time limit.
Thus, \Fbf is not used 
neither for larger instance size,
nor for the new benchmark.
Other experiments show that \Fbf can only solve 
5 over 10 instances for $n\!=\!15$.
\Fti is able to optimally solve 
Biskup and Feldmann's instances up to size $50$
in less than 40 seconds.
\Fetdx is able to optimally solve 
instances up to size $30$ in around 40 seconds.
In contrast, 
ten minutes are required to optimally solve 
$40$-task instances and
\Fetdx fails to solve $50$-task instances within the time limit.
However, 
other experiments show that, 
under a time limit of 10\!~000~seconds,
\Fetdx solves 9 over the 10 instances 
for $n\!=\!50$,
with an average computation time of 4721 seconds.
\Fdx is able to optimally solve
all the instances up to size $50$ within the time limit.
Other experiments conducted without CPLEX features
show that \Fdx can be faster:
22 seconds for $n\!=\!40$,
215 seconds for $n\!=\!50$ and
4063 seconds for $n\!=\!60$.\\

\tblCompPiLongunres
As shown in Table~\ref{tab_comp_pi_long_unres}, the efficiency of \Fti greatly depends on the value of parameter $\pmax$, which was expected since the number of variables is related to the length of the horizon, 
\ie $2\,p(J)$.
While it only takes 40 seconds in average to solve all the $50$-task Biskup and Feldmann's instances, \Fti  solves 8 over the 10 $50$-task instances in the new benchmark
for $\pmax\!=\!100$, within 690 seconds in average. 
In addition, \Fti fails at solving any instance for $\pmax\!=\!300$ due to memory limitations.
CPLEX could not even provide a solution or a lower bound in this case. 
We can notice that for $20$-task instances, 
the computation time required is at least 360 seconds for $\pmax\!=\!200$ and $\pmax\!=\!300$.
\Fetdx is able to optimally solve all instances up to size $30$ regardless of $\pmax$ value, faster that \Fti. 
The same observation holds for \Fdx up to size $40$ regardless of $\pmax$ value.\\

To sum up for the unrestrictive case,
\Fti gives the bests results 
for the Biskup and Feldmann's benchmark.
However, this formulation is sensitive to the
total length of the processing times (\ie $p(J)$),
and is unable to solve the $50$-task instances 
with long processing times ($\pmax\!=\!300$).
In contrast,
the results obtained with \Fetdx and \Fdx 
do not significantly get worse with 
processing time increase.

\subsection{Formulations for the general case}
In this section the problem is solved using formulations \Faetdx
as well as \Fbf and \Fti.
Table~\ref{tab_comp_bf_res} presents the results obtained
on the Biskup and Feldmann's benchmark,
while Table~\ref{tab_comp_pi_long_res} 
presents those obtained on long processing times instances,
for $\pmax \!=\!200$.\\

\tblCompBFres
As shown in  Table~\ref{tab_comp_bf_res},
\Fbf is unable to solve 
any restrictive instance for $n\!=\!20$ 
within the time limit.
Thus, \Fbf is not used 
neither for larger instance size,
nor for the new benchmark.
Other experiments show that \Fbf
cannot solve $15$-task instance 
when $h\!=\!0.2$ and $h\!=\!0.4$.
When $h\!=\!0.6$ ( resp. $h\!=\!0.8$),
\Fbf solves 5 over the 10 instances for $n\!=\!15$,
using in average 2278 seconds (resp. 1575 seconds).
\Fti is able to optimally solve
all the Biskup and Feldmann's restrictive instances.
Moreover, the computation times are similar 
to those obtained for the unrestrictive instances:
less than 25 seconds for $30$-task instances.
\Faetdx is able to optimally solve   
all the instances up to size 20 
as well as the $30$-task instances when $h\!=\!0.2$.
However,
the computation time is much larger than for \Fti:
around 2 minutes for $n\!=\!20$ and
20 minutes for $n\!=\!30$ when $h\!=\!0.2$.\\
\tblCompPiLongres

As shown in Table~\ref{tab_comp_pi_long_res}, for long processing time instances with $\pmax\!=\!200$, 
\Fti optimally solves almost all the instances within the time limit up to $n\!=\!20$.
It is important to notice that, for similar size, \Fti computation time is significantly larger for long processing times instances than for Biskup and Feldmann's ones: for $n\!=\!20$, at least 116 seconds against a few seconds. 
In contrast, \Faetdx optimally solves 
all instances up to size 20 
along with $30$-task instances when $h\!=\!0.2$.
Note that, for these instances, 
\Faetdx is rather faster than \Fti for $n\!=\!20$.
However,
\Faetdx fails to solve instances with $n\!=\!30$.\\

For general case instances, we obtain 
the same conclusion drawn for the unrestrictive instances. 
\Fti is faster than \Faetdx 
for Biskup and Feldmann's instances,
while this is not the case 
for long processing times instances. \\

Other experiments show that \Faetdx 
used on unrestrictive Biskup and Feldmann's instances 
(\ie with $h\!=\!1$) 
is less efficient than \Fetdx:
77 seconds in average for the $20$-task instances,
and more than 1300 seconds for the six optimally solved $30$-task instances.
The following paragraph will exploit this remark.\\

$\bullet$ \textit{What is really an unrestrictive instance?}\\
We have defined a due date as unrestrictive 
as soon as $d \!\geqslant\! \sum p_j$,
since it is the common definition.
But according to~\cite{benchmark},
a due date must be said unrestrictive 
if solving the problem for an arbitrary due date 
gives a solution for this due date.
This definition raises two issues.
First, since for some instances 
there exist several optimal solutions 
whose early tasks do not have the same total length,
this definition depends on the algorithm,
or even on the execution of the algorithm.
Secondly, this definition requires 
an optimal solution
to be found to say if the instance is unrestrictive or not.
Therefore the prior definition is more convenient.
But this remark leads to the following \Fdx-\Faetdx procedure 
to solve a general instance.\\
\renewcommand{\arraystretch}{1.2}
\begin{tabular}{r@{ }p{17.5cm}}
1.& Solving the instance without considering $d$ using the formulation \Fdx. \\
2.& Testing if the total duration of early tasks is
smaller than the due date \ie $\sum\limits_{\delta_j=1}p_j \!\leqslant\! d$.\\[-0.3cm]
&If it is the case, 
then the solution obtained is optimal.\\
&Otherwise solving the instance considering $d$ using \Faetdx\\
\end{tabular}\\
\renewcommand{\arraystretch}{1}
On average on the Biskup and Feldmann's benchmark,
the total length of the early tasks 
in the optimal solutions is 60\% of the total length.
That means that in this benchmark,
instances with $h \!\geqslant\! 0.6$ (\ie $d \!\geqslant\! 0.6\, p(J)$)
are mostly unrestrictive as defined by~\cite{benchmark}.
For these instances, the \Fdx-\Faetdx procedure
can be relevant (but we do not present corresponding numerical results).

\subsection{\modif{Linear relaxations analysis for \Fdx}
}
\label{sub_sec_relax}
Table~\ref{tbl_DX-LP} 
shows that the lower bound provided by the linear relaxation \FdxLP of \Fdx 
is far from the optimal value (see the third column).
Note that other experiments show that
\Fetdx provides the same lower bound.
We try to strengthen this lower bound 
by adding CPLEX cuts and/or 
the triangle inequalities introduced by~\cite{Padberg_89}.
\tblDXLP

For $n \!\leqslant\!30$, 
adding the CPLEX cuts provides a better lower bound
than adding the triangle inequalities,
and combining both of them 
does not provide a better lower bound.
Conversely, for $n \!\geqslant\!40$,
adding the triangle inequalities provides a much better lower bound
than adding the CPLEX cuts,
and combining both of them 
provides almost the same bound 
as adding only triangle inequalities,
but reduces the running time. 
For instance, for $60$-task instances,
adding triangle inequalities 
reduces the gap from $92.8\%$ to $23.5\%$,
and combining them with the CPLEX cuts
reduces the running time 
from $746$ seconds to $337$ seconds.

These observations lead to look for other valid inequalities 
for the quadratic polytope defined and studied by~\cite{Padberg_89},
in order to strengthen our formulations.
Indeed, 
as triangle inequalities, such inequalities can improve
the lower bounds given by the linear relaxation of \Fdx,
but also \Fetdx and \Faetdx, where $\delta$ and $x$ variables 
satisfy the same inequalities.
These observations also drive 
to deal with the related algorithmic aspects.
Indeed, since directly adding such inequalities in the model 
increases significantly the computation times, 
we should define how to manage these inequalities,
for instance by providing a cutting-plane based algorithm. 

\newpage
\section{Conclusion\\[-0.6cm]}
In this paper, \modif{thanks to our theoretical contributions on the non-overlapping inequalities}, we proposed three new formulations
based on earliness/tardiness variables
to solve the common due date scheduling problem.
Our formulations allow to solve 
unrestrictive instances with up to $50$ tasks
and general instances up to $20$ tasks within few minutes.
\modif{While scheduling literature proposes pseudo-polynomial methods strongly dependant on the total length of processing times, our formulation size does not rely on this value.  }
Even if our results for the Biskup and Feldmann's benchmark 
are far from those presented by~\cite{Sourd_09}, 
our MIP formulations outperform 
the compact MIP formulation based on linear ordering variables.
\modif{In addition,
for instances with long processing times,
our formulations outperform the time-indexed formulation in the unrestrictive case.}
A key part in our work is
the theoretical study of the non-overlapping inequalities,
in particular Lemmas~\ref{lm_cle_1} and~\ref{lm_cle_2}
and the scheme of proof used for 
Theorems~\ref{thm_etdx_complet} and~\ref{thm_etdx_correct}
(resp$.$~\ref{thm_aetdx_complet} and~\ref{thm_aetdx_correct}),
which should allow to extend our approach to 
other related scheduling problems.

Further works will focus on
the earliness-tardiness scheduling problem
with parallel machines, 
where each machine imposes the same due date for all the tasks.
Another issue is to address
the single machine common due date scheduling problem
with machine unavailability constraints.
For both problems formulations similar to \Faetdx can be derived.

An interesting issue is to study 
the polyhedra associated to such formulations,
to strengthen them using facet defining inequalities, as triangle inequalities,
which can be used in any formulation using $\delta$ and $x$
variables to describe a cut in a graph.


\bibliographystyle{plain}

\bibliography{biblio}
\newpage
\section*{Appendix : Proof of Theorem \ref{thm_aetdx_correct}}

Let us set, for any task $j$ in $J$, 
$C^*_j = (\dma) - e'_j +t'_j$.

The first step of the proof is to show that
$C^*$ gives the completion times of 
the schedule encoded by $X^*$ using $\theta'$ or $\widetilde{\theta'}$
\ie that $X^* \!\!=\! \theta'(C^*)$ or $X^* \!\!=\! \widetilde{\theta'}(C^*)$.\\
First we derive from inequalities~(\ref{e0}')-(\ref{t1}')
that $\forall j \!\in\! \Td(\delta),\, e'_j \!=\!0$ 
and  $\forall j \!\in\! \Ed(\delta),\, t'_j \!=\!0$.\\
Since $\delta$ and $\gamma$ are in $\{0,1\} ^J$,
and $X^*$ satisfies~\eqref{x.1}-\eqref{x.4}, \eqref{g1}-\eqref{g2} and (\ref{S2}'),
Lemma~\ref{lm_gamma} ensures that 
there exists $j_0 \!\in\! \Td(\delta)$  
such that $\gamma \!=\! \mathbb{1}_{\{j_0\}}$,
$t'_\jj \!\!=p_\jj$,
and  $\forall j \!\in\! \Td(\delta),\, j \!\neq\! \jj,\,$
$t'_j \geqslant t'_\jj \!\!+ p_j$.
Since $\jj$ is in $\Td(\delta)$,
$e'_\jj \!\!= 0$ and $C^*_\jj \!\!- p_\jj = \dma$.
Then for any other task $j$ in $\Td(\delta)$,
$C^*_j - p_j = (C^*_\jj \!\!- p_\jj)+t'j -p_j
\geqslant (C^*_\jj \!\!- p_\jj)+ t'_\jj 
= C^*_\jj > C^*_\jj \!\!- p_\jj$.
We deduce that
$C^*_\jj \!\!- p_\jj =\! \min\limits_{j \in \Td(\delta)} C^*_j \!-\! p_j$,
and then
$a \!=\! d -\!\! \min\limits_{j \in \Td(\delta)} C^*_j \!-\! p_j$.

The question is is whether $\Td(\delta) \!=\! T(C^*)$
or $\Td(\delta) \!=\! \Ttc(C^*)$.
Indeed, if $\Td(\delta) \!=\! \Tc(C^*)$, 
the value of $a$ is the one expected with the encoding $\theta'$,
whereas if $\Td(\delta) \!=\! \Ttc(C^*)$,
it is the one expected with $\widetilde{\theta'}$.

For any task $j \!\neq\! \jj$ in $\Td(\delta)$,
$C^*_j = \dma + t'j > \dma + t'_\jj \!\!= \dma +p_\jj $.
Since $\gamma_\jj \!\!=1$,
inequality~(\ref{a2}) gives $a\leqslant p_\jj$,
thus $C^*_j >d$.
We deduce that $\Td(\delta) \backslash \{\jj\} \!\subseteq\! \Tc(C^*)$.
Conversely, for a task $j$ in $\Tc(C^*)$,
$C^*_j > d$, which is equivalent to $t'_j -e'_j > a$.
Since $a \geqslant 0$ by inequality~(\ref{a0}),
$t'_j > e'_j$, which would be impossible if $j$ was in $\Ed(\delta)$,
according to inequalities~(\ref{e0}') and~(\ref{t1}').
We deduce that $\Tc(C^*) \subseteq \Td(\delta)$.
Two cases have to be considered.

$\enskip \rightarrow$ 
If $a \!<\! p_\jj$, then $C^*_\jj \!\!>\! d$,
 \ie $\jj \!\in\! \Tc(C^*)$,
and then
$\Td(\delta) \!=\! \Tc(C^*)$ and
$\Ed(\delta) \!=\! \Ec(C^*)$.

$\enskip \rightarrow$ 
If $a = p_\jj$, then $C^*_\jj \!\!= d$ 
and $\jj \!\in\! \Tt(C^*)$,
we deduce that $\Td(\delta) \!\subseteq\! \Ttc(C^*)$.
For $j$ in $\Ttc(C^*)$, 
either  $C_j \!=\!d$ 
or $j\!\in\! \Tc(C^*) \!\subseteq\! \Td(\delta)$, that is 
$t'_j = e'_j \!+ a = e'_j \!+p_\jj \! > e'_j$,
and necessarily $j \!\in\! \Td(\delta)$.
We conclude that 
$\Td(\delta) \!=\! \Ttc(C^*)$ and
$\Ed(\delta) \!=\! \Etc(C^*)$.

For the remainder of the proof, we assume that we are in the first case.
Then $E$ (resp. $T$) will denote $\Ec(C^*)$ (resp. $\Tc(C^*)$),
and we will use the encoding $\theta'$.
To handle the second case,
it suffices to replace 
$\Ec(C^*)$ by $\Etc(C^*)$,
$\Tc(C^*)$ by $\Ttc(C^*)$,
and $\theta'$  by $\widetilde{\theta'}$
in the second step,
and using that $\jj$ is an on-time task 
in the third step.

We can rewrite 
$\delta$ as $\mathbb{1}_{\Ec(C^*)}$,
and thus $b$ as $a\mathbb{1}_{\Ec(C^*)}$,
since $b \!=\! a \delta$
by inequalities~\eqref{ap0}-\eqref{ap3} 
and Lemma~\ref{lm_ap}.
Using inequalities~(\ref{e0}'-\ref{t1}'),
it is easy to show that $e' \!=\big( [\dma \!-\! C^*_j]^+ \big)_{j \in J}$
and $t'\! = \big ([C^*_j \!-\! (\dma)]^+ \big)_{j \in J}$.
Then we can conclude that 
$(e',t',a,b)\!=\! \theta'(C^*)$,
that is that $C^*$ and $(e',t',a,b)$ encode the same schedule,
which will be denoted by \Sopt.

The second step is to show that \Sopt is feasible, 
by proving that
$C^*$ satisfies~(\ref{pos}) and (\ref{non-ch}).\\
For a task $j$ in $E$,
inequality~(\ref{a1}) ensures that 
$p_j \leqslant \dma - e'_j \!= C^*_j$.
For a task $j$ in $T$,
inequality~(\ref{a1}) ensures that 
$ a \leqslant d$, then 
$C_j \!= \dma + t'_j \!\geqslant t'_j$.
For another, 
we deduce that $t'_j \!\geqslant p_j$
from inequality~(\ref{S2}') associated to $\{j\}$.
Thus $C^*$ satisfies~(\ref{pos}). 

To show that $C^*$ satisfies~(\ref{non-ch}) using Lemma~\ref{lm_8bis},
it remains to show that vectors
$(e' \!+\! p)_{\slash E}$ 
and  $t'_{\slash T}$ 
satisfy (\ref{pos}) and (\ref{non-ch}).
Since inequalities~(\ref{S1}') and~(\ref{S2}') are satisfied,
we know from Lemma~\ref{lm_S_Q} that 
$(e' \!+\! p)_{\slash E}$ and  $t'_{\slash T}$
satisfy inequalities~(\ref{Q}).
On one hand, 
these inequalities for the singletons
ensure that both vectors satisfy~(\ref{pos}).
On the other hand,
inequalities~(\ref{Q})
will allow us to show that 
both vectors satisfy~(\ref{non-ch}).

Let us assume that $(e' \!+\! p)_{\slash E}$ does not satisfy~(\ref{non-ch}).
Then there exist two tasks $i$ and $j$ in $E$ such that
$e'_i \!+\! p_i \!\leqslant\! e'_j \!+\! p_j \!<\!(e'_i \!+\! p_i) \!+\! p_j$.
Three cases have to be considered.

$\enskip \rightarrow$ If $e'_j \!+\! p_j \!\geqslant\! p(J)$,
then $e'_j \!+\! p_j \!\geqslant\! p(E)$.
Applying Lemma~\ref{lm_cle_2},
we can construct a vector $e'^-$ such that
$X^-\!=(e'^-\!,t',\delta,x,a,b,\gamma)$ is in $\PaetdxE$
and $\hab(e'^-\!,t',a,b) < \hab(e',t',a,b)$
since $\alpha \!\in\! (\RpeJ\!\!$,
which contradicts the minimality of $(e',t',a,b)$.

$\enskip \rightarrow$ If $e'_j \!+\! p_j \!=\! \dma$,
we can derive the same contradiction 
since $\dma \!\geqslant\! p(E)$ from inequality~(\ref{a3}).

$\enskip \rightarrow$ If $e'_j \!+\! p_j \!<\! p(J)$
and $e'_j \!+\! p_j \!<\! \dma$,
then applying Lemma~\ref{lm_cle_1} to $(e' \!+\! p)_{\slash E}$,
we can construct two vectors $e'^{+-}$ and $e'^{-+}$ such that
$X^{+-}\!=(e'^{+-}\!,t',\delta,x,a,b,\gamma)$ 
and  $X^{-+}\!=(e'^{-+}\!,t',\delta,x,a,b,\gamma)$ 
are in $\PaetdxE$ and that $X^*$ 
is the middle point of the segment $[X^{+-},X^{-+}]$.
which contradicts the extremality of $X^*$.

Similarly, let us assume that $t'_{\slash T}$ 
does not satisfy~(\ref{non-ch}).
Then there exist two tasks $i$ and $j$ in $T$ such that
$t'_i \!\leqslant\! t'_j \!<\!t'_i \!+\! p_j$.
Since $\forall k \!\in\! T(\delta),\, k \!\neq\! \jj,\,t'_k \geqslant t'_\jj \!\!+ p_k$,
we deduce that $i \!\neq \jj$.
Then for tasks $i$ and $j$,
inequalities~(\ref{t1}') and~(\ref{g3}) are equivalent,
and $t'_i$ and $t'_j$ are only bounded from above by $p(J)$.
Then two cases have to be considered:

$\enskip \rightarrow$ If $t'_j \!\geqslant\! p(J)$,
then $t'_j \!\geqslant\! p(T)$.
Applying Lemma~\ref{lm_cle_2},
we can derive a contradiction to the minimality of $(e',t',a,b)$.

$\enskip \rightarrow$ If $t'_j \!<\! p(J)$,
Applying Lemma~\ref{lm_cle_1}
we can derive a contradiction to the extremality of $X^*$.

Finally, \Sopt is feasible.

The third step is to show that \Sopt
is a \dZblocSing.
We first prove that \Sopt is a \blocSing
using the same method as in the proof of Theorem~\ref{thm_etdx_correct}.
Assuming that \Sopt is not a \blocSing,
we construct a better schedule \Smieux 
by tightening tasks arround $d$.
Using Theorem~\ref{thm_aetdx_complet},
there exists $\widehat{X} \!\in\!  \intdg(\Paetdx)$
encoding \Smieux, and it contradicts he minimality of $(e',t',a,b)$.
\\
Thus \Sopt is a \blocSing.
Now we have to show that
\Sopt starts at time $0$ or holds an on-time task.
Let us assume that it is not the case,
then setting $\varepsilon \!=\!\frac{1}{2} \min(p_\jj \!\!-\! a,\, a,\, s)$
where $s$ denotes the starting time of \Sopt,
we have $\varepsilon \!>\! 0$. 
Setting $a^- \!\!=\! a \!-\! \varepsilon$
and $X^- \!=\! (e',t',\delta,x,a^-,a^-\delta,\gamma)$
(resp.  $a^+  \!\!=\! a \!+\! \varepsilon$ 
and $X^+ \!=\! (e',t',\delta,x,a^+,a^+\delta,\gamma)$),
$X^-$ (resp. $X^+$) encodes using $\theta'$ 
the schedule obtained by shifting backward (resp. forward)
by $\varepsilon$ time unit all the tasks.
By definition of $\varepsilon$, 
$X^-$ (resp. $X^+$) still satisfies 
inequalities~(\ref{a1}), (\ref{a2}), (\ref{a0}) and (\ref{a3}),
thus $X^-\!\!\in\! \Paetdx$ (resp. $X^+\!\!\in\! \Paetdx$).
Since $X^*$ is the middle of $[X^-,X^+]$,
that contradicts the extremality of $X^*$.

We deduce that $X^*$ encodes a \dZblocSing.

\end{document}

%% file: notations.tex
\renewcommand{\S}{$\mathcal{S\,}$\xspace}
\newcommand{\Sopt}{$\mathcal{S^*}$\xspace}
\newcommand{\Smieux}{$\widehat{\mathcal{S}}$\xspace}

\newcommand{\pc}[1]{p\!*\!y\,(#1)}
\newcommand{\pck}[1]{p\!*\!y\,(#1)}

\newcommand{\M}{p(J)}

\newcommand{\Rp}{\mathbb{R}_+}
\newcommand{\RpJ}{{(\Rp)}^J}
\newcommand{\Rpe}{\mathbb{R}_+^*}
\newcommand{\RpeJ}{{(\Rpe)}^J}

\newcommand{\extr}[1]{\text{extr}(#1)}
\newcommand{\extre}[1]{\text{extr}^*(#1)}

\newcommand{\K}{J^{<}}

\newcommand{\Q}{non-overlapping Queyranne's inequalities\xspace}

\DeclareMathOperator*{\argmin}{argmin}

\newcommand{\unn}{\llbracket 1 , n\rrbracket}

\newcommand{\Ec}{\mathsfit{E}}
\newcommand{\Tc}{\mathsfit{T}}
\newcommand{\Ed}{E}
\newcommand{\Td}{T}
\newcommand{\dma}{d \!-\! a}
\newcommand{\dmat}{d \!-\! \ati}
\newcommand{\ati}{\widetilde{a}}
\newcommand{\Et}{\widetilde{E}}
\newcommand{\Tt}{\widetilde{T}}
\newcommand{\Etc}{\mathsfit{\widetilde{E}}}
\newcommand{\Ttc}{\mathsfit{\widetilde{T}}}

\newcommand{\Pdx}{P^2}

\newcommand{\Petdx}{P^1}
\newcommand{\PetdxE}{\intd(\Petdx)}

\newcommand{\Paetdx}{P^3}
\newcommand{\PaetdxE}{\intdg(\Paetdx)}

\newcommand{\Pdelta}{P^{\,\delta^*}}
\newcommand{\PdeltaE}{P^{\,\delta^*\!\!,\,E}}
\newcommand{\PdeltaT}{P^{\,\delta^*\!\!,\,T}}

\newcommand{\Fetdx}{(F1)\xspace}
\newcommand{\FetdxLP}{(F1-LP)\xspace}

\newcommand{\Fetdxextr}{(F1-extr)\xspace}
\newcommand{\Fetdxint}{(F1-int)\xspace}
\newcommand{\Fdx}{(F2)\xspace}

\newcommand{\FdxLP}{(F2-LP)\xspace}

\newcommand{\Faetdx}{(F3)\xspace}

\newcommand{\Fbf}{(F$_\text{LO}$)\xspace}

\newcommand{\val}[1]{value#1}
\newcommand{\Fti}{(F$_\text{TI}$)\xspace}

\newcommand{\fab}{f_{\alpha,\beta}}
\newcommand{\gab}{g_{\alpha,\beta}}
\newcommand{\hab}{h_{\alpha,\beta}}

\newcommand{\intd}{\text{int}_\delta}
\newcommand{\intdg}{\text{int}_{\delta,\gamma}}

\newcommand{\jj}{{j_0}}
\newcommand{\ii}{{i_0}}
\newcommand{\jt}{{j_t}}
\newcommand{\js}{{j_s}}

\newcommand{\gamun}{\Gamma^{c^1\!,q^1}}
\newcommand{\gamde}{\Gamma^{c^2\!,q^2}}
\newcommand{\gamcq}{\Gamma^{\,c,q}}

\newcommand{\widebar}[1]{\overline{#1}}

%% file: dominances.tex
\newcommand{\blocSing}{block\xspace}
\newcommand{\blocPlur}{blocks\xspace}

\newcommand{\dSing}{$d$-schedule\xspace}
\newcommand{\dPlur}{$d$-schedules\xspace}
\newcommand{\dblocSing}{$d$-block\xspace}
\newcommand{\dblocPlur}{$d$-blocks\xspace}
\newcommand{\dZblocSing}{$d$-or-left-block\xspace}
\newcommand{\dZblocPlur}{$d$-or-left-blocks\xspace}



%% file: blocs_inegalites.tex

\newcommand{\Cd}{(\delta)}
\newcommand{\Cg}[1]{(\gamma.#1)}
\newcommand{\Cl}[1]{(l.#1)}
\newcommand{\Cr}[1]{(r.#1)}
\newcommand{\Cx}[1]{(x.#1)}
\newcommand{\Ce}[1]{(e.#1)}
\newcommand{\Ct}[1]{(t.#1)}
\newcommand{\Cep}[1]{(e'.#1)}
\newcommand{\Ctp}[1]{(t'.#1)}
\newcommand{\Ca}[1]{(a.#1)}
\renewcommand{\Cap}[1]{(a'.#1)}

\newcommand{\Se}{(S1)}
\newcommand{\St}{(S2)}
\newcommand{\Sex}{(S1)}
\newcommand{\Stx}{(S2)}
\newcommand{\Sep}{(\hat{S1})}
\newcommand{\Sepx}{(\hat{S1})}
\newcommand{\Stp}{(\hat{S2})}
\newcommand{\Stpx}{(\hat{S2})}
\newcommand{\Sepz}{(\hat{S3})}
\newcommand{\Sepzx}{(\hat{S3})}

\renewcommand{\arraystretch}{0.9}
\tikzstyle{traits}=[expli!40!white,densely dotted]

\newcommand{\x}{0}
\newcommand{\y}{0}
\small

\newcommand{\blocegros}{
\begin{tikzpicture}
\draw[traits] (\x,\y) rectangle (\x+4.4,\y+1.2);
\draw[traits] (\x+3.5,\y) -- (\x+3.5,\y+1.2);
\draw (\x+1.9,\y+0.6) node {
\begin{tabular}{@{}r@{ }l}
 $\forall j\!\in\! J,$&$ e_j\geqslant 0$\\
&$ e_j\leqslant p(J)  \delta_j$\\
\end{tabular}
};
\draw [expli](\x+3.95,\y+0.9) node {$\Ce{1}$};
\draw [expli](\x+3.95,\y+0.4) node {$\Ce{2}$};
\end{tikzpicture}
}

\newcommand{\bloce}{
\begin{tikzpicture}
\draw[traits] (\x,\y) rectangle (\x+4.3,\y+0.8);
\draw[traits] (\x+3.3,\y) -- (\x+3.3,\y+0.8);
\draw (\x+1.9,\y+0.4) node {
\begin{tabular}{@{}r@{ }l}
 $\forall j\!\in\! J,$&$ e_j\leqslant p(J)  \delta_j$\\
\end{tabular}
};
\draw [expli](\x+3.8,\y+0.4) node {$\Ce{1}$};
\end{tikzpicture}
}

\newcommand{\bloceprimegros}{
\begin{tikzpicture}
\draw[traits] (\x,\y) rectangle (\x+5.2,\y+1.2);
\draw[traits] (\x+4.2,\y) -- (\x+4.2,\y+1.2);
\draw (\x+2.3,\y+0.6) node {
\begin{tabular}{@{}r@{ }l}
 $\forall j\!\in\! J,$&$ e'_j\geqslant 0$\\[0.1cm]
&$ e'_j\leqslant (d\!-\!p_j)  \delta_j$\\
\end{tabular}
};
\draw [expli](\x+4.7,\y+0.9) node {$\Cep{1}$};
\draw [expli](\x+4.7,\y+0.35) node {$\Cep{2}$};
\end{tikzpicture}
}

\newcommand{\bloceprime}{
\begin{tikzpicture}
\draw[traits] (\x,\y) rectangle (\x+5.1,\y+0.8);
\draw[traits] (\x+4.1,\y) -- (\x+4.1,\y+0.8);
\draw (\x+2.2,\y+0.4) node {
\begin{tabular}{@{}r@{ }l}
 $\forall j\!\in\! J,$&$ e'_j\leqslant (d\!-\!p_j)  \delta_j$\\
\end{tabular}
};
\draw [expli](\x+4.6,\y+0.4) node {$\Cep{1}$};
\end{tikzpicture}
}


\newcommand{\bloctgros}{
\begin{tikzpicture}
\draw[traits] (\x,\y) rectangle (\x+5.4,\y+1.2);
\draw[traits] (\x+4.5,\y) -- (\x+4.5,\y+1.2);
\draw (\x+2.4,\y+0.6) node {
\begin{tabular}{@{}r@{ }l}
$\forall j\!\in\! J,$&$ t_j\geqslant 0$\\
&$ t_j\leqslant p(J)  (1\!-\!\delta_j)$\\
\end{tabular}
};
\draw [expli](\x+4.95,\y+0.9) node {$\Ct{1}$};
\draw [expli](\x+4.95,\y+0.4) node {$\Ct{2}$};
\end{tikzpicture}
}

\newcommand{\bloct}{
\begin{tikzpicture}
\draw[traits] (\x,\y) rectangle (\x+5.1,\y+0.8);
\draw[traits] (\x+4.2,\y) -- (\x+4.2,\y+0.8);
\draw (\x+2.3,\y+0.4) node {
\begin{tabular}{@{}r@{ }l}
$\forall j\!\in\! J,$
&$ t_j\leqslant p(J)  (1\!-\!\delta_j)$\\
\end{tabular}
};
\draw [expli](\x+4.7,\y+0.4) node {$\Ct{1}$};
\end{tikzpicture}
}

\newcommand{\bloctprimegros}{
\begin{tikzpicture}
\draw[traits] (\x,\y) rectangle (\x+5.2,\y+1.2);
\draw[traits] (\x+4.2,\y) -- (\x+4.2,\y+1.2);
\draw (\x+2.3,\y+0.6) node {
\begin{tabular}{@{}r@{ }l}
$\forall j\!\in\! J,$&$ t'_j\geqslant 0$\\[0.1cm]
&$ t'_j\leqslant p(J)  (1\!-\!\delta_j)$\\
\end{tabular}
};
\draw [expli](\x+4.7,\y+0.9) node {$\Ctp{1}$};
\draw [expli](\x+4.7,\y+0.35) node {$\Ctp{2}$};
\end{tikzpicture}
}

\newcommand{\bloctprime}{
\begin{tikzpicture}
\draw[traits] (\x,\y) rectangle (\x+5.1,\y+0.8);
\draw[traits] (\x+4.1,\y) -- (\x+4.1,\y+0.8);
\draw (\x+2.2,\y+0.4) node {
\begin{tabular}{@{}r@{ }l}
$\forall j\!\in\! J,$&$ t'_j\leqslant p(J) (1\!-\!\delta_j)$\\
\end{tabular}
};
\draw [expli](\x+4.6,\y+0.4) node {$\Ctp{1}$};
\end{tikzpicture}
}

\newcommand{\blocdelta}{
\begin{tikzpicture}
\draw[traits] (\x,\y) rectangle (\x+4.3,\y+0.8);
\draw[traits] (\x+3.6,\y) -- (\x+3.6,\y+0.8);
\draw (\x+1.9,\y+0.4) node {
$\forall j\!\in\! J,\enskip 0\leqslant\!\delta_j \!\leqslant 1$
};
\draw [expli](\x+3.95,\y+0.4) node {$\Cd$};
\end{tikzpicture}
}

\newcommand{\blocgamma}{
\begin{tikzpicture}
\draw[traits] (\x,\y) rectangle (\x+7,\y+3.1);
\draw[traits] (\x+6,\y) -- (\x+6,\y+3.1);
\draw (\x+3.2,\y+1.5) node {
\begin{tabular}{@{}l@{ }l}
$\forall j\!\in\! J,$& {$\gamma_j \geqslant 0$}\\
& {$\gamma_j \leqslant 1$}\\[0.1cm]
& {$\delta_j \leqslant 1 \!-\! \gamma_j$}\\[0.1cm]
& {$t'_j \leqslant p_j + (1 \!-\! \gamma_j) (p(J)  \!-\! p_j)$}\\[0.25cm]
\multicolumn{2}{@{}l}{$\sum\limits_{j\in J}\gamma_j \!=\!1$}\\
\end{tabular}
};
\draw [expli](\x+6.5,\y+2.8) node {$\Cg{0}$};
\draw [expli](\x+6.5,\y+2.35) node {$\Cg{1}$};
\draw [expli](\x+6.5,\y+1.8) node {$\Cg{2}$};
\draw [expli](\x+6.5,\y+1.27) node {$\Cg{3}$};
\draw [expli](\x+6.5,\y+0.65) node {$\Cg{4}$};
\end{tikzpicture}
}

\newcommand{\blocl}{
\begin{tikzpicture}
\draw[traits] (\x,\y) rectangle (\x+6.4,\y+2);
\draw[traits] (\x+5.5,\y) -- (\x+5.5,\y+2);
\draw (\x+2.9,\y+1) node {
\begin{tabular}{@{}r@{ }l}
$\forall (i,j)\hspace{-0.1cm}\in\hspace{-0.1cm}\K,$
&$ l_{i,j} \geqslant 0$ \\
&$l_{i,j} \leqslant \delta_i$\\
&$l_{i,j} \leqslant \delta_j$\\
&$l_{i,j} \geqslant (\delta_i \!+\! \delta_j) -1$\\
\end{tabular}
 };
\draw [expli](\x+5.95,\y+1.65) node {$\Cl{1}$};
\draw [expli](\x+5.95,\y+1.2) node {$\Cl{2}$};
\draw [expli](\x+5.95,\y+0.75) node {$\Cl{3}$};
\draw [expli](\x+5.95,\y+0.3) node {$\Cl{4}$};
\end{tikzpicture}
}

\newcommand{\blocr}{
\begin{tikzpicture}
\draw[traits] (\x,\y) rectangle (\x+6.4,\y+2);
\draw[traits] (\x+5.5,\y) -- (\x+5.5,\y+2);
\draw (\x+2.9,\y+1) node {
\begin{tabular}{@{}r@{ }l}
$\forall (i,j)\hspace{-0.1cm}\in\hspace{-0.1cm}\K,$&
$ r_{i,j} \geqslant 0$ \\
&$r_{i,j} \leqslant  (1\!-\!\delta_i)$\\
&$r_{i,j} \leqslant (1\!-\!\delta_j)$\\
&$r_{i,j} \geqslant 1- (\delta_i \!+\! \delta_j) $\\
\end{tabular}
 };
\draw [expli](\x+5.95,\y+1.65) node {$\Cr{1}$};
\draw [expli](\x+5.95,\y+1.2) node {$\Cr{2}$};
\draw [expli](\x+5.95,\y+0.75) node {$\Cr{3}$};
\draw [expli](\x+5.95,\y+0.3) node {$\Cr{4}$};
\end{tikzpicture}
}

\newcommand{\blocx}{
\begin{tikzpicture}
\draw[traits] (\x,\y) rectangle (\x+6.4,\y+2);
\draw[traits] (\x+5.4,\y) -- (\x+5.4,\y+2);
\draw (\x+2.9,\y+1) node {
\begin{tabular}{@{}r@{ }l}
$\forall (i,j)\hspace{-0.1cm}\in\hspace{-0.1cm}\K,$
&$x_{i,j} \geqslant \delta_i \!-\!\delta_j$ \\
&$x_{i,j} \geqslant \delta_j \!-\!\delta_i$ \\
&$x_{i,j} \leqslant \delta_i \!+\!\delta_j$ \\
&$x_{i,j} \leqslant  2- (\delta_i\!+\!\delta_j)$\\
\end{tabular}
 };
\draw [expli](\x+5.95,\y+1.65) node {$\Cx{1}$};
\draw [expli](\x+5.95,\y+1.2) node {$\Cx{2}$};
\draw [expli](\x+5.95,\y+0.75) node {$\Cx{3}$};
\draw [expli](\x+5.95,\y+0.3) node {$\Cx{4}$};
\end{tikzpicture}
}

\newcommand{\bloca}{
\begin{tikzpicture}
\draw[traits] (\x,\y) rectangle (\x+5.6,\y+2.6);
\draw[traits] (\x+4.6,\y) -- (\x+4.6,\y+2.6);
\draw (\x+2.4,\y+1.3) node {
\begin{tabular}{@{}l@{ }l}
\multicolumn{2}{@{}l}{$a \geqslant 0$}\\[0.12cm]
\multicolumn{2}{@{}l}{$\sum\limits_{j\in J}p_j \delta_j \leqslant d\!-\!a$}\\[0.45cm]
$\forall j\!\in\! J,$&$ a\leqslant d - e'_j - p_j\delta_j$\\[0.1cm]
&$ a \leqslant p_j \!+\! (1 \!-\! \gamma_j) d$\\
\end{tabular}
};
\draw [expli](\x+5.1,\y+2.25) node {$\Ca{0}$};
\draw [expli](\x+5.1,\y+1.65) node {$\Ca{1}$};
\draw [expli](\x+5.1,\y+0.8) node {$\Ca{2}$};
\draw [expli](\x+5.1,\y+0.3) node {$\Ca{3}$};
\end{tikzpicture}
}

\newcommand{\blocaprime}{
\begin{tikzpicture}
\draw[traits] (\x,\y) rectangle (\x+5.6,\y+2.3);
\draw[traits] (\x+4.6,\y) -- (\x+4.6,\y+2.3);
\draw (\x+2.3,\y+1.1) node {
\begin{tabular}{@{}r@{ }l}
$\forall j\hspace{-0.1cm}\in\hspace{-0.1cm}J,$
& $ a'_j \geqslant 0$ \\[0.1cm]
& $ a'_j \leqslant a$ \\[0.1cm]
& $ a'_j \leqslant \delta_j d$ \\[0.1cm]
& $ a'_j \geqslant a - (1\!-\!\delta_j) d$ 
\end{tabular}
 };
\draw [expli](\x+5.1,\y+1.9) node {$\Cap{0}$};
\draw [expli](\x+5.1,\y+1.4) node {$\Cap{1}$};
\draw [expli](\x+5.1,\y+0.9) node {$\Cap{2}$};
\draw [expli](\x+5.1,\y+0.4) node {$\Cap{3}$};
\end{tikzpicture}
}

\newcommand{\blocS}{
\begin{tikzpicture}
\draw[traits] (\x,\y) rectangle (\x+9.1,\y+2.2);
\draw[traits] (\x+8.2,\y) -- (\x+8.2,\y+2.2);
\draw (\x+4.3,\y+1.1) node {
\begin{tabular}{@{}r@{ }l}
$\forall S\! \in\! \mathcal{P}^*(J),$&$ \sum\limits_{i\in S} p_i e_i \geqslant 
\hspace{-0.4cm} \sum\limits_{(i,j)\in S^<}\hspace{-0.3cm}  p_i p_j l_{i,j}$\\[0.7cm]
&$ \sum\limits_{i\in S} p_i t_i \geqslant 
\hspace{-0.4cm} \sum\limits_{(i,j)\in S^<}\hspace{-0.3cm}  p_i p_j r_{i,j} +  \sum\limits_{i\in S} p_i^2 (1\!-\!\delta_i) $\\
\end{tabular}
 };
\draw [expli](\x+8.65,\y+1.8) node {\Se};
\draw [expli](\x+8.65,\y+0.8) node {\St};
\end{tikzpicture}
}

\newcommand{\blocSp}{
\begin{tikzpicture}
\draw[traits] (\x,\y) rectangle (\x+9.1,\y+2.2);
\draw[traits] (\x+8.2,\y) -- (\x+8.2,\y+2.2);
\draw (\x+4.3,\y+1.1) node {
\begin{tabular}{@{}r@{ }l}
$\forall S\! \in\! \mathcal{P}^*(J),$&$ \sum\limits_{i\in S} p_i e'_i \geqslant 
\hspace{-0.4cm} \sum\limits_{(i,j)\in S^<}\hspace{-0.3cm}  p_i p_j l_{i,j}$\\[0.7cm]
&$ \sum\limits_{i\in S} p_i t'_i \geqslant 
\hspace{-0.4cm} \sum\limits_{(i,j)\in S^<}\hspace{-0.3cm}  p_i p_j r_{i,j} +  \sum\limits_{i\in S} p_i^2 (1\!-\!\delta_i) $\\
\end{tabular}
 };
\draw [expli](\x+8.65,\y+1.8) node {\Se};
\draw [expli](\x+8.65,\y+0.8) node {\St};
\end{tikzpicture}
}

\newcommand{\blocSx}{
\begin{tikzpicture}
\draw[traits] (\x,\y) rectangle (\x+10.1,\y+2.2);
\draw[traits] (\x+9.2,\y) -- (\x+9.2,\y+2.2);
\draw (\x+5,\y+1.1) node {
\begin{tabular}{@{}r@{ }l}
$\forall S\! \in\! \mathcal{P}^*(J),$&$ \sum\limits_{i\in S} p_i e_i \geqslant 
\hspace{-0.4cm} \sum\limits_{(i,j)\in S^<}\hspace{-0.3cm}  p_i p_j \frac{\delta_i\!+\!\delta_j - x_{i,j}}{2}$\\[0.5cm]
&$ \sum\limits_{i\in S} p_i t_i \geqslant 
\hspace{-0.4cm} \sum\limits_{(i,j)\in S^<}\hspace{-0.3cm}  p_i p_j \frac{2-(\delta_i\!+\!\delta_j) - x_{i,j}}{2} +  \sum\limits_{i\in S} p_i^2 (1\!-\!\delta_i) $\\
\end{tabular}
 };
\draw [expli](\x+9.65,\y+1.7) node {\Sex};
\draw [expli](\x+9.65,\y+0.3) node {\Stx};
\end{tikzpicture}
}

\newcommand{\blocSxp}{
\begin{tikzpicture}
\draw[traits] (\x,\y) rectangle (\x+10.1,\y+2.2);
\draw[traits] (\x+9.2,\y) -- (\x+9.2,\y+2.2);
\draw (\x+5,\y+1.1) node {
\begin{tabular}{@{}r@{ }l}
$\forall S\! \in\! \mathcal{P}^*(J),$&$ \sum\limits_{i\in S} p_i e'_i \geqslant 
\hspace{-0.4cm} \sum\limits_{(i,j)\in S^<}\hspace{-0.3cm}  p_i p_j \frac{\delta_i\!+\!\delta_j - x_{i,j}}{2}$\\[0.5cm]
&$ \sum\limits_{i\in S} p_i t'_i \geqslant 
\hspace{-0.4cm} \sum\limits_{(i,j)\in S^<}\hspace{-0.3cm}  p_i p_j \frac{2-(\delta_i\!+\!\delta_j) - x_{i,j}}{2} +  \sum\limits_{i\in S} p_i^2 (1\!-\!\delta_i) $\\
\end{tabular}
 };
\draw [expli](\x+9.65,\y+1.7) node {\Sex};
\draw [expli](\x+9.65,\y+0.3) node {\Stx};
\end{tikzpicture}
}

\newcommand{\blocSa}{
\begin{tikzpicture}
\draw[traits] (\x,\y) rectangle (\x+9.5,\y+3.5);
\draw[traits] (\x+8.7,\y) -- (\x+8.7,\y+3.5);
\draw (\x+4.6,\y+1.7) node {
\begin{tabular}{@{}r@{ }l}
$\forall S\!\in\! \mathcal{P}^*\!(J),$
&$ \sum\limits_{i\in S} p_i e'_i \geqslant 
\hspace{-0.4cm} \sum\limits_{(i,j)\in S^<}\hspace{-0.3cm}  p_i p_j l_{i,j}$\\[0.7cm]
&$ \sum\limits_{i\in S} p_i t'_i \geqslant 
\hspace{-0.4cm} \sum\limits_{(i,j)\in S^<}\hspace{-0.3cm}  p_i p_j r_{i,j} +  \sum\limits_{i\in S} p_i^2 (1\!-\!\delta_i) $\\
&$ \sum\limits_{i\in S} p_i (d\!-\! a \!-\! e'_i \!+\!t'_i) \!\geqslant\!
 \frac{1}{2}\!\left[\!\big(\hspace{-0.1cm}\sum\limits _{j\in S}\! p_j\big)^2 \hspace{-0.1cm}+\hspace{-0.2cm} \sum\limits_{j\in S}\!p_j^2\right] $\\
\end{tabular}
 };
\draw [expli](\x+9.1,\y+3) node {$\Sep$};
\draw [expli](\x+9.1,\y+1.9) node {$\Stp$};
\draw [expli](\x+9.1,\y+0.8) node {$\Sepz$};
\end{tikzpicture}
}

\newcommand{\blocSax}{
\begin{tikzpicture}
\draw[traits] (\x,\y) rectangle (\x+10.1,\y+3.5);
\draw[traits] (\x+9.3,\y) -- (\x+9.3,\y+3.5);
\draw (\x+4.8,\y+1.7) node {
\begin{tabular}{@{}r@{ }l}
$\forall S\!\in\! \mathcal{P}^*\!(J),$
&$ \sum\limits_{i\in S} p_i e'_i \geqslant 
\hspace{-0.4cm} \sum\limits_{(i,j)\in S^<}\hspace{-0.3cm}  p_i p_j \frac{\delta_i\!+\!\delta_j - x_{i,j}}{2}$\\[0.7cm]
&$ \sum\limits_{i\in S} p_i t'_i \geqslant 
\hspace{-0.4cm} \sum\limits_{(i,j)\in S^<}\hspace{-0.3cm}  p_i p_j \frac{2-(\delta_i\!+\!\delta_j) - x_{i,j}}{2} \hspace{-0.1cm}+\hspace{-0.2cm}
 \sum\limits_{i\in S} p_i^2 (1\!-\!\delta_i) $\\
&$ \sum\limits_{i\in S} p_i (d\!-\! a \!-\! e'_i \!+\!t'_i) \!\geqslant\!
 \frac{1}{2}\!\left[\!\big(\hspace{-0.1cm}\sum\limits _{j\in S}\! p_j\big)^2 \hspace{-0.1cm}+\hspace{-0.2cm} \sum\limits_{j\in S}\!p_j^2\right] $\\
\end{tabular}
 };
\draw [expli](\x+9.7,\y+3) node {$\Sepx$};
\draw [expli](\x+9.7,\y+1.9) node {$\Stpx$};
\draw [expli](\x+9.7,\y+0.8) node {$\Sepzx$};
\end{tikzpicture}
}

%% file: figures_ordos.tex
\newcommand*{\tache}[3]{
	\draw[rounded corners] (#3-#2,+0.1) rectangle  (#3,0.6);
	\draw (#3-#2/2,+0.32) node {#1};
}
\newcommand*{\tacheH}[4]{
	\draw[rounded corners] (#3-#2,0.2) rectangle  (#3,0.1+#4);
	\draw (#3-#2/2,0.4) node {#1};
}

\newcommand*{\tacheExpli}[3]{
	\draw[rounded corners,expli] (#3-#2,+0.1) rectangle  (#3,0.6);
	\draw[expli](#3-#2/2,+0.32) node {#1};
}
\newcommand*{\tacheHExpli}[4]{
	\draw[rounded corners,expli] (#3-#2,0.1) rectangle  (#3,0.1+#4);
	\draw[expli] (#3-#2/2,0.32) node {#1};
}
\newcommand*{\tacheCout}[4]{
	\draw[rounded corners] (#3-#2,+0.1) rectangle  (#3,0.6);
	\draw (#3-#2/2,+0.32) node {#1};
	\draw[color=expli] (#3-0.3,-0.8) node[above=5pt] {$\downarrow$} node {#4};
}

\newcommand{\figEcarte}{
\begin{tikzpicture}[yscale=0.7]
\draw [->](-0.2,0)--(+5.4,0);
\fill [rounded corners,orange!50!white](2,0.1) rectangle (5,0.8);
\fill [rounded corners,orange](2.5,0.1) rectangle (5.5,0.8);
\draw [black, rounded corners] (2,0.1) rectangle (5,0.8);
\fill [pattern color=rose!60!white,pattern=dots, rounded corners] (0.5,0.3) rectangle (3.5,1);
\draw [rose!60!white, rounded corners] (0.5,0.3) rectangle (3.5,1);
\fill [thick, pattern color=rose,pattern= crosshatch dots, rounded corners] (0,0.3) rectangle (3,1);
\draw [rose,thick,rounded corners] (0,0.3) rectangle (3,1);
\draw [black, thick, rounded corners] (2.5,0.1) rectangle (5.5,0.8);
\draw[latex-|] (0,-0.2)--(0.5,-0.2) node [midway,below] {\small $\varepsilon / p_i$};
\draw[|-latex] (5,-0.2)--(5.5,-0.2) node [midway,below] {\small $\varepsilon / p_j$};
\draw[] (-0.5,0) node{(b)};
\end{tikzpicture}
}

\newcommand{\figResserre}{
\begin{tikzpicture}[yscale=0.7]
\draw [->](0.2,0)--(5.4,0);
\fill [rounded corners,orange!50!white](2,0.1) rectangle (5,0.8);
\draw [black, rounded corners] (2,0.1) rectangle (5,0.8);
\fill [pattern color=rose!60!white,pattern=dots, rounded corners] (0.5,0.3) rectangle (3.5,1);
\draw [rose!60!white, rounded corners] (0.5,0.3) rectangle (3.5,1);
\fill [rounded corners,orange](1.5,0.1) rectangle (4.5,0.8);
\draw [black, thick, rounded corners] (1.5,0.1) rectangle (4.5,0.8);
\fill [thick, pattern color=rose,pattern= crosshatch dots, rounded corners] (1,0.3) rectangle (4,1);
\draw [rose,thick,rounded corners] (1,0.3) rectangle (4,1);
\draw[|-latex] (0.5,-0.2)--(1,-0.2) node [midway,below] {\small $\varepsilon / p_i$};
\draw[latex-|] (4.5,-0.2)--(5,-0.2) node [midway,below] {\small $\varepsilon / p_j$};
\draw[] (-0.1,0) node{(a)};
\end{tikzpicture}
}

\newcommand{\figJ}{
\begin{tikzpicture}
\newcommand{\T}{8}
\draw [->](-0.2,0)--(\T+0.4,0);
\draw (0,0) node {$|$} node [below=6pt] {$0$};
\draw (\T,0) node {$|$} node [below=6pt] {$\M$};
\fill [rounded corners,pattern color=vert,pattern=north east lines](0.4,0.1) rectangle (1.2,0.6);
\draw [rounded corners](0.4,0.1) rectangle (1.2,0.6);
\fill [rounded corners,pattern color=vert,pattern=north east lines](2,0.1) rectangle (3.5,0.6);
\draw [rounded corners](2,0.1) rectangle (3.5,0.6);
\fill [rounded corners,pattern color=vert,pattern=north east lines](3.5,0.1) rectangle (4.5,0.6);
\draw [rounded corners](3.5,0.1) rectangle (4.5,0.6);
\fill [rounded corners,orange!80!white](4.5,0.1) rectangle (7.5,0.6);
\draw [rounded corners](4.5,0.1) rectangle (7.5,0.6);
\fill [rounded corners,orange!80!white](6.5,-0.1) rectangle (7.5,-0.6);
\draw [rounded corners](6.5,-0.1) rectangle (7.5,-0.6);
\draw [rounded corners,rose, thick](6.5,0.3) rectangle (8.5,0.8);
\draw [rounded corners,rose, thick](7,-0.25) rectangle (7.95,0.25);
\end{tikzpicture}
}

\newcommand{\figCEalpha}[7]{
\begin{tikzpicture}
\draw [->](-6.2,0)--(7+0.4,0);
\draw (-6,0) node[below=6pt]{$0$} node{$|$};
\draw (0,0) node[below=6pt]{$d$} node{$|$};
\foreach \x in {-6,-5,...,7}  \draw (\x,0) node  {$\shortmid$};
\tacheCout{$J_1$}{5}{#1}{#2}
\tacheCout{$J_2$}{3}{#3}{#4}
\tacheCout{$J_3$}{2}{#5}{#6}
\draw[color=expli] (9,0.2) node{$\rightarrow$ #7};
\end{tikzpicture}
}

\newcommand{\figCEdom}{
\begin{tikzpicture}[yscale=0.8]
\small
\draw(-3,0) node{$\mathcal{{S}}$};
\newcommand{\T}{11};
\draw [->](-2.2,0)--(\T+0.4,0);
\draw (0,0) node[below=10pt]{$d$} node{$|$};
\draw[<->] (-2,-0.3)--(0,-0.3) node [midway,below] {$a$};
\foreach \x in {-2,-1,...,\T}  \draw (\x,0) node  {$\shortmid$};
\tache{$J_7$}{3}{1}
\draw[color=expli] (0.5,-1) node[above=6pt] {$\downarrow$} node {$5*1$};
\tache{$J_8$}{4}{11}
\draw[color=expli] (10.5,-1) node[above=6pt] {$\downarrow$} node {$8*11$};
\foreach \x in {1,2,...,6}  \tache{$J_\x$}{1}{\x+1};
\draw [color=expli,decorate,decoration={brace,amplitude=10pt}](7,-0.2) -- (1,-0.2);
\draw[color=expli] (4,-1) node {$4*(2\!+\!3\!+\!4\!+\!5\!+\!6\!+\!7)$};
\draw[color=expli] (12.5,-1) node{$\rightarrow 201$};
\end{tikzpicture}
}

\newcommand{\figCEdompur}{
\begin{tikzpicture}[yscale=0.8]
\small
\newcommand{\T}{11};
\draw [->](-2.2,0)--(\T+0.4,0);
\draw (-0,0) --(-0,-0.4) node [below] {$d$};
\draw (-2,0) --(-2,-0.4) node [below] {$0$};
\draw[<->] (-1.95,-0.3)--(-0.05,-0.3) node [midway,below] {$a$};
\foreach \x in {-2,-1,...,\T}  \draw (\x,0) node  {$\shortmid$};
\tache{$J_7$}{3}{1}
\tache{$J_8$}{4}{11}
\foreach \x in {1,2,...,6}  \tache{$J_\x$}{1}{\x+1};
\end{tikzpicture}
}

\newcommand{\figCEdombis}{
\begin{tikzpicture}[yscale=0.8]
\small
\draw(-3,0) node{$\mathcal{\hat{S}}$};
\newcommand{\T}{11};
\draw [->](-2.2,0)--(\T+0.4,0);
\draw (0,0) node[below=10pt]{$d$} node{$|$};
\draw[<->] (-2,-0.3)--(0,-0.3) node [midway,below] {$a$};
\foreach \x in {-2,-1,...,\T}  \draw (\x,0) node  {$\shortmid$};
\tache{$J_7$}{3}{11}
\draw[color=expli] (1.5,-1) node[above=6pt] {$\downarrow$} node {$8*2$};
\tache{$J_8$}{4}{2}
\draw[color=expli] (10.5,-1) node[above=6pt] {$\downarrow$} node {$5*11$};
\foreach \x in {1,2,...,6}  \tache{$J_\x$}{1}{\x+2};
\draw [color=expli,decorate,decoration={brace,amplitude=10pt}](8,-0.2) -- (2,-0.2);
\draw[color=expli] (5,-1) node {$4*(3\!+\!4\!+\!5\!+\!6\!+\!7\!+\!8)$};
\draw[color=expli] (12.5,-1) node{$\rightarrow 203$};
\end{tikzpicture}
}

\newcommand{\figCEdombispur}{
\begin{tikzpicture}[yscale=0.8]
\small
\newcommand{\T}{11};
\draw [->](-2.2,0)--(\T+0.4,0);
\draw (-0,0) --(-0,-0.4) node [below] {$d$};
\draw (-2,0) --(-2,-0.4) node [below] {$0$};
\draw[<->] (-1.95,-0.3)--(-0.05,-0.3) node [midway,below] {$a$};
\foreach \x in {-2,-1,...,\T}  \draw (\x,0) node  {$\shortmid$};
\tache{$J_7$}{3}{11}
\tache{$J_8$}{4}{2}
\foreach \x in {1,2,...,6}  \tache{$J_\x$}{1}{\x+2};
\end{tikzpicture}
}

\newcommand{\figCEdomter}{
\begin{tikzpicture}[yscale=0.8]
\small
\draw(-3,0) node{$\mathcal{{S}}$};
\newcommand{\T}{9};
\draw [->](-2.2,0)--(\T+0.4,0);
\draw (0,0) node[below=10pt]{$d$} node{$|$};
\draw[<->] (-2,-0.3)--(0,-0.3) node [midway,below] {$a$};
\foreach \x in {-2,-1,...,\T}  \draw (\x,0) node  {$\shortmid$};
\tache{$J_5$}{3}{8}
\draw[color=expli] (1.5,-1) node[above=6pt] {$\downarrow$} node {$5\!*\!2$};
\tache{$J_4$}{4}{2}
\draw[color=expli] (7.5,-1) node[above=6pt] {$\downarrow$} node {$3\!*\!8$};
\foreach \x in {1,2,3}  \tache{$J_\x$}{1}{\x+2};
\draw [color=expli,decorate,decoration={brace,amplitude=10pt}](5,-0.2) -- (2,-0.2);
\draw[color=expli] (3.6,-1) node {$2*(3\!+\!4\!+\!5)$};
\draw[color=expli] (10,-1) node{$\rightarrow 58$};
\end{tikzpicture}
}

\newcommand{\figCEdomterpur}{
\begin{tikzpicture}[yscale=0.8]
\small
\newcommand{\T}{9};
\draw [->](-2.2,0)--(\T+0.4,0);
\draw (-0,0) --(-0,-0.4) node [below] {$d$};
\draw (-2,0) --(-2,-0.4) node [below] {$0$};
\draw[<->] (-1.95,-0.3)--(-0.05,-0.3) node [midway,below] {$a$};
\foreach \x in {-2,-1,...,\T}  \draw (\x,0) node  {$\shortmid$};
\tache{$J_5$}{3}{8}
\tache{$J_4$}{4}{2}
\foreach \x in {1,2,3}  \tache{$J_\x$}{1}{\x+2};
\end{tikzpicture}
}

\newcommand{\figCEdomquater}{
\begin{tikzpicture}[yscale=0.8]
\small
\draw(-3,0) node{$\mathcal{\hat{S}}$};
\newcommand{\T}{9};
\draw [->](-2.2,0)--(\T+0.4,0);
\draw (0,0) node[below=10pt]{$d$} node{$|$};
\draw[<->] (-2,-0.3)--(0,-0.3) node [midway,below] {$a$};
\foreach \x in {-2,-1,...,\T}  \draw (\x,0) node  {$\shortmid$};
\tache{$J_5$}{3}{1}
\draw[color=expli] (0.5,-1) node[above=6pt] {$\downarrow$} node {$3\!*\!1$};
\tache{$J_4$}{4}{8}
\draw[color=expli] (7.5,-1) node[above=6pt] {$\downarrow$} node {$5\!*\!8$};
\foreach \x in {1,2,3}  \tache{$J_\x$}{1}{\x+1};
\draw [color=expli,decorate,decoration={brace,amplitude=10pt}](4,-0.2) -- (1,-0.2);
\draw[color=expli] (2.5,-1) node {$2*(2\!+\!3\!+\!4)$};
\draw[color=expli] (10,-1) node{$\rightarrow 61$};
\end{tikzpicture}
}

\newcommand{\figCEdomquaterpur}{
\begin{tikzpicture}[yscale=0.8]
\small
\newcommand{\T}{9};
\draw [->](-2.2,0)--(\T+0.4,0);
\draw (-0,0) --(-0,-0.4) node [below] {$d$};
\draw (-2,0) --(-2,-0.4) node [below] {$0$};
\draw[<->] (-1.95,-0.3)--(-0.05,-0.3) node [midway,below] {$a$};
\foreach \x in {-2,-1,...,\T}  \draw (\x,0) node  {$\shortmid$};
\tache{$J_5$}{3}{1}
\tache{$J_4$}{4}{8}
\foreach \x in {1,2,3}  \tache{$J_\x$}{1}{\x+1};
\end{tikzpicture}
}

\newcommand{\figChevauchement}[1]{
\begin{tikzpicture}
\draw [->](0.2,0)--(6+0.4,0);
\tacheH{$J_i$}{3}{3.5}{0.6}
\draw (3.5,0)  node[below=15pt]{$#1_i$} node[below]{$|$} node{$|$};
\tache{$J_j$}{2}{5}
\draw (5,0) node[below=15pt]{$#1_j$} node[below]{$|$} node{$|$};
\draw[latex-latex] (3,-0.2)--(5,-0.2) node [midway,below] {$p_j$};
\draw (7.3,0) node{ou };
\draw [->](8.2,0)--(14.4,0);
\tacheH{$J_i$}{3}{13.5}{0.6}
\draw (13.5,0)  node[below=15pt]{$#1_i$} node[below]{$|$} node{$|$};
\tache{$J_j$}{2}{11}
\draw (11,0) node[below=15pt]{$#1_j$} node[below]{$|$} node{$|$};
\draw[latex-latex] (10.5,-0.2)--(13.5,-0.2) node [midway,below] {$p_i$};
\end{tikzpicture}
}

\newcommand{\figastr}{
\begin{tikzpicture}[yscale=0.8,decoration={brace}]
\small
\draw [->](-5.4,0)--(6,0);
\draw (-5,-0.2) --(-5,0.8) node [above] {$0$};
\tache{$\js$}{4}{2.5}
\tache{$i$}{2}{-3}
\tache{}{1.5}{-1.5}
\tache{}{1.5}{5}
\tache{$j$}{1}{3.5}
\draw (0,-0.2) --(0,0.8) node [above] {$d$};
\draw[very thick,red](-1.5,-0.4) --(-1.5,0.8)node [above] {$d\!-\!a$};
\draw[<->,>=latex] (-1.5,0.8)--(0,0.8) node [midway,above] {$a$};
\draw[<-|,>=latex] (-3,-0.3)--(-1.55,-0.3) node [midway,below] {$e'_i$};
\draw[|->,>=latex] (-1.45,-0.3)--(3.5,-0.3) node [midway,below] {$t'_j$};
\end{tikzpicture}
}

\newcommand{\figaotun}{
\begin{tikzpicture}[yscale=0.8,,decoration={brace}]
\small
\draw [->](-5.4,0)--(6,0);
\draw (-5,-0.2) --(-5,0.8) node [above] {$0$};
\tache{$\jt$}{1.5}{0}
\tache{$i$}{2}{-3}
\tache{}{1.5}{-1.5}
\tache{}{1.5}{5}
\tache{$j$}{1}{3.5}
\tache{}{2.5}{2.5}
\draw (0,-0.2) --(0,0.8) node [above] {$d \!=\! d\!-\!a$};
\draw[<-|,>=latex] (-3,-0.3)--(-0.05,-0.3) node [midway,below] {$e'_i$};
\draw[|->,>=latex] (+0.05,-0.3)--(3.5,-0.3) node [midway,below] {$t'_j$};
\end{tikzpicture}
}

\newcommand{\figaotdeux}{
\begin{tikzpicture}[yscale=0.8,decoration={brace}]
\small
\draw [->](-5.4,0)--(6,0);
\draw (-5,-0.2) --(-5,0.8) node [above] {$0$};
\tache{$\jt$}{1.5}{0}
\tache{$i$}{2}{-3}
\tache{}{1.5}{-1.5}
\tache{}{1.5}{5}
\tache{$j$}{1}{3.5}
\tache{}{2.5}{2.5}
\draw (0,-0.2) --(0,0.8) node [above] {$d$};
\draw[very thick,red](-1.5,-0.4) --(-1.5,0.8)node [above] {$d\!-\!a$};
\draw[<->,>=latex] (-1.5,0.8)--(0,0.8) node [midway,above] {$a$};
\draw[<-|,>=latex] (-3,-0.3)--(-1.55,-0.3) node [midway,below] {$e'_i$};
\draw[|->,>=latex] (-1.45,-0.3)--(3.5,-0.3) node [midway,below] {$t'_j$};
\end{tikzpicture}
}

\newcommand{\figaotdeuxbis}{
\begin{tikzpicture}[yscale=0.8,decoration={brace}]
\small
\draw [->](-5.4,0)--(6,0);
\draw (-5,-0.2) --(-5,0.8) node [above] {$0$};
\tache{$\jt$}{1.5}{0}
\tache{$i$}{2}{-3}
\tache{}{1.5}{-1.5}
\draw (0,-0.2) --(0,0.8) node [above] {$d$};
\draw[very thick,red](-1.5,-0.4) --(-1.5,0.8)node [above] {$d\!-\!a$};
\draw[<->,>=latex] (-1.5,0.8)--(0,0.8) node [midway,above] {$a$};
\draw[<-|,>=latex] (-3,-0.3)--(-1.55,-0.3) node [midway,below] {$e'_i$};
\draw[|->,>=latex] (-1.45,-0.3)--(0,-0.3) node [midway,below] {$t'_\jt$};
\end{tikzpicture}
}

\newcommand{\figpba}{
\begin{tikzpicture}[yscale=0.8,decoration={brace}]
\small
\draw [->](-7.4,0)--(6,0);
\draw (-7,-0.2) --(-7,0.8) node [above] {$0$};
\foreach \x in {-6,-5,...,5}  \draw (\x,0) node  {$\shortmid$};
\draw[rounded corners] (-2,0.1) rectangle  (0,0.7);
\draw (-1,0.4) node {$i$};
\draw[rounded corners] (-3, -0.4) rectangle  (0,0.2);
\draw (-1.5, -0.1) node {$j$};
\draw[rounded corners] (-2, -0.9) rectangle  (0,-0.3);
\draw (-1, -0.6) node {$k$};
\tache{}{1.5}{5}
\tache{}{1}{3.5}
\tache{}{2.5}{2.5}
\draw (0,-1.1) --(0,0.9) node [above] {$d$};
\draw[very thick,red](-7,-1.1) --(-7,0.9)node [above=0.3cm] {$d\!-\!a$};
\draw[<->,>=latex] (-7,0.8)--(0,0.8) node [midway,above] {$a$};
\draw[|->,>=latex] (-6.95,-1)--(0,-1) node [midway,below] {$t'_j \!=\! t'_i \!=\! t'_k$};
\end{tikzpicture}
}

%% file: figures_autres.tex
%
%

\newcommand{\figConesBorne}{
\begin{tikzpicture}[scale=0.6]
\definecolor{c}{RGB}{224, 17, 95}
\definecolor{c_fonce}{RGB}{224, 17, 95}
\definecolor{cc}{RGB}{78, 179, 211}
\definecolor{cc_fonce}{RGB}{77, 150, 168}
\definecolor{conv_clair}{RGB}{165, 209, 82}
\definecolor{conv_fonce}{RGB}{121, 166, 45}
\newcommand{\p}{1.5};
\newcommand{\pp}{2.5};
\renewcommand{\l}{1.4}
\newcommand{\h}{1}
\renewcommand{\M}{10}
\newcommand{\m}{1.2}
\newcommand{\n}{0.4}

\draw[->] (0,-0.3) node[below] {0} -- (0,\M+\m) node[left] {$C_2$};
\draw[->] (-0.3,0) node[left] {0}  -- (\M+\m,0 )node[below] {$C_1$};

\draw[conv_fonce,very thick,dashed] (\p,\M+\n)-- (\p,\M+\m);
\draw[conv_fonce,very thick] (\p,\M+\n)-- (\p,\p+\pp)
	-- (\p+\pp,\pp)
	-- (\M+\n,\pp);
\draw[conv_fonce,very thick,dashed] (\M+\n,\pp)-- (\M+\m,\pp);

\draw (0,\pp) node {\_} node [left=3pt] {$p_2$};
\draw[dotted, cc]  (0,\pp) 
	-- (\p+\pp,\pp) node [cc_fonce] {$\bullet$}
	-- (\p+\pp,0);
\draw (\p+\pp,0) node {$\shortmid $} node [below=3pt] {$p_1$+$p_2$};

\draw (\p,0) node {$\shortmid $} node [below=3pt] {$p_1$};
\draw[dotted, c]  (\p,0) 
	-- (\p,\pp+\p) node [c] {$\bullet$}
	-- (0,\p+\pp);
\draw (0,\p+\pp) node {\_} node [left=3pt] {$p_1$+$p_2$};

\fill[pattern color=c!25!white,pattern= vertical lines](\p+0.1,\p+\pp+0.1)--(\p+0.1,\M+\m)--(\M+\m-\pp,\M+\m);
\draw[c,dashed] (\p+0.1,\M+\n)-- (\p+0.1,\M+\m);
\draw[c] (\M+\n-\pp,\M+\n)-- (\p+0.1,\p+\pp+0.1) -- (\p+0.1,\M+\n);
\draw[c,dashed] (\M+\n-\pp,\M+\n)-- (\M+\m-\pp,\M+\m);

\fill[pattern color=cc!25!white,pattern= horizontal lines](\p+\pp+0.1,\pp+0.1)--(\M+\m,\pp+0.1)--(\M+\m,\M+\m-\p);
\draw[cc,dashed] (\M,\pp+0.1)-- (\M+\m,\pp+0.1);
\draw[cc] (\M+\n,\pp+0.1)-- (\p+\pp+0.1,\pp+0.1) -- (\M+\n,\M+\n-\p);
\draw[cc,dashed] (\M+\n,\M+\n-\p)-- (\M+\m,\M+\m-\p);

\draw[thick,c_fonce,->] (\p+0.1,\p+\pp+0.1) -- (\p+0.1,\p+\pp+1);
\draw[thick,c_fonce,->] (\p+0.1,\p+\pp+0.1) -- (\p+0.8,\p+\pp+0.8);
\draw[thick,cc_fonce,->] (\p+\pp+0.1,\pp+0.1) -- (\p+\pp+1,\pp+0.1) ;
\draw[thick,cc_fonce,->] (\p+\pp+0.1,\pp+0.1) -- (\p+\pp+0.8,\pp+0.8);

\draw[thick,color=gray](0,\M) node [left]{$M$}--(\M+\n,\M);
\draw[thick,color=gray](\M,0) node [below]{$M$}--(\M,\M+\n);
\draw[thick,color=gray](\M,\M) node{$\bullet$};

\draw[color=gray!40!white](\M+2.4,\M+1.5) rectangle (\M+14,0);
\draw (\M+3.8,\M+0.8) node {Legend:};
\renewcommand{\x}{\M+3.2}
\renewcommand{\y}{\M-1}
\fill[pattern color=c!25!white,pattern= vertical lines] (\x,\y) rectangle (\x+\l,\y+\h);
\draw[color=c,thick](\x,\y) rectangle (\x+\l,\y+\h);
\draw[black](\x+\l+4.8,\y+0.4*\h) node{
\begin{tabular}[t]{@{}p{5cm}@{}}
Cone of feasible schedules where 1 is executed before 2\\
\end{tabular}
};

\renewcommand{\y}{\M-3}
\fill[pattern color=cc!25!white,pattern= horizontal lines] (\x,\y) rectangle (\x+\l,\y+\h);
\draw[color=cc,thick](\x,\y) rectangle (\x+\l,\y+\h);
\draw[black](\x+\l+4.8,\y+0.4*\h) node{
\begin{tabular}[t]{@{}p{5cm}@{}}
Cone of feasible schedules where 2 is executed before 1
\end{tabular}
};
\renewcommand{\y}{\M-5}
\fill[pattern color=cc!25!white,pattern= horizontal lines] (\x,\y)-- (\x+\l,\y) -- (\x+\l,\y+\h);
\draw[color=cc,thick](\x,\y)-- (\x+\l,\y) -- (\x+\l,\y+\h);
\fill[pattern color=c!25!white,pattern= vertical lines] (\x,\y)-- (\x,\y+\h) -- (\x+\l,\y+\h);
\draw[color=c,thick](\x,\y)-- (\x,\y+\h) -- (\x+\l,\y+\h);
\draw[white](\x,\y)-- (\x+\l,\y+\h);
\draw[black](\x+\l+1,\y+0.5*\h) node{Q};
\renewcommand{\y}{\M-7}
\fill[pattern color=cc!25!white,pattern= horizontal lines] (\x,\y)-- (\x+\l,\y) -- (\x+\l,\y+\h);
\fill[pattern color=c!25!white,pattern= vertical lines] (\x,\y)-- (\x,\y+\h) -- (\x+\l,\y+\h);
\draw[color=white,line width=6pt](\x,\y) -- (\x+\l,\y+\h);
\draw[color=conv_fonce,line width=2pt](\x,\y) rectangle (\x+\l,\y+\h);
\draw[black](\x+\l+1.8,\y+0.5*\h) node{conv(Q)};
\renewcommand{\y}{\M-9}
\draw(\x,\y) rectangle (\x+\l,\y+\h);
\draw[black](\x+\l+4.8,\y+0.5*\h) node{Area without feasible schedule};

\end{tikzpicture}
}


\newcommand{\figMiroir}{
\begin{tikzpicture}
\small
\newcommand{\T}{4.2};
\newcommand{\rT}{-4.7};
\draw [expli,](\rT,0)--(-0.2,0);
\draw [expli,->](0.2,0)--(\T+0.8,0);
\draw [densely dotted] (0,-0.9)--(0,2.2);
\draw (-0.2,0) node[below=6pt]{$d$} node{$|$};
\draw (0.2,0) node[below=6pt]{$0$} node{$|$};
\tache{$J_{i}$}{1.5}{-3}{0}
\draw (-3,0) node{$|$};
\draw[<->] (-3,-0.2)--(-0.2,-0.2);
\draw (-1.5,-0.5) node{$e_i$};
\draw[<->] (-3,-0.2)--(-4.5,-0.2);
\draw (-3.8,-0.5) node{$p_i$};
\tache{$J_{t}$}{2}{-0.2}{0}
\tache{$J_{t}$}{2}{2.2}{0}
\tache{$J_{i}$}{1.5}{4.5}{0}
\draw (4.5,0) node{$|$};
\draw[<->] (0.2,-0.2)--(4.5,-0.2);
\draw (2.8,-0.6) node{$C_i=p_i\text{+}e_i$};
\draw[dashed, >=latex, ->] (-3.8,0.7) to[bend left=30] (3.8,0.7);
\draw[dashed, >=latex, ->] (-1.2,0.7) to[bend left=90] (1.2,0.7);
\end{tikzpicture}
}

\newcommand{\figGraphSepa}{
\begin{tikzpicture}[scale=1.5]
\colorlet{c_m}{black!50!white}
\node[] (1) at (0,2){};
\node[] (2) at (1.5,1){};
\node[] (3) at (-1.5,1){};
\node[] (4) at (0.9,-0.5){};
\node[] (5) at (-0.9,-0.5){};
\foreach \k in {1,...,5}{\foreach \l in {1,...,\k}
\draw[c_m, line width=1.05pt] (\k.center) -- (\l.center);}
\draw [c_m](5) -- (4) node[midway,below]{$w_{i,j} \!=\! q_{i,j}$};

\definecolor{cn}{RGB}{255, 0, 127}
\node[] (6) at (3,0.8) {};
\draw[cn] (6) node[right] {$n\!+\!1$};
\foreach \k in {1,...,5}{\draw[color=cn] (6.center) -- (\k.center);}
\draw[cn] (4) -- (6) node[midway,below,sloped]{$w_{\{i,n+1\}} \!=\![k_i]^-$};
\draw[cn] (6) node {$\bullet$};

\definecolor{cz}{RGB}{255, 164, 20}

\node[] (0) at (-3,0.6){};
\draw[cz] (0) node[left] {$0$};
\foreach \k in {1,...,5}{\draw[color=cz] (0.center) -- (\k.center);}
\draw[cz] (0) -- (5) node[midway,below,sloped]{$w_{\{0,j\}} \!=\![k_j]^+$};
\draw[cz] (0) node {$\bullet$};

\foreach \k in {1,...,5}{\draw[c_m] (\k) node {$\bullet$};}

\draw [] (5) node[below] {$j$};
\draw [] (4) node[below] {$i$};
\end{tikzpicture}
}

%% file: tables_article.tex

\renewcommand{\l}{0.2cm}
\renewcommand{\ll}{0.2cm}

\newcommand{\modiftable}{\color{blue}    }
\newcommand{\vide}{ & & }
\newcommand{\bug}{x&x&x}

\newcommand{\tblCompBFunres}{
\begin{table}[h!]
\renewcommand{\arraystretch}{1.2}
\centering
\small
\begin{tabular}{c*{4}{p{3pt}@{}c@{\hspace{0.2cm}}c@{\hspace{0.2cm}}c@{}p{3pt}}}
\multirow{3}{*}{$n$} 
&& \multicolumn{3}{c}{\Fbf}
&&& \multicolumn{3}{c}{\Fti}
&&& \multicolumn{3}{c}{\Fetdx}
&&& \multicolumn{3}{c}{\Fdx}\\[0.1cm]
\cline{3-5}\cline{8-10}\cline{13-15}\cline{18-20}\\[-0.3cm]
&&  \#opt & avg-T & gap
&&& \#opt & avg-T & gap
&&& \#opt & avg-T & gap
&&& \#opt & avg-T & gap\\[0.1cm]
\hline\hline
10 && 10&9&-     &&& 10& 1&-  &&& 10&  0 &-    &&& 10&  3 &- & \\
\hline
20 &&  0&-&144\% &&& 10& 4&-  &&& 10&  2 &-    &&& 10&  3 &- \\
\hline
30 && \vide      &&& 10&15&-  &&& 10& 44 &-    &&& 10&  7 &- \\
\hline
40 && \vide      &&& 10&40&-  &&& 10& 637&-    &&& 10& 106&- \\
\hline
50 && \vide      &&& 10&41&-  &&& 1&1388&16\%  &&& 10&1315&- \\
\end{tabular}\\[-0.2cm]
\caption{Solving Biskup and Feldmann's unrestrictive instances 
using \Fbf, \Fti, \Fetdx and \Fdx}
\label{tab_comp_bf_unres}
\end{table}
}

\newcommand{\tblCompPiLongunres}{
\begin{table}[h!]
\renewcommand{\arraystretch}{1.2}
\centering
\small
\begin{tabular}{c@{\hspace{0.5cm}}c@{\hspace{0.6cm}}*{3}{p{4pt}@{}ccc@{}p{4pt}}}
\multirow{3}{*}{$p_{max}$} &
\multirow{3}{*}{$n$} 
&& \multicolumn{3}{c}{\Fti}
&&& \multicolumn{3}{c}{\Fetdx}
&&& \multicolumn{3}{c}{\Fdx}\\[0.1cm]
\cline{4-6}\cline{9-11}\cline{14-16}\\[-0.35cm]
&&& \#opt & avg-T & gap
&&& \#opt & avg-T & gap
&&& \#opt & avg-T & gap\\[0.1cm]
\hline\hline
100 &10 && 10& 6 &-    &&& 10&  0 &-    &&& 10&  3 &- &\\
    &20 && 10& 74&-    &&& 10&  3 &-    &&& 10&  3 &- \\ 
    &30 && 10&186&-    &&& 10& 68 &-    &&& 10& 13 &- \\
    &40 && 10&494&-    &&&  8&1335&4\%  &&& 10& 294&- \\
    &50 &&  8&690&0\%  &&&  0&  - &22\%  &&&  9&1743& 2\% \\[0.1cm]
\hline\hline
200 &10 && 10& 15 &-    &&& 10&  0 &-    &&& 10&  3 &- \\
    &20 && 10& 361&-    &&& 10&  3 &-    &&& 10&  3 &- \\
    &30 && 10& 886&-    &&& 10& 56 &-    &&& 10& 12 &- \\
    &40 &&  7&1322&0\%  &&&  6&1173&8\%  &&& 10& 359&- \\
    &50 &&  7&1289&2\%  &&&  1&1859&29\%  &&&  4& 1738&6\% \\[0.1cm]
\hline\hline
300 &10 && 10& 27 &-    &&& 10&  0 &-      &&& 10&  3 &- \\
    &20 && 10& 380&-    &&& 10&  4 &-      &&& 10&  3 &- \\
    &30 &&  6&1508&3\%  &&& 10& 87 &-      &&& 10& 15 &- \\
    &40 &&  8&2533&0\%  &&&  9&1572&11\%  &&& 10& 210&- \\
    &50 &&  \bug        &&&  0& -  &29\%  &&& 5&2662&5\% \\[0.1cm]
\end{tabular}\\[-0.2cm]
\caption{Solving unrestrictive instances generated with $p_{max}$ 100, 200, 300
using \Fti, \Fetdx and \Fdx}
\label{tab_comp_pi_long_unres}
\end{table}
}

\newcommand{\tblBF}{
\begin{table}[h!]
\renewcommand{\arraystretch}{1.1}
\centering
\small
\begin{tabular}{cc p{\l} cccc p{\l} p{\ll}cc}
\multirow{2}{*}{$n$} & 
\multirow{2}{*}{\#opt}&&
\multicolumn{4}{c}{ Optimally solved instances}
&&\multicolumn{3}{c}{ Not opt. solved instances}\\
\cline{4-7}\cline{9-11}
&&& avg-T & min-T & max-T &\nd 
&&&gap & \ndd \\
\hline\hline
10& 10  
&& 9 & 5 & 23 &11230 
&&&-&-\\
\hline
15& 5 
&&1821 & 255 & 3002 & 2762970 
&&&10\% & 5315709\\
\hline
20 & 0
&&-&-&-&- 
&&& 144\% & 1268881\\
\end{tabular}\\[-0.2cm]
\caption{Solving the unrestrictive case using \Fbf}
\label{tab_Fbf}
\end{table}
}

\newcommand{\tblETDX}{
\begin{table}[h!]
\renewcommand{\ll}{0.3cm}
\renewcommand{\arraystretch}{1.1}
\centering
\small
\begin{tabular}{cc p{\l} cccc p{\l} p{\ll}cc}
\multirow{2}{*}{$n$} & 
\multirow{2}{*}{\#opt}&&
\multicolumn{4}{c}{ Optimally solved instances}
&&\multicolumn{3}{c}{ Not opt. solved instances}\\
\cline{4-7}\cline{9-11}
&&& avg-T & min-T & max-T &\nd 
&&&gap & \ndd \\
\hline\hline
10& 10  
&& $<1$ & $<1$ & $<1$ & 4
&&&-&-\\
\hline
20& 10  
&& 5 & 4 & 5 & 105
&&&-&-\\
\hline
30& 10   
&& 50 & 34  & 96 & 690
&&&-&-\\
\hline
40& 10  
&& 477 & 226 & 706 & 3871
&&&-&-\\
\hline\hline
50& 9 
&& 4721 & 2626  & 9771  & 15375 
&&& 2\% & 33450 \\
\end{tabular}\\[-0.2cm]
\caption{Solving the unrestrictive case using \Fetdx}
\label{tab_Fetdx_F}
\end{table}
}

\newcommand{\tblDX}{
\begin{table}[h!]
\renewcommand{\ll}{0.4cm}
\renewcommand{\arraystretch}{1.1}
\centering
\small
\begin{tabular}{cc p{\l} cccc p{\l} p{\ll}cc}
\multirow{2}{*}{$n$} & 
\multirow{2}{*}{\#opt}&&
\multicolumn{4}{c}{ Optimally solved instances}
&&\multicolumn{3}{c}{ Not opt. solved instances}\\
\cline{4-7}\cline{9-11}
&&& avg-T & min-T & max-T &\nd 
&&&gap & \ndd \\
\hline\hline
10& 10  
&& 4 & 3 & 5 &  10
&&&-&-\\
\hline
20& 10  
&& 3 & 3 & 4 & 197
&&&-&-\\
\hline
30& 10   
&& 5 & 3 & 7 & 1380
&&&-&-\\
\hline
40& 10  
&& 22 & 11 & 43 & 8545
&&&-&-\\
\hline
50& 10 
&& 215 & 69 & 466 &  53210
&&&-&- \\
\hline
\hline
60& 10 
&& 4063 & 262 & 9543 &  455873
&&&-&-\\
\end{tabular}\\[-0.2cm]
\caption{Solving the unrestrictive case using \Fdx}
\label{tab_Fdx}
\vspace*{-0.2cm}
\end{table}
}


\newcommand{\tblCompBFres}{
\begin{table}[h!]
\renewcommand{\arraystretch}{1.2}
\centering
\small
\begin{tabular}{cc*{4}{p{4pt}@{}ccc@{}p{4pt}}}
\multirow{3}{*}{$n$} 
&\multirow{3}{*}{$h$} 
&& \multicolumn{3}{c}{\Fbf}
&&& \multicolumn{3}{c}{\Fti}
&&& \multicolumn{3}{c}{\Faetdx}\\[0.1cm]
\cline{4-6}\cline{9-11}\cline{14-16}\\[-0.3cm]
&&&  \#opt & avg-T & gap
&&& \#opt & avg-T & gap
&&& \#opt & avg-T & gap\\[0.1cm]
\hline\hline
10 & 0.2 && 10&1&-     &&& 10& 1&-  &&& 10&  0 &-  &\\
   & 0.4 && 10&1&-     &&& 10& 1&-  &&& 10&  1 &-  &\\
   & 0.6 && 10&1&-     &&& 10& 1&-  &&& 10&  1 &-  & \\
   & 0.8 && 10&1&-     &&& 10& 1&-  &&& 10&  1 &-  & \\[0.1cm]
\hline
20 & 0.2 &&  0&-&437\%  &&& 10& 3&-  &&& 10& 36 &-  &\\
   & 0.4 &&  0&-&245\%  &&& 10& 4&-  &&& 10& 116&-  &\\
   & 0.6 &&  0&-&159\%  &&& 10& 4&-  &&& 10& 125&-  & \\
   & 0.8 &&  0&-&145\%  &&& 10& 3&-  &&& 10& 118&-  & \\[0.1cm]
\hline
30 & 0.2 && \vide     &&& 10&17&-  &&& 10&1255& -   &\\
   & 0.4 && \vide     &&& 10&22&-  &&&  3&1620&6\%  &\\
   & 0.6 && \vide     &&& 10& 9&-  &&&  4& 962&8\%  &\\
   & 0.8 && \vide     &&& 10&13&-  &&&  5&1405&9\%  &\\[0.1cm]
\end{tabular}\\[-0.2cm]
\caption{Solving Biskup and Feldmann's restrictive instances 
using \Fbf, \Fti and \Faetdx}
\label{tab_comp_bf_res}
\end{table}
}

\newcommand{\tblCompPiLongres}{
\begin{table}[h!]
\renewcommand{\arraystretch}{1.2}
\centering
\small
\begin{tabular}{c@{\hspace{0.5cm}}c@{\hspace{0.5cm}}c@{\hspace{0.6cm}}*{2}{p{4pt}@{}ccc@{}p{4pt}}}
\multirow{3}{*}{$p_{max}$}
& \multirow{3}{*}{$n$} 
& \multirow{3}{*}{$h$} 
&& \multicolumn{3}{c}{\Fti}
&&& \multicolumn{3}{c}{\Faetdx}\\[0.1cm]
\cline{5-7}\cline{10-12}\\[-0.35cm]
&&&& \#opt & avg-T & gap
&&& \#opt & avg-T & gap\\[0.1cm]
\hline\hline
200 &10 &0.2 && 10& 22 &-    &&& 10&  1 &-  &\\
    &   &0.4 && 10& 24 &-    &&& 10&  1 &-  &\\
    &   &0.6 && 10& 15 &-    &&& 10&  1 &-  &\\
    &   &0.8 && 10& 14 &-    &&& 10&  1 &-  &\\[0.1cm]
\hline
200 &20 &0.2 && 10& 116&-     &&& 10& 30 &-  &\\
    &   &0.4 &&  9& 343&1\%   &&& 10& 91 &-  &\\
    &   &0.6 && 10& 299&-     &&& 10& 93 &-  &\\
    &   &0.8 && 10& 333&-     &&& 10& 89 &-  &\\[0.1cm]
\hline
200 &30 &0.2 &&  8& 821&2\%    &&& 10&1377&-  &\\
    &   &0.4 &&  7&1293&3\%    &&&  4&1143&4\%  &\\
    &   &0.6 && 10& 803&-      &&&  7&1479&5\%  &\\
    &   &0.8 && 10& 740&-      &&&  7&1166&6\%  &\\[0.1cm]
\hline
\end{tabular}\\[-0.2cm]
\caption{Solving restrictive instances generated with $p_{max}\!=\!200$
using \Fti and  \Faetdx }
\label{tab_comp_pi_long_res}
\end{table}
}

\newcommand{\tblAETDX}{
\begin{table}[h!]
\renewcommand{\arraystretch}{1.1}
\renewcommand{\ll}{0.2cm}
\centering
\small
\begin{tabular}{ccc p{\l} cccc p{\l} p{\ll}cc}
\multirow{2}{*}{$n$} & 
\multirow{2}{*}{$h$} & 
\multirow{2}{*}{\#opt}&&
\multicolumn{4}{c}{ Optimally solved instances}
&&\multicolumn{3}{c}{ Not opt. solved instances}\\
\cline{5-8}\cline{10-12}
&&&& avg-T & min-T & max-T &\nd 
&&&gap & \ndd \\
\hline\hline
10& 0.2& 10  
&& $<1$ & $<1$ & $<1$ &  107
&&&-&-\\
& 0.4& 10  
&& $<1$ & $<1$ & $<1$ &  278
&&&-&-\\
& 0.6& 10  
&& 1 & $<1$ & 2 &  289
&&&-&-\\
& 0.8& 10  
&& 1 & $<1$ & 2 &  272
&&&-&-\\
& 1.0& 10  
&& 2 & $<1$ & 10 &  253
&&&-&-\\
\hline
20& 0.2& 10  
&& 25 & 17 & 33 &  2919
&&&-&-\\
& 0.4& 10  
&& 73 & 28 & 145 &  8360
&&&-&-\\
& 0.6& 10  
&& 86 & 38 & 159 &  8957
&&&-&-\\
& 0.8& 10  
&& 74 & 28 & 144 &  7662
&&&-&-\\
& 1.0& 10  
&& 77 & 27 & 136 &  7851
&&&-&-\\
\hline
30& 0.2& 10  
&& 966 & 107 & 2142 & 35085  
&&&-&-\\
& 0.4& 3  
&& 1582 & 381 & 1514 & 62017  
&&& 7\% & 83820 \\
& 0.6& 7 
&& 1750 & 392 & 3494 & 56221
&&& 15\% & 44827\\
& 0.8& 5  
&& 1120 & 444 & 2929 & 46641
&&& 10\% & 74973\\
& 1.0& 6  
&& 1383 & 517 & 3372 & 48450
&&& 10\% & 71672\\
\end{tabular}\\
\caption{Solving the general case using \Faetdx}
\label{tab_Faetdx}
\end{table}
}

\newcommand{\tblBFgen}{
\begin{table}[h!]
\renewcommand{\arraystretch}{1.1}
\renewcommand{\ll}{0.3cm}
\centering
\small
\begin{tabular}{ccc p{\l} cccc p{\l} p{\ll}cc}
\multirow{2}{*}{$n$} & 
\multirow{2}{*}{$h$} & 
\multirow{2}{*}{\#opt}&&
\multicolumn{4}{c}{ Optimally solved instances}
&&\multicolumn{3}{c}{Not opt. solved instances}\\
\cline{5-8}\cline{10-12}
&&&& avg-T & min-T & max-T &\nd 
&&&gap & \ndd \\
\hline\hline
10& 0.2& 10  
&& 14 & 6 & 23 & 30790  
&&&-&-\\
& 0.4& 10  
&& 18 & 5 & 19 & 36011
&&&-&-\\
& 0.6& 10  
&& 10 & 4 & 18 & 12645  
&&&-&-\\
& 0.8& 10  
&& 8 & 4 & 17 & 10015  
&&&-&-\\
\hline
15& 0.2& 0  
&& - & - & - &  -
&&&  73\% & 3372585\\
& 0.4& 0  
&& - & - & - &  -
&&& 59\% &3562377\\
& 0.6& 5  
&& 2278 & 1422 & 3105 & 2722027  
&&& 24\% &4452452\\
& 0.8& 5
&& 1575 & 338 & 2281 & 2308746  
&&& 9\% & 5438115\\
\hline
20& 0.2& 0  
&& - & - & - &  -
&&& 437\% & $1162593$\\
& 0.4& 0  
&& - & - & - &  -
&&& 245\% & 1278346\\
& 0.6& 0  
&& - & - & - &  -
&&& 159\% & 1324055\\
& 0.8& 0  
&& - & - & - &  -
&&& 145\% & 1302895\\
\end{tabular}\\[-0.2cm]
\caption{Solving the general case using \Fbf}
\label{tab_Fbf_gen}
\end{table}
}


\newcommand{\tblDXLP}{
\begin{table}[h!]
\renewcommand{\arraystretch}{1.1}
\newcommand{\e}{0.2cm}
\centering
\small
\begin{tabular}{cp{\e}ccp{\e}ccp{\e}ccp{\e}p{0.4cm}c@{}c}
&& 
\multicolumn{2}{c}{\FdxLP}&&
\multicolumn{2}{c}{\FdxLP}&&
\multicolumn{2}{c}{\FdxLP}&&
\multicolumn{3}{c}{\FdxLP}\\[-0.1cm]
&& 
\multicolumn{2}{c}{}&&
\multicolumn{2}{c}{+ Cplex Cuts}&&
\multicolumn{2}{c}{+ Triangle}&&
\multicolumn{3}{c}{+ Triangle + Cplex Cuts}\\[-0.1cm]
$n$  &&time & gap && time & gap && time & gap &&&time & gap \\
\hline\hline
10&& 0.14 & 41.1\% && 2.72 & 0.00\%  && 0.05 & 3.29\% &&& 1.61 & 0.00\% \\
\hline
20&& 0.03 & 67.9\% && 3.19 & 0.00\%  && 0.52 & 13.2\% &&& 2.11 & 10.3\%  \\
\hline
30&& 0.12 & 77.0\% && 4.86 & 3.72\%  && 0.52 & 19.4\% &&& 11.7 & 18.1\% \\
\hline
40&& 0.29 & 82.9\% && 9.86 & 26.7\%  && 31.9 & 21.5\% &&& 48.3 & 20.9\%\\
\hline
50&& 0.62 &  86.1\% && 26.6 & 42.1\% && 177 & 22.5\% &&& 145 & 22.4\%\\
\hline
60&& 0.74 &  92.8\% && 375 & 44.9\%  && 746 & 23.5\% &&& 337  & 23.5\%\\
\end{tabular}\\[-0.2cm]
\caption{Improvement of the lower bound by adding Cplex cuts and triangle inequalities}
\label{tbl_DX-LP}
\end{table}
}

%% file: article_DAM_v_finale_joli.bbl
\begin{thebibliography}{10}

\bibitem{Baker_Scudder_90}
Kenneth~R. Baker and Gary~D. Scudder.
\newblock Sequencing with earliness and tardiness penalties: {A} review.
\newblock {\em Operations Research}, 38(1):22--36, 1990.

\bibitem{Balas}
Egon Balas.
\newblock On the facial structure of scheduling polyhedra.
\newblock {\em Mathematical Programming}, 24:179--218, 1985.

\bibitem{Barahona_Mahjoub_86}
Francisco Barahona and Ali~Ridha Mahjoub.
\newblock On the cut polytope.
\newblock {\em Mathematical Programming}, 36(2):157--173, 1986.

\bibitem{benchmark_url}
Dirk Biskup and Martin Feldmann.
\newblock \uppercase{ORLIB} common due date scheduling.
\newblock http://people.brunel.ac.uk/~mastjjb/jeb/orlib/schinfo.html, 1998.

\bibitem{benchmark}
Dirk Biskup and Martin Feldmann.
\newblock Benchmarks for scheduling on a single machine against restrictive and
  unrestrictive common due dates.
\newblock {\em Computers {\&} Opeartions Research}, 28(8):787--801, 2001.

\bibitem{lemon}
Coin-OR.
\newblock \uppercase{LEMON}, library for efficient modeling and optimization in
  networks.
\newblock http://lemon.cs.elte.hu/, 2003.

\bibitem{Correa_Schul_2005}
Jos{\'{e}}~R. Correa and Andreas~S. Schulz.
\newblock Single-machine scheduling with precedence constraints.
\newblock {\em Mathematics of Operations Research}, 30(4):1005--1021, 2005.

\bibitem{Dyer_Wolsey_90}
Martin~E. Dyer and Laurence~A. Wolsey.
\newblock Formulating the single machine sequencing problem with release dates
  as a mixed integer program.
\newblock {\em Discrete Applied Mathematics}, 26(2-3):255--270, 1990.

\bibitem{Fortet}
R.~Fortet.
\newblock L'alg\`ebre de \textsc{B}oole et ses applications en recherche
  op\'erationelle.
\newblock {\em Cahiers du Centre d'\'Etudes en Recherche Op\'erationnelle},
  4:5, 1959.

\bibitem{Gomory_et_Hu_61}
R.~E. Gomory and T.~C. Hu.
\newblock Multi-terminal network flows.
\newblock {\em Journal of the Society for Industrial and Applied Mathematics},
  9(4):551--570, 1961.

\bibitem{GLS}
Martin Gr{\"{o}}tschel, L{\'{a}}szl{\'{o}} Lov{\'{a}}sz, and Alexander
  Schrijver.
\newblock The ellipsoid method and its consequences in combinatorial
  optimization.
\newblock {\em Combinatorica}, 1(2):169--197, 1981.

\bibitem{Hall_Kubiak_et_Sethi}
Nicholas~G. Hall, Wieslaw Kubiak, and Suresh~P. Sethi.
\newblock Earliness-tardiness scheduling problems, {II:} deviation of
  completion times about a restrictive common due date.
\newblock {\em Operations Research}, 39(5):847--856, 1991.

\bibitem{Hall_et_Posner}
Nicholas~G. Hall and Marc~E. Posner.
\newblock Earliness-tardiness scheduling problems, {I:} weighted deviation of
  completion times about a common due date.
\newblock {\em Operations Research}, 39(5):836--846, 1991.

\bibitem{HVDV}
J.A. Hoogeveen and S.L. van~de Velde.
\newblock Scheduling around a small common due date.
\newblock {\em European Journal of Operational Research}, 55(2):237 -- 242,
  1991.

\bibitem{Kanet}
John~J. Kanet.
\newblock Minimizing the average deviation of job completion times about a
  common due date.
\newblock {\em Naval Research Logistics Quarterly}, 28:643--651, Dec 1981.

\bibitem{survey_2000}
John~J. Kanet and V.~Sridharan.
\newblock Scheduling with inserted idle time: Problem taxonomy and literature
  review.
\newblock {\em Operations Research}, 48(1):99--110, 2000.

\bibitem{survey_17}
Arthur Kramer and Anand Subramanian.
\newblock A unified heuristic and an annotated bibliography for a large class
  of earliness-tardiness scheduling problems.
\newblock {\em Journal of Scheduling}, online, 2017.

\bibitem{Padberg_89}
Manfred Padberg.
\newblock The boolean quadric polytope: Some characteristics, facets and
  relatives.
\newblock {\em Mathematical Programming}, 45(1-3):139--172, 1989.

\bibitem{min_cut}
Jean{-}Claude Picard and H.~Donald Ratliff.
\newblock Minimum cuts and related problems.
\newblock {\em Networks}, 5(4):357--370, 1975.

\bibitem{Queyranne}
Maurice Queyranne.
\newblock Structure of a simple scheduling polyhedron.
\newblock {\em Mathematical Programming}, 58:263--285, 1993.

\bibitem{Queyranne_Schulz_94}
Maurice Queyranne and Andreas~S. Schulz.
\newblock Polyhedral approaches to machine scheduling.
\newblock Technical Report 408, TU Berlin, 1994, revised 1996.

\bibitem{Queyranne_Wang_91}
Maurice Queyranne and Yaoguang Wang.
\newblock Single-machine scheduling polyhedra with precedence constraints.
\newblock {\em Mathematics of Operations Research}, 16(1):1--20, 1991.

\bibitem{Smith}
Wayne~E. Smith.
\newblock Various optimizers for single-stage production.
\newblock {\em Naval Research Logistics Quarterly}, 3(1‐2):59--66, 1956.

\bibitem{Sourd_09}
Francis Sourd.
\newblock New exact algorithms for one-machine earliness-tardiness scheduling.
\newblock {\em {INFORMS} Journal on Computing}, 21(1):167--175, 2009.

\bibitem{Tanaka_Araki_13}
Shunji Tanaka and Mituhiko Araki.
\newblock An exact algorithm for the single-machine total weighted tardiness
  problem with sequence-dependent setup times.
\newblock {\em Computers {\&} Opeartions Research}, 40(1):344--352, 2013.

\end{thebibliography}
